\documentclass[%
 aip,
 jcp,
 reprint,
 graphicx,
 amsmath,amssymb,
longbibliography, %disable this for citations without title!
nofootinbib %footinbib         % controls if footnote is with references or on the same page
]{revtex4-1}

\usepackage[%
                  bookmarks,%
                  bookmarksopen=false,% Klappt die Bookmarks in Acrobat aus
                  pdfauthor={Autor},%
                  pdftitle={Titel},%
                  colorlinks=true,%ture=kein kasten false=kasten
                  linkcolor=blue,%red green blue cyan magenta yellow black white
                  citecolor=red,%
                  urlcolor=cyan,%
                ]{hyperref}
                
\usepackage{refcount}
\usepackage{dcolumn}% Align table columns on decimal point
\usepackage{bm}% bold math
\usepackage{braket}
\usepackage[T1]{fontenc}
\usepackage[utf8]{inputenc}
\usepackage[english]{babel}
\usepackage{comment}
\usepackage{amssymb} %geschwungene Buchstaben \mathscr 
\usepackage{caption}
\usepackage{subcaption}

\usepackage{siunitx}
\DeclareSIUnit\hartree{\text{E}\ensuremath{_{\text{h}}}}
\DeclareSIUnit\Bau{\text{B}\ensuremath{_0}}
\DeclareSIUnit\gauss{\text{G}}

\usepackage{float}
\usepackage{booktabs}
\usepackage{multirow, makecell}
\usepackage{graphicx}
\newcommand{\abs}[1]{\lvert#1\rvert}
\usepackage{chemfig}
\usepackage{chemexec}
\usepackage{tikz}
\usetikzlibrary{positioning, calc}
\usepackage{pgfplots}
\pgfplotsset{compat=1.17}
\usepackage{derivative}
\usepackage[abs]{overpic}

\newcommand{\ped}[1]{\textormath{\textsubscript{#1}}{_{\mathrm{#1}}}} %This needs babel package!
\newcommand{\app}[1]{\textormath{\textsuperscript{#1}}{^{\mathrm{#1}}}}%This needs babel package!

\draft % marks overfull lines with a black rule on the right

\begin{document}

\preprint{AIP/?????}
\title[]{Unitary Coupled-Cluster theory for the treatment
of molecules in strong magnetic fields}

\author{Laura Grazioli}
 \email{laura.grazioli@enpc.fr}
 \affiliation{CERMICS, École nationale des ponts et chaussées, 6 et 8 avenue Blaise Pascal, Cité Descartes, 77455 Marne la Vallée Cedex 2, France}
  \affiliation{Department Chemie, Johannes Gutenberg-Universität Mainz, Duesbergweg 10-14, D-55128 Mainz, Germany}%
  \affiliation{Fachrichtung Chemie, Universit{\"a}t des Saarlandes, Campus B2 2, D-66123 Saarbr{\"u}cken, Germany \\}

\author{Marios-Petros Kitsaras}
 \email{kitsaras@irsamc.ups-tlse.fr}
 \affiliation{Laboratoire de Chimie et Physique Quantiques - UMR5626, CNRS, Université de Toulouse - Bat. 3R1b4 - 118 route de Narbonne, F-31062, Toulouse, France}
  \affiliation{Department Chemie, Johannes Gutenberg-Universität Mainz, Duesbergweg 10-14, D-55128 Mainz, Germany}%
  \affiliation{Fachrichtung Chemie, Universit{\"a}t des Saarlandes, Campus B2 2, D-66123 Saarbr{\"u}cken, Germany \\}
  
\author{Stella Stopkowicz}%
 \email{stella.stopkowicz@uni-saarland.de}
 \affiliation{Fachrichtung Chemie, Universit{\"a}t des Saarlandes, Campus B2 2, D-66123 Saarbr{\"u}cken, Germany \\}%
 \affiliation{Hylleraas Centre for Quantum Molecular Sciences, Department of Chemistry, University of Oslo, P.O. Box 1033, N-0315 Oslo, Norway}

\date{\today}
%\thanks{}

\begin{abstract}
In Coupled-Cluster (CC) theory, unphysical complex energies may arise in the presence of strong magnetic fields, near conical intersections, or in systems exhibiting complex Abelian point group symmetries. This issue originates from the non-Hermitian nature of the CC energy expression. A promising solution is provided by unitary Coupled-Cluster (UCC) theory, which retains the advantages of an exponential parameterization while ensuring real-valued energy eigenvalues. In this work, we present an implementation of finite-field second-order (ff-UCC2) and third-order (ff-UCC3) UCC theory. We assess the performance of these truncation levels in comparison to conventional finite-field CC methods, using the methylidyne  ion, water, and boric acid.
\end{abstract}


\maketitle %\maketitle must follow title, authors, abstract and \pacs
%%%%%%%%%%%%%%%%%%%%%%%%%%%%%%%%%%%%%%
\section{Introduction}\label{sec:introduction}
%%%%%%%%%%%%%%%%%%%%%%%%%%%%%%%%%%%%%%

The study of chemical systems in extreme conditions is an active field of interest in quantum chemistry.\cite{tyagi2017materials,goldman2019computational} One such  condition is the presence of strong magnetic fields. 
In this work, we focus on the so-called mixing regime, where the field-induced interactions are comparable to the Coulomb forces. In this regime, the magnetic field cannot be treated merely as a perturbation, and thus a non-perturbative approach is required. 
The development of finite-field approaches is therefore essential to describe such environments accurately. \cite{Hampe2017,Stopkowicz2017,holzer2019gw,Hampe2019,Hampe2020,hollands2023dz} Magnetic fields in the mixing regime can be found on magnetic white dwarf stars, where field strengths up to 100000 T have been observed.\cite{Franzon2015, Franzon2017,Bera2014,Boshkayev2014,Terada2008,Reimers1996} 
White dwarf stars are the endpoint of stellar evolution forming after the hydrogen reserves have been exhausted.\cite{Koester1990} In their atmospheres, a variety of elements has been detected, including hydrogen, helium, carbon, silicon, and various metals that typically originate from planetary debris.\cite{Greenstein1984,Henry1985,Dufour2007} As white dwarfs no longer generate energy via nuclear fusion, they cool over time, allowing the formation of molecules such as H$_2$, CH, and C$_2$ in their atmospheres.\cite{Berdyugina2007,Jordan2008,Xu2013} As approximately 97\% of all stars evolve into white dwarfs at the end of their lifetimes, these stellar remnants are highly common. Notably, around 25\% of white dwarfs exhibit strong magnetic fields.\cite{Landstreet2012,Bagnulo2020} The investigation of magnetic white-dwarf atmospheres is crucial for understanding the evolution of these stellar remnants. Since magnetic fields of comparable strength cannot be generated under laboratory conditions on Earth, theoretical predictions of atomic and molecular behavior in strong magnetic fields are essential for the interpretation and assignment of their observed spectra.

Systems in strong magnetic fields have been investigated since the 1980s, \cite{Rosner1984,Schmelcher1988,Schmelcher1988a,Schmelcher1988b,Schmelcher1991,Jordan1998} with a focus on a range of field strengths beyond the perturbative regime — particularly where magnetic interactions are comparable to, but do not yet dominate, other forces. These studies were primarily theoretical, culminating in a significant advancement with the spectral assignment of helium in a strongly magnetic white dwarf. \cite{Jordan1998} These theoretical finite-field studies were mostly conducted at the full configuration interaction (FCI) level of theory. However, the description of atoms and molecules in a strong magnetic field through FCI theory is feasible only for systems with few electrons.\cite{Becken1999,AlHujaj2004_Li, AlHujaj2004_Be} 
The prohibitive computational cost of FCI for larger systems necessitates the adoption of approximate electronic-structure methods. Early studies were often confined to systems that exhibit cylindrical symmetry. In general, the presence of a magnetic field introduces complex-valued wave-function parameters, requiring specialized implementations. For molecular systems, an additional challenge arises from the gauge-origin dependence of the Hamiltonian. While wave function methods yield gauge-origin independent results in the basis-set limit, the commonly adopted finite-basis set representation does not. A widely adopted solution to this problem is the use of \textit{gauge-including atomic orbitals} (GIAOs).\cite{London37,Hameka58,Ditchfield72,Wolinski90}
By now, non-linear systems as well as non-parallel orientations with respect to an external magnetic field have been investigated using various quantum-chemical methods, including  Hartree-Fock (HF) theory,\cite{Soncini2004,Tellgren2008} FCI,\cite{Lange2012} coupled-cluster (CC) theory\cite{Stopkowicz2015} and (current) density functional theory.\cite{Tellgren2014,Furness2015}
\\
High accuracy is essential for the interpretation of white dwarf spectra. Therefore, when FCI becomes computationally infeasible, finite-field extensions of standard CC theory\cite{cizek1980coupled,vcivzek1991origins,bartlett2007coupled} offer a practical alternative, balancing accuracy with lower computational cost.
The coupled-cluster (CC) wave function is parameterized using an exponential ansatz. In practice, truncation of the excitation space is necessary, leading to commonly used methods such as CC with Singles and Doubles (CCSD),\cite{Purvis1982} CC with Singles, Doubles, and Triples (CCSDT),\cite{Lee1984} perturbative approximations like CC2 and CC3,\cite{christiansen1995second,koch1997cc3} and non-iterative approaches such as the \textit{gold-standard} CCSD(T) method.\cite{stanton1997ccsd} For excited-state calculations, Equation-of-Motion CC (EOM-CC) theory is often employed.\cite{Stanton1993_eom} 
The aforementioned CC methods have also been extended to the finite-field regime.\cite{Stopkowicz2015,Hampe2017,Hampe2020,Kitsaras2024}
While CC methods are highly effective, they also have inherent limitations. Their standard formulation is non-Hermitian, and energies are obtained non-variationally through a projection procedure. As a result, complex energies can arise, for example, near conical intersections,\cite{Khn2007} and in the presence of finite magnetic fields.\cite{Stopkowicz2017,Thomas2021} Notably, ref.~\onlinecite{Thomas2021} shows that complex energies are the norm rather than the exception under magnetic fields, and only for atoms, linear molecules, and specific symmetries will the energies remain real.
Unlike electronic resonances, where non-Hermitian quantum mechanics provides a physical interpretation of the imaginary part of the energy as the lifetime of metastable states,\cite{Bravaya2013,Jagau2015,Jagau2017,Benda2018} here, the imaginary component lacks physical meaning and reflects a limitation of the theory. One potential solution to this issue is adopting an alternative wave-function parameterization that preserves Hermiticity. 
The unitary transformation of the Hamiltonian results in a Hermitian energy expression, ensuring real eigenvalues. Since any unitary operator can be written in exponential form, the connection to CC theory is, in principle, straightforward. This idea was first introduced by Kutzelnigg,\cite{kutzelnigg1977pair} with further developments by Bartlett and co-workers.\cite{Bartlett1989} More recently, the unitary CC approach has been explored in the context of quantum computing,\cite{Romero2018,Lee2018,Evangelista2019,Anand2022} including a recent theoretical extension to magnetic fields.\cite{Culpitt2023} However, this direction has so far not been pursued within conventional quantum-chemical frameworks in the finite-field regime.
In this context, the unitary operator is applied by expanding the exponential form. However, this introduces complications: the Baker–Campbell–Hausdorff (BCH) expansion of the transformed Hamiltonian results in a non-terminating infinite series. This contrasts with standard CC theory, where the similarity-transformed Hamiltonian expansion naturally self-truncates, yielding an exact expression within the chosen excitation manifold. Since no such self-truncation occurs for unitary CC parameterizations, an external criterion must be introduced to define an appropriate truncation.
To address the infinite series in the Baker–Campbell–Hausdorff expansion, different truncation strategies have been proposed. One such approach is the UCC($n$) formalism by Bartlett and co-workers,\cite{Bartlett1989} which relies on a perturbative truncation of the nested commutators. Another makes use of the Zassenhaus expansion;\cite{Taube2006} however, truncating this series generally compromises either variationality or size-extensivity. Recent developments\cite{liu2021unitary,Liu2022} have been based on the truncation after a given rank of commutators \cite{Bauman2022,Neuscamman2009} and seem to improve results for systems which do not have a smoothly-converging Møller-Plesset series at low orders. 
Furthermore, a scheme based on the perturbative truncation of the Bernoulli expansion, coined UCC\textit{n}, has been explored by Liu \textit{et al.}\cite{Liu2018} 
Further truncation schemes have also been described in the literature.\cite{liu2021unitary,Liu2022}
Recently, the UCC\textit{n} scheme has been shown to converge more rapidly toward the UCCSD limit and to yield more reliable results than the original UCC($n$) approach, particularly for molecular systems away from equilibrium geometries.\cite{UCC_deprince} Furthermore, a connection between UCC\textit{n} and the algebraic diagrammatic construction (ADC) scheme has been established,\cite{asthana2023quantum} and growing interest in this approximation for computing energies and properties has emerged in recent years.\cite{Hodecker2020,Thielen2021,Hodecker2022,Dreuw2023} In the present work, we extend the UCC\textit{n} approach to the finite-field regime. 
We investigate finite-field UCC2 and UCC3 for the description of atoms and molecules in strong magnetic fields, with a particular focus on assessing their performance relative to standard finite-field coupled-cluster theory. 

This manuscript is organized as follows. In section~\ref{sec:theory}, a brief overview of unitary coupled cluster (UCC) theory is provided, with a focus on the specific truncation scheme employed. Section~\ref{sec:implementation} outlines the implementation details, and section~\ref{sec:validation} describes the validation strategies used. In Section~\ref{sec:results}, we present ground- and excited-state energies as functions of the orientation and strength of an external magnetic field. The methylidynium ion, a relevant candidate for molecules in the atmospheres of strongly magnetized white dwarfs, serves as an astrophysically motivated example. Water provides a point of reference for comparison with studies on complex energies in finite magnetic fields,\cite{Thomas2021} while boric acid, with its complex Abelian point group, highlights the emergence of complex excitation energies within the EOM-CC framework.\cite{Kitsarasthes}
Finally, in section~\ref{sec:concl}, we summarize our conclusions and outline potential directions for future work.

\section{Theory}\label{sec:theory}
\subsection{Hamiltonian in a magnetic field}
In a uniform finite magnetic field, the molecular Hamiltonian is given as
\begin{equation}
    \hat{H}=\hat{H}_0+\frac{1}{2}\mathbf{B\cdot \hat{L}}_\text{O}+\mathbf{B\cdot \hat{S}}+\frac{1}{8}\sum_i^N(B^2r_{i\text{O}}^2-(\mathbf{B\cdot r}_{i\text{O}})^2),
    \label{eqn:magn_h}
\end{equation}
where $\hat{H}_0$ is the non-relativistic Hamiltonian in the field-free case. The sum runs over the number of electrons.
In order to ensure that observables remain gauge-origin independent also for approximate wave functions, the so-called \textit{gauge including atomic orbitals} (GIAOs) can be used. \cite{London37,Hameka58,Ditchfield72,Wolinski90}
Furthermore, the presence of the angular momentum $\mathbf{L}_\text{O}$ leads in general to a complex wave function. As already mentioned in the introduction, approximate non-Hermitian parameterizations of the wave function, like CC, often lead to complex energies. 
Therefore, a Hermitian formalism is needed in the setting of strong magnetic fields in order to ensure real eigenvalues.

\subsection{Unitary Coupled-Cluster Theory}\label{sec:ucc}
In unitary CC (UCC) theory, the ansatz for the ground state wave function is given by a unitary exponential operator acting on the reference state, usually the HF state $\ket{0}$
\begin{equation}
    \ket{\Psi\ped{UCC}}=e^{\tilde{\sigma}}\ket{0},
\end{equation}
where $\tilde{\sigma}=\hat{\sigma}-\hat{\sigma}^\dagger$ and
\begin{equation}
\begin{split}
    &\hat{\sigma}=\hat{\sigma}_1+\hat{\sigma}_2+\hat{\sigma}_3+\dots \\ &\hat{\sigma}_n=\frac{1}{(n!)^2}\sum \sigma^{abc\dots}_{ijk\dots}\{a^\dagger i b^\dagger j c^\dagger k\dots\}.
    \end{split}
\end{equation}
The indices $i,j,k, \dots$ and $a,b,c,\dots$ refer to occupied and virtual orbitals, respectively.
For a normalized wave function, the energy expectation value is given by
\begin{equation}
\begin{split}
    E&=\bra{\Psi\ped{UCC}}\hat{H}\ket{\Psi\ped{UCC}}\\
    &=\bra{0}e^{-\tilde{\sigma}}\hat{H}e^{\tilde{\sigma}}\ket{0}=\bra{0}\hat{\bar{H}}\ket{0},
    \end{split}
\end{equation}
where $\hat{\bar{H}}=e^{-\tilde{\sigma}}\hat{H}e^{\tilde{\sigma}}$ is the unitarily transformed Hamiltonian. This Hermitian form ensures the energies to be real. 

In analogy to standard CC theory, UCC theory is size extensive and the amplitude equations are obtained by projection onto 
excited determinants $\{\Phi_\mu\}$
\begin{align}
    \bra{\Phi_\mu}e^{-\tilde{\sigma}}\hat{H}e^{\tilde{\sigma}}\ket{0}=0.
\end{align}

However, unlike standard CC theory, the expansion of the transformed Hamiltonian is not self-truncating. The truncation criterion must therefore be chosen with care. Here, we follow the so-called \textit{Bernoulli expansion}, as described in ref.~\onlinecite{Liu2018}.  In the Bernoulli expansion, the Fock operator occurs in only one single commutator, making the equations more compact than the BCH expansion. 
Note that if not truncated, the two expansions are equivalent.

\subsection{EOM-UCC for excited states}
The description of excited states is obtained adapting the formalism of EOM-CC to the UCC framework. The most intuitive parameterization consists in applying an excitation operator on the UCC ground-state wave function, giving $\ket{\Psi_k}=\hat{R}e^{\tilde{\sigma}}\ket{0}$, with $\hat{R}$ defined 
as 
\begin{equation}
\hat{R}=\sum_{ia}r_i^a\{\hat{a}^\dagger \hat{i}\}+ \sum_{i<j,a<b}r_{ij}^{ab}\{\hat{a}^\dagger \hat{i} \hat{b}^\dagger \hat{j}\}+\dots
\end{equation}
Alternatively, the excitation operator $\hat{R}$ can be applied to the reference state, before the unitary transformation, yielding $\ket{\Psi_k}=e^{\tilde{\sigma}}\hat{R}\ket{0}$. Unlike for CC theory, for UCC these two formulations are not equivalent, as the exponential operator and the excitation operator $\hat{R}$ do not commute, i.e. $[e^{\tilde{\sigma}},\hat{R}]\neq 0$. 
In principle, both parameterizations could be used as a starting point. We note, however, that the so-called \textit{killer condition} 
\begin{equation}
    \hat{O}_k^\dagger\ket{\Psi_\text{GS}}=0 \qquad \forall k,
\end{equation}
needs to be fulfilled.\cite{asthana2023quantum,kim2023two}
In the above equation,
$\hat{O}_k^\dagger=\ket{\Psi_\text{GS}}\bra{\Psi_k}$ is a de-excitation operator to the ground-state and $k$ labels the excited states. Therefore, the killer condition $\ket{\Psi_\text{GS}}\braket{\Psi_k|\Psi_\text{GS}}=0 \;\; \forall k$ is satisfied if the excited states are orthogonal or biorthogonal to the ground state.

For UCC, the state $\bra{\Psi_k}$ is given by the adjoint of the state $\ket{\Psi_k}$.
For the first formulation of the EOM ansatz for UCC, i.e., $\ket{\Psi_k}=\hat{R} e^{\tilde{\sigma}}\ket{0}$ and $\bra{\Psi_k}=\bra{0}e^{-\tilde{\sigma}}\hat{R} $, the killer condition is not satisfied, as
\begin{equation}
\braket{\Psi_k|\Psi_\text{GS}}=\bra{0}e^{-\tilde{\sigma}}\hat{R}^\dagger e^{\tilde{\sigma}}\ket{0}.
\end{equation}
The overlap $\braket{\Psi_k|\Psi_\text{GS}}$ does not vanish in the general case, as $[\hat{R}, e^{\tilde{\sigma}}]\neq 0$.
As suggested in  ref.~\onlinecite{asthana2023quantum}, the killer condition is fulfilled by the ansatz $e^{\tilde{\sigma}}\hat{R}\ket{0}$. This is equivalent to defining a  similarity-transformed operator $\tilde{R}=e^{\tilde{\sigma}}\hat{R} e^{-\tilde{\sigma}}$ that acts on the UCC wave function:  
\begin{equation}
\begin{split}
    \ket{\Psi_k}&=\tilde{R} e^{\tilde{\sigma}}\ket{0}\\
    &=e^{\tilde{\sigma}}\hat{R} e^{-\tilde{\sigma}}e^{\tilde{\sigma}}\ket{0}=e^{\tilde{\sigma}}\hat{R} \ket{0}.
    \end{split}
\end{equation}
With this ansatz, the killer condition reads
\begin{equation}
\braket{\Psi_k|\Psi_\text{GS}}=\bra{0}e^{-\tilde{\sigma}} e^{\tilde{\sigma}}\hat{R}\ket{0}=\bra{0}\hat{R}\ket{0}=0.
\end{equation}
The Schrödinger equation for excited states therefore is
\begin{equation}
    \hat{H}\ket{\Psi_k}=E_k\ket{\Psi_k} \qquad \hat{H}e^{\tilde{\sigma}}\hat{R} \ket{0}=E_ke^{\tilde{\sigma}}\hat{R} \ket{0},
\end{equation}
and, by left-multiplying with $e^{-\tilde{\sigma}}$, we obtain the CI-like eigenvalue-problem
\begin{equation}
    \bar{H}\hat{R}\ket{0}=E_k \hat{R}\ket{0}.
\end{equation}
The excited states are found via diagonalization of the transformed Hamiltonian matrix.

UCC is characterized by its Hermitian formalism and, unlike for CC theory, the left eigenstates are simply parameterized by the adjoint operator $\hat{R}^\dagger$, as
\begin{equation}
    \bra{\Psi}=\bra{0}\hat{R}^\dagger e^{-\tilde{\sigma}}.\label{eqn:ucc_bra}
\end{equation}
The orthonormality condition for different UCC excited states reads
\begin{equation}
    \braket{\Psi_k|\Psi_l}=\bra{0}\hat{R}_k^\dagger \hat{R}_l\ket{0}=\delta_{kl}. \label{eqn:orthon_ucc}
\end{equation}

\subsection{The UCC\textit{n} methods}
In this work, the excitation space includes single and double excitations $\hat{\sigma}=\hat{\sigma}_1+\hat{\sigma}_2$, defining a UCC-analogon to the CCSD method, UCCSD. It has been shown that UCCSD recovers a similar amount of correlation energy as standard CCSD.\cite{Evangelista2011} The transformed Hamiltonian matrix has the following block structure
\begin{equation}
        \Bar{H}=\begin{pmatrix}
            \bar{H}_\text{SS} &\bar{H}_\text{SD}\\
            \bar{H}_\text{DS}&\bar{H}_\text{DD}
        \end{pmatrix},
\end{equation}
where S and D refer to single and double excitations, respectively.
As discussed in section \ref{sec:ucc}, the expansion of the similarity-transformed Hamiltonian matrix does not truncate. Within UCC$n$, the truncation scheme is designed on the basis of perturbation theory: 
The $\sigma_2$-amplitudes are of first order, while the $\sigma_1$-amplitudes are of second order in perturbation theory. 
Truncation is then performed at a given order $n$ in perturbation theory for the amplitude equations. Hence, in the UCC$n$ methods,  $n=2,3,\dots$ is the order at which the truncation is performed. In this work, we adopt the finite-field versions of the second- and third-order approximated methods, UCC2 and UCC3. \cite{Liu2018}

In the following, we present the UCC2 equations and refer the reader to ref.~\onlinecite{Liu2018} for the corresponding UCC3 expressions. 
\footnote{Note that in eq. 64 of ref.~\onlinecite{Liu2018}, the terms $f_{ai}+f_{ab}\sigma_i^b-f_{ij}\sigma_j^a$ and in eq. 65 the terms $\braket{ab||ij}+P(ab)f_{bc}\sigma_{ij}^{ac}-P(ij)f_{kj}\sigma_{ik}^{ab}$ are missing.}
For UCC2, the blocks of the Hamiltonian matrix are approximated as $H\ped{SS}^{(2)}$, $H\ped{SD}^{(1)}$, $H\ped{DS}^{(1)}$, $H\ped{DD}^{(0)}$, where the exponents mark the order in perturbation theory. The terms occurring in these blocks are given by
\begin{equation}
\begin{split}
    \bar{H}\app{UCC2}_{ij}=&f_{ij}+(\frac{1}{4}\bra{ik}\ket{ab}\sigma_{jk}^{ab}+ h.c.)+f_{ia}\sigma_j^a+f_{aj}\sigma_i^{a*},\\
    \bar{H}\app{UCC2}_{ab}=&f_{ab}-(\frac{1}{4}\bra{ij}\ket{bc}\sigma_{ij}^{ac}+ h.c.)-f_{ib}\sigma_i^a-f_{ai}\sigma_i^{b*},\\
    \bar{H}\app{UCC2}_{ia,bj}=&\bra{ia}\ket{bj}+(\frac{1}{2}\bra{ac}\ket{jk}\sigma_{ik}^{bc*}+h.c.),\\
    \bar{H}\app{UCC2}_{ci,ab}=&\bra{ci}\ket{ab}-f_{cj}\sigma_{ji}^{ab*},\\
    \bar{H}\app{UCC2}_{jk,ia}=&\bra{jk}\ket{ia}+f_{bi}\sigma_{kj}^{ab*},\\
    \bar{H}\app{UCC2}_{ibc,ajk}=&0, \qquad \bar{H}\app{UCC2}_{ij,kl}=0, \qquad \bar{H}\app{UCC2}_{ab,cd}=0,\\
    \bar{H}\app{UCC2}_{ai}=&\frac{1}{2}\bra{aj}\ket{cb}\sigma_{ij}^{cb}-\frac{1}{2}\bra{kj}\ket{ib}\sigma_{jk}^{ba}+f_{ai}+f_{jb}\sigma_{ij}^{db}\\
    &+f_{ab}\sigma_{i}^{b}-f_{ji}\sigma_{j}^{a},\\
    \bar{H}\app{UCC2}_{abij}=&\bra{ab}\ket{ij}+\frac{1}{2}\bra{kl}\ket{ij}\sigma_{kl}^{ab}+\frac{1}{2}\bra{ab}\ket{cd}\sigma_{ij}^{cd}\\
    &+P(ij)P(ab)\bra{ak}\ket{ic}\sigma_{jk}^{bc}+P(ab)f_{ac}\sigma_{ij}^{cb}\\
    &-P(ij)f_{ki}\sigma_{kj}^{ab}.
    \end{split}
\end{equation}
$P(ij), P(ab)$ are the antisymmetric permutation operators for the indices \textit{ij} and \textit{ab}, respectively. $f_{pq}$, where $p,q,\dots$ refer to generic indices, are  elements of the Fock matrix. $\bra{pq}\ket{rs}$ are the antisymmetrized two-electron integrals in Dirac notation, $\bra{pq}\ket{rs}=\braket{pq|rs}-\braket{pq|sr}$.\\
Similarly to the CC2 method, UCC2 scales as $\sim N^5$ with system size. Furthermore, the $r_{ij}^{ab}$ amplitudes for the double excitations are completely determined by the amplitudes $r_i^a$. Therefore, the EOM-UCC2 matrix elements can be written as a non-linear set of equations that only depend on the single-excitation amplitudes. 
It can therefore be expected that the EOM-UCC2 framework is not suitable for the description of states dominated by a double-excitation character. \\

For UCC3, the blocks of the Hamiltonian matrix are approximated as $H\ped{SS}^{(3)}$, $H\ped{SD}^{(2)}$, $H\ped{DS}^{(2)}$, $H\ped{DD}^{(1)}$. 

Note that UCC3 is an approximation to UCCSD and does not, contrary to CC3, contain triple excitations. UCC3 scales as CCSD, i.e., as $\sim N^6$ with system size, but with a larger prefactor.\cite{Liu2018}

The energy expression is given as
\begin{equation}
\begin{split}
    E\ped{UCC2/UCC3}&=\bigg(F_{ia}\sigma_i^a+\frac{1}{8}\braket{ij||ab}\sigma_{ij}^{ab}\bigg)+h.c.
    \end{split}
\end{equation}
and holds for both UCC2 as well as UCC3. 
The corresponding diagrams for UCC2 and UCC3 are given in the \hyperlink{si}{supplementary material}. 
\section{Diagrammatic rules}\label{sec:diag}
In this section, we discuss the rules to derive the diagrams for UCC2 and UCC3. Note that we discuss only the differences with respect to standard CC theory. 
A  complete explanation of the diagrammatic rules in CC may be found for example in  ref.~\onlinecite{Shavitt2009}.

The terms required for the UCC3 method (see also ref.~\onlinecite{Liu2018}) are given as  
\begin{align}
    \Bar{H}^0&=F+V,\\
    \Bar{H}^1&=[F,\tilde{\sigma}_1+\tilde{\sigma}_2]+\frac{1}{2}[V,\tilde{\sigma}_1+\tilde{\sigma}_2]+\frac{1}{2}[V\ped{R},\tilde{\sigma}_1+\tilde{\sigma}_2], \label{eqn:H1}\\
    \Bar{H}^2&=\frac{1}{12}[[V\ped{ND},\tilde{\sigma}_2],\tilde{\sigma}_2]+\frac{1}{4}[[V,\tilde{\sigma}_2]\ped{R},\tilde{\sigma}_2] \nonumber\\
    &+\frac{1}{4}[[V\ped{R},\tilde{\sigma}_2]\ped{R},\tilde{\sigma}_2].\label{eqn:H2}
\end{align}
The `ND' (non-diagonal) part of operators or contractions of operators is given by pure excitations or de-excitations up to the level of the chosen excitation space, while the `R' (rest) parts are the remaining components of the operator, i.e. operators not consisting of pure excitations or de-excitations only. For example, the terms contributing to the $V\ped{ND}$ operator are $\braket{ab||ij}\{\hat{a}^\dagger\hat{b}^\dagger \hat{j}\hat{i}\}$ and $\braket{ij||ab}\{\hat{i}^\dagger\hat{j}^\dagger \hat{b}\hat{a}\}$. All other terms in $V$, as for example $\braket{ak||ij}\{\hat{a}^\dagger\hat{k}^\dagger \hat{j}\hat{i}\}$, belong to the \textit{rest} part $V\ped{R}$. The CC diagrammatic rules can be used to determine the prefactors; however, to apply them consistently, one must account for the differing coefficients in front of the commutators in the two expansions. Therefore, on top of the known rules, the following four additional steps are needed:
\begin{enumerate}
    \item in case the diagram in question involves $\hat{V}$, determine whether it belongs to the \textit{non-diagonal} or the \textit{rest} part.
    \item determine whether the contractions $[\hat{V},\tilde{\sigma}]$ belong to the \textit{non-diagona}l or the \textit{rest} part. This classification is essential for identifying which terms in eqs.~\ref{eqn:H1} and \ref{eqn:H2} the diagram contributes to.
    \item consider the prefactors of the terms identified with rule 2.  
    Calculate the ratio between the prefactors of the corresponding terms in the Bernoulli and the BCH expansions, respectively. This ratio needs to be multiplied to the prefactor determined via the standard diagrammatic rules.
    \item for terms belonging to eq.~\ref{eqn:H2}, consider whether the Hamiltonian is connected to only one or both $\tilde{\sigma}$ operators. If it is connected to only one of them, a further factor of $\frac{1}{2}$ is required, as only half of the terms in the commutator contribute to the diagram.
\end{enumerate}
A couple of examples is given in the appendix. The diagrammatic representation of the UCC equations can be found in the \hyperlink{si}{supplementary material}. 
\section{Implementation}\label{sec:implementation}
The equations for UCC2 and UCC3 ground and excited states have been implemented in the QCUMBRE\cite{Hampe2017,QCUMBRE} program package.  The SCF calculations are performed via an interface with the Mainz INTegral (MINT) package \cite{mint} of the CFOUR program. \cite{matthews2020coupled} The implementation uses complex algebra, as the finite-field setting implies a potentially complex wave function. Point-group symmetry is implemented to speed up calculations. For the diagonalizations needed for the solution of the amplitude equations and the excited-state equations, the  finite-field modified versions of the Davidson scheme are used, as described in ref.~\onlinecite{Kitsarasthes}.

Matrix multiplications are performed through calls to efficient Basic Linear Algebra Subprograms (BLAS) like ZGEMM\cite{Lawson1979}. 
 
Intermediate contractions have been defined to keep the cost scaling  as $N^5$ and $N^6$ for UCC2 and UCC3, respectively.  
The amplitude equations are solved iteratively.  Similar to standard CC theory,  the initial guess for the $\sigma_1$ amplitudes are assumed to be zero, while the $\sigma_2$ amplitudes are initialized with their leading contribution, i.e. the MP2 amplitudes $\sigma_{ij}^{ab}=\frac{\braket{ab||ij}}{\epsilon_i+\epsilon_j-\epsilon_a-\epsilon_b}$. 

\section{Validation}\label{sec:validation}
The implementation for the field-free case has been verified by comparing results with calculations provided by the authors of ref.~\onlinecite{Liu2018}, both for the ground state and the excited states. Our implementation has then been adapted to the finite-field case. The symmetry implementation has been validated by comparison to calculations run in $C_1$ symmetry. \\
The implementation of the ff-Hamiltonian itself was already present in the QCUMBRE package. Furthermore, the Hermiticity of the equations was tested.

\section{Results}\label{sec:results}
In this section we investigate the performance of UCC in the finite-field context. The methylidynium ion exemplifies an astrophysically relevant system for strongly magnetized white-dwarf atmospheres. The corresponding finite-field calculations are analyzed to investigate the accuracy of EOM-UCC2 and EOM-UCC3 for the calculation of states with single and double-excitation character. 

In subsection \ref{sec:h2o}, the water molecule serves as a reference for benchmarking complex energies in finite magnetic fields. It is used to investigate the physical interpretation of the imaginary part of the CCSD energy. 
Boric acid, discussed in subsection \ref{sec:Boh3}, due to its nontrivial Abelian point group, illustrates the manifestation of complex excitation energies within the EOM-CC framework.

\subsection{Computational details}
All calculations have been performed using the QCUMBRE program package. The underlying SCF calculations were performed using the complex SCF implementation within the CFOUR\cite{gauss2021,cfour} program package interfaced to QCUMBRE. The integrals are computed using the Mainz INTegral (MINT) package  of the CFOUR program. \cite{mint} All calculations have been performed with uncontracted (unc) basis sets, to ensure the necessary flexibility to account for the anisotropy introduced by the magnetic field.

\subsection{Methylidyne ion}\label{sec:CH+_en}
In this section, we investigate the performance of the ff-UCC methods using the methylidyne cation in a strong magnetic field as an example and using ff-CC results from a previous study\cite{Hampe2020} as a reference.
In ref.~\onlinecite{Hampe2020}, the electronic energies of the methylidyne ion in an increasingly strong magnetic field between 0 and 1 B$_0$ were investigated in steps of 0.05 B$_0$. Various orientations with respect to the bond axis, i.e., $0^\circ, 30^\circ, 60^\circ$, and $90^\circ $ have been considered. The calculations were performed with the unc-cc-pVDZ basis set. \cite{pritchard2019a,feller1996a,schuchardt2007a,dunning1989a,kendall1992a,prascher2011a} For the molecular geometry, the ground state was optimized at the CCSD/unc-cc-pVDZ level in absence of the magnetic field. 

\begin{figure*}
    \centering
    \begin{subfigure}{0.95\columnwidth}
    \includegraphics[width=1.\linewidth]{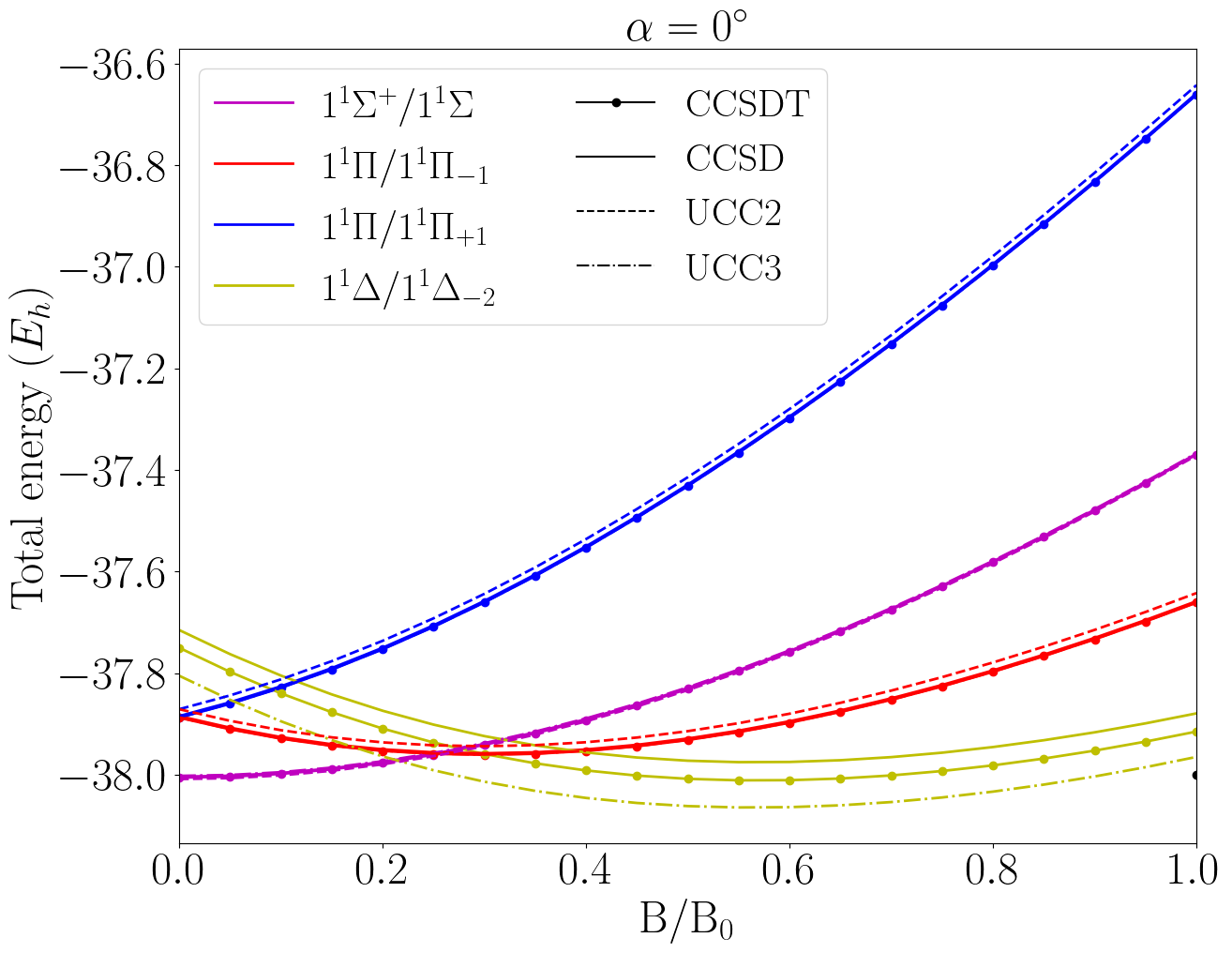}
    \caption{}
    \label{fig:alfa0}
    \end{subfigure}
    \begin{subfigure}{0.95\columnwidth}
    \includegraphics[width=1.\linewidth]{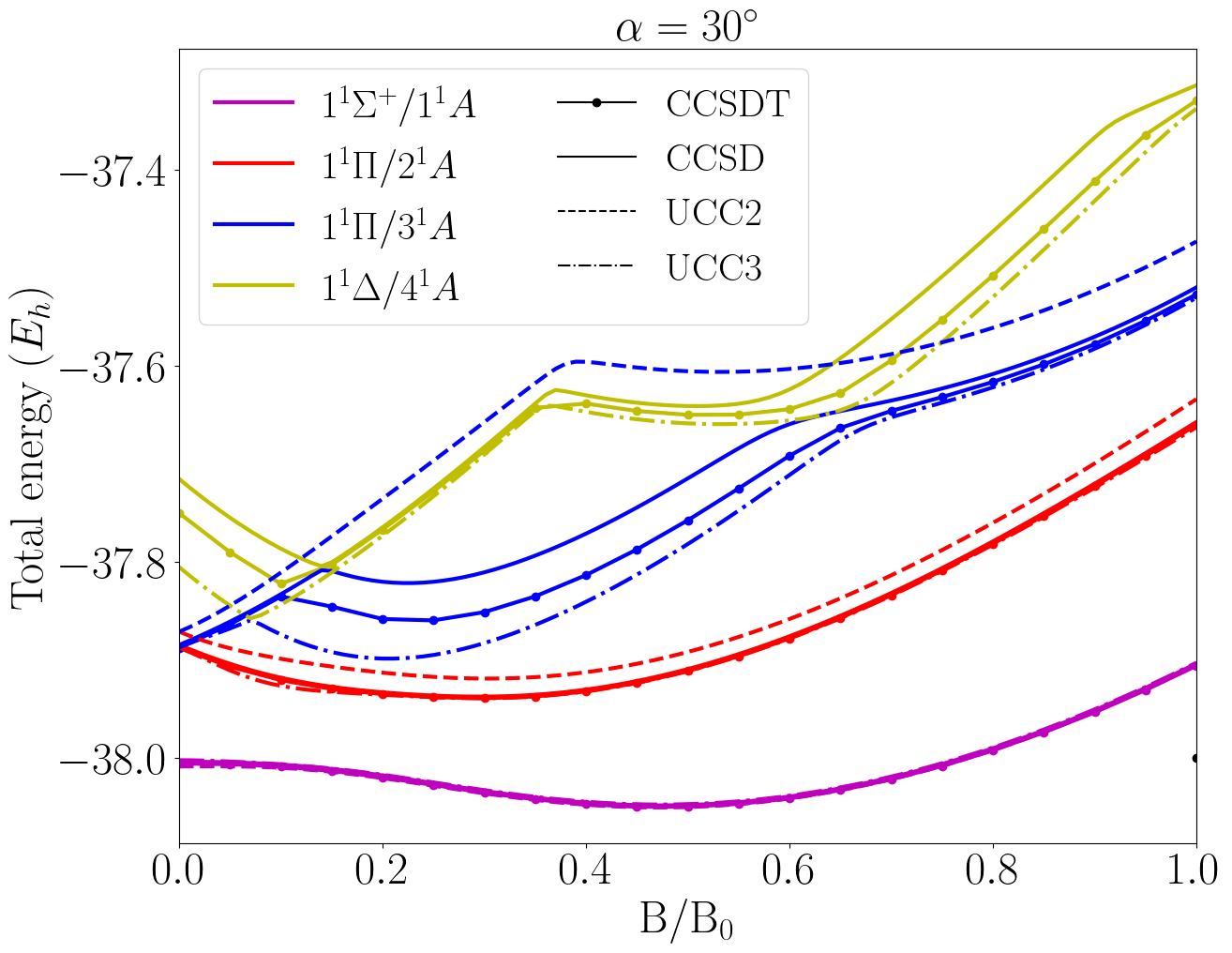}
    \caption{}
    \label{fig:alfa30}
    \end{subfigure}
    \centering
    \begin{subfigure}{0.95\columnwidth}
    \includegraphics[width=1.\linewidth]{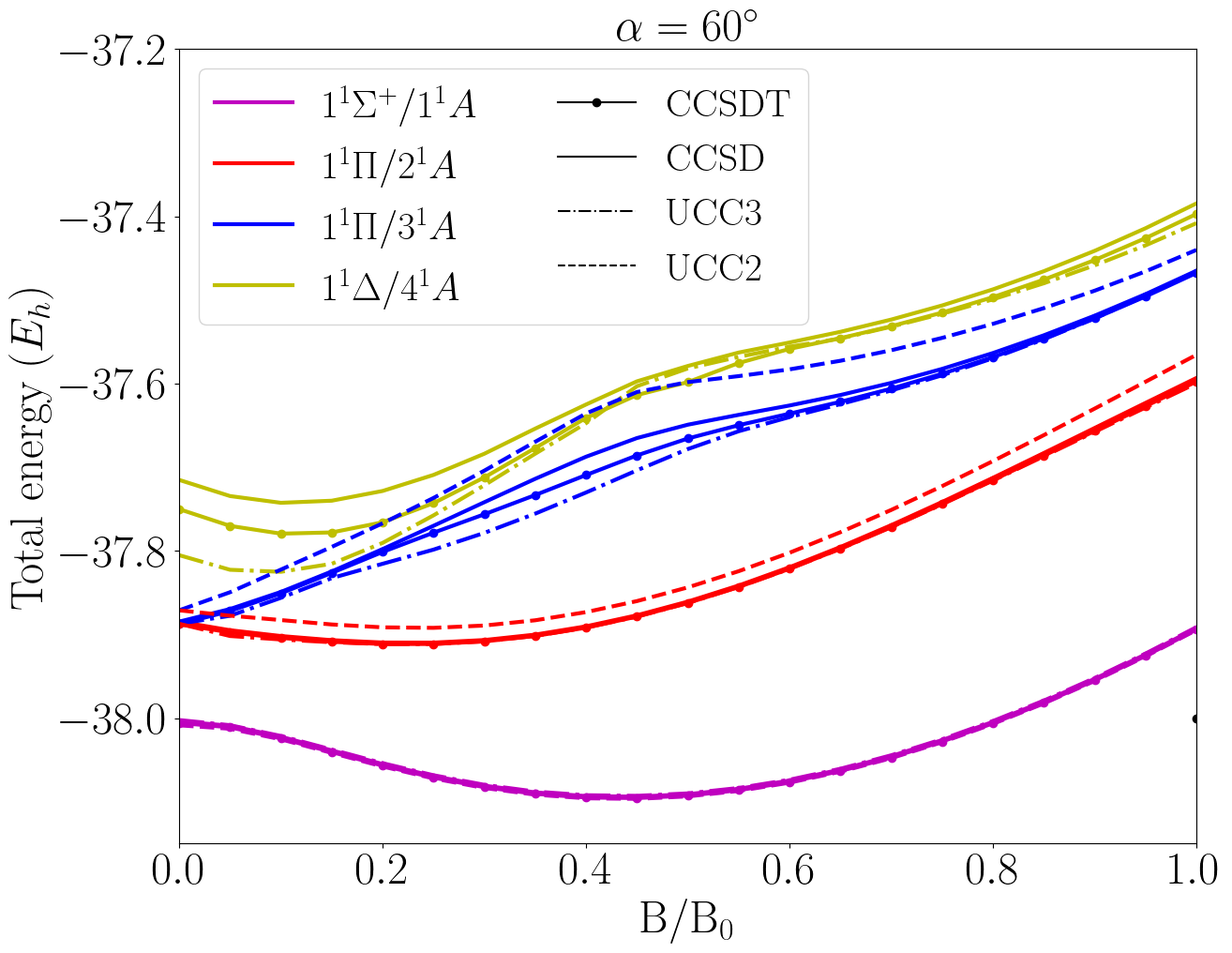}
    \caption{}
    \label{fig:alfa60}
    \end{subfigure}
    \begin{subfigure}{0.95\columnwidth}
    \includegraphics[width=1.\linewidth]{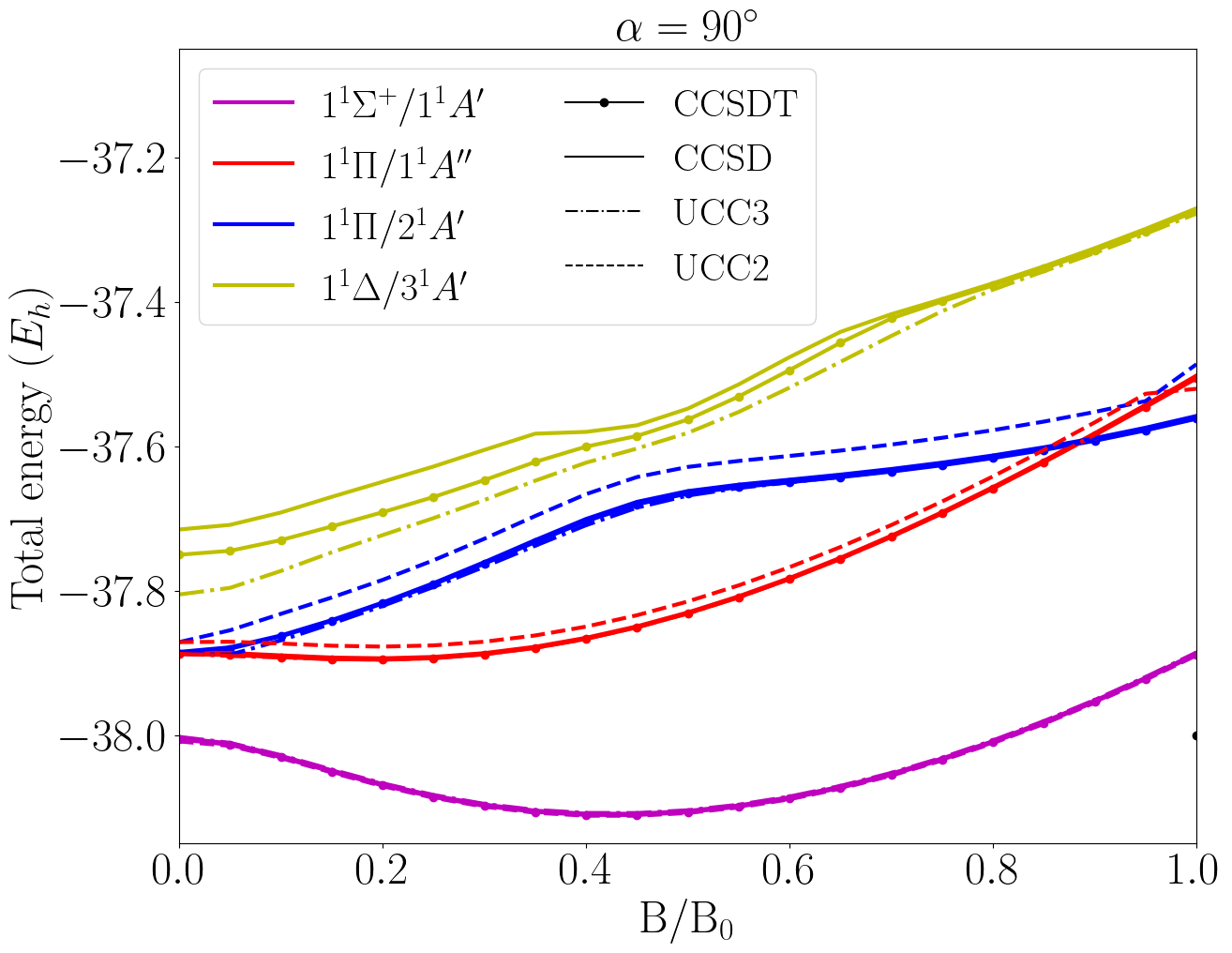}
    \caption{}
    \label{fig:alfa90}
    \end{subfigure}
     \caption{Ground state and low-lying excited singlet states of CH$^+$ in a magnetic field B between 0 and 1 B$_0$ and orientations of  $\alpha=0^\circ, 30^\circ, 60^\circ, 90^\circ$ with respect to the bond axis. The symmetry labels are composed of two terms, the first one referring to the $C_{\infty v}$ point group for the field-free case and the second one to the point group within the magnetic field, respectively.}
    \label{fig:CH+_angles}
\end{figure*}
\begin{table}
    \centering
    \begin{tabular}{|l|r|r|r|}
    \hline
    $\alpha=0$& $\Delta E\ped{CCSD}$/m$E_\text{h}$ &$ \Delta E\ped{UCC3}$/m$E_\text{h}$&$ \Delta E\ped{UCC2}$/m$E_\text{h}$\\
    \hline
    $1^1\Sigma^+/1^1\Sigma$&1.71&3.13&-2.42\\
    $1^1\Pi/1^1\Pi_{-1}$&2.95 &0.47 &17.55\\
    $1^1\Pi/1^1\Pi_{+1}$ & 2.95 &0.47 &17.55\\
    $1^1\Delta/1^1\Delta_{-2}$&35.59 &-53.23& - \\
    \hline
    $\alpha=\pi/6$& &\\
    \hline
    $1^1\Sigma^+/1^1A$ &1.98 &2.77&-0.38\\
    $1^1\Pi/2^1A$ &2.30 &-1.05&20.91 \\
    $1^1\Pi/3^1A$ &23.78 &-19.28&100.37 \\
    $1^1\Delta/4^1A$ &22.75&-16.63& -\\
    \hline
    $\alpha=\pi/3$& &\\
    \hline
    $1^1\Sigma^+/1^1A$ &1.82 &2.81& -1.00\\
    $1^1\Pi/2^1A$ &2.36 &-0.60&21.26\\
    $1^1\Pi/3^1A$ &8.29 &-7.85&43.98 \\
    $1^1\Delta/4^1A$ &19.96&-11.94&- \\
    \hline
    $\alpha=\pi/2$&&\\
    \hline
    $1^1\Sigma^+/1^1A^\prime$ &1.79 &2.83&-1.21\\
    $1^1\Pi/1^1A^{\prime\prime}$ &2.11&2.70& 15.80\\
    $1^1\Pi/2^1A^\prime$ &2.93 &-1.93& 36.82\\
    $1^1\Delta/3^1A^\prime$ &20.99&-23.23&-\\
    \hline
    \end{tabular}
    \caption{Mean energy differences (mEh) of the ground and three lowest excited singlet states per symmetry for CCSD and UCC3 relative to CCSDT reference values for CH$^+$ in a magnetic field. The orientation of the magnetic field is varied at different angles with respect to the bond axis and labeled by $\alpha$. The mean value is computed via $\Delta E\ped{Method}=E\ped{Method}-E\ped{CCSDT}$ over the range of varying magnetic field strengths between 0 and 1 B$_0$ and taking the arithmetic average.  The symmetry labels are composed of two terms, the first one referring to the $C_{\infty v}$ point group for the field-free case and the second one to the point group within the magnetic field, respectively}
    \label{tab:mean_values}
\end{table}
The $^1\Sigma^+$ state was taken as a reference for the EOM-CC calculations. This state is characterized by the single closed-shell configuration $1\sigma^22\sigma^23\sigma^2$. Starting from this reference state, the three lowest-lying singlet excited states are considered. In absence of an external magnetic field, these states are given by the two degenerate 1$^1\Pi$ states (with the configuration $1\sigma^22\sigma^23\sigma^11\pi^1$) and the degenerate 1$^1\Delta$ state (with the configuration $1\sigma^22\sigma^21\pi^2$). The latter possesses a predominant double-excitation character with respect to the ground state. \\ 
Since CCSD poorly describes states with strong double-excitation character, we include CCSDT results alongside CCSD to better assess the performance of the ff-UCC methods. 
Comparable accuracy between CCSD and UCC3 is anticipated, given that both methods operate within the same excitation manifold. 
Fig.~\ref{fig:CH+_angles} illustrates the energy evolution of the respective states as a function of magnetic field strength, while Tab.~\ref{tab:mean_values} summarizes the mean energy deviations relative to the CCSDT reference values.

The presence of a magnetic field generally induces symmetry reduction, with the resulting point group dependent on the field’s orientation relative to the bond axis. States belonging to different irreducible representations (IRREPs) are allowed to cross (see for example the blue and red lines in Fig.~\ref{fig:alfa90}), while avoided crossings occur for states of same symmetry (see for example the blue and the yellow line in Figs.~\ref{fig:alfa30}, \ref{fig:alfa60} and \ref{fig:alfa90}). For all orientations, the computational ground state $^1\Sigma$ exhibits a quadratic dependence on the magnetic field strength. For all non-parallel orientations, mixing with higher-lying states is observed.

For all orientations, UCC3 correctly reproduces the qualitative shape of the CCSDT curves. Also, in particular the $^1\Sigma$ reference state is very well described. Unlike for ff-CCSD which overestimates the ff-CCSDT energies for all states, orientations, and field strengths studied here, the behaviour of UCC3 is more complicated. \\
The most significant deviations from the ff-CCSDT reference results are observed for the $1^1\Delta$ state of the field-free case. 
These deviations stem from the inadequate description of the $1^1\Delta$ state — relative to the $^1\Sigma^+$ reference — when the cluster operator is limited to single and double excitations. A similar behavior for the corresponding ff-CCSD results has been reported in Ref.~\onlinecite{Hampe2020}.

For UCC3, however, the energies are underestimated.  \\ 
Consistent with the observation for ff-CC2 in ref.~\onlinecite{Kitsaras2024}, UCC2 is unable to describe doubly-excited states, as explained in sec.~\ref{sec:theory}.  Hence, for the parallel orientation ($\alpha=0$), the $1^1\Delta/1^1\Delta_{-2}$ state cannot be targeted with the UCC2 method.
For the non-parallel orientations of $30^\circ$ and $60^\circ$, the  $1^1\Delta/4^1A$ state mixes with the $^1\Pi/3^1A$ state. Similarly to the observations in ref.~\onlinecite{Hampe2020}, the double-excitation character is partially passed from the $1^1\Delta/4^1A$ state to the $^1\Pi/3^1A$ state around the field strengths at which the avoided crossings occur. Hence, the errors in the predicted energies are larger for field strengths at which the respective state is dominated by a substantial double excitation character. 

From the data collected in table \ref{tab:mean_values}, the average difference of the computed ground-state energies with respect to the CCSDT reference values is about 3 m$E\ped{h}$ and 2 m$E\ped{h}$ for UCC3 and CCSD, respectively. States that are described by a single excitation with respect to the reference state show small deviations from the CCSDT reference results.
The UCC3 description of the first excited state has a similarly high accuracy for all orientations. 
In particular, for the states originating from the $^1 \Pi$ state, in the parallel case UCC3 has a higher accuracy than CCSD, with an average energy difference of about 0.5 m$E\ped{h}$ (about 13.6 meV) for UCC3 and almost 3 m$E\ped{h}$ (about 81.6 meV) for CCSD. 
\\
Average deviations have larger values for states with a partial double-excitation character by one order of magnitude. 
For the non-parallel orientations, we observe that the UCC3 results have slightly smaller deviations from CCSDT than the corresponding CCSD results. However, both are of the same order of magnitude.\\
On the other hand, UCC2 overestimates the electronic energies of the states originating from $1^1\Pi$, with large differences between 18 and 100 m$E\ped{h}$ observed for all orientations.
We note that the deviations from CCSDT are, depending on state and orientation, either positive or negative for UCC3 while they are only positive in the case of CCSD.

Both CCSD and UCC3 consistently face challenges in accurately describing states dominated by double excitations, while they reproduce CCSDT results well for singly-excited states. This limitation stems from the inherent approximations of the methods and was therefore anticipated.
UCC2 provides a qualitatively correct approximation for states dominated by a single-excitation character, while it leads to qualitatively wrong results in cases with a significant double-excitation character. 

\subsection{Water molecule}\label{sec:h2o}
\begin{figure}
    \centering
    \includegraphics[width=.8\linewidth]{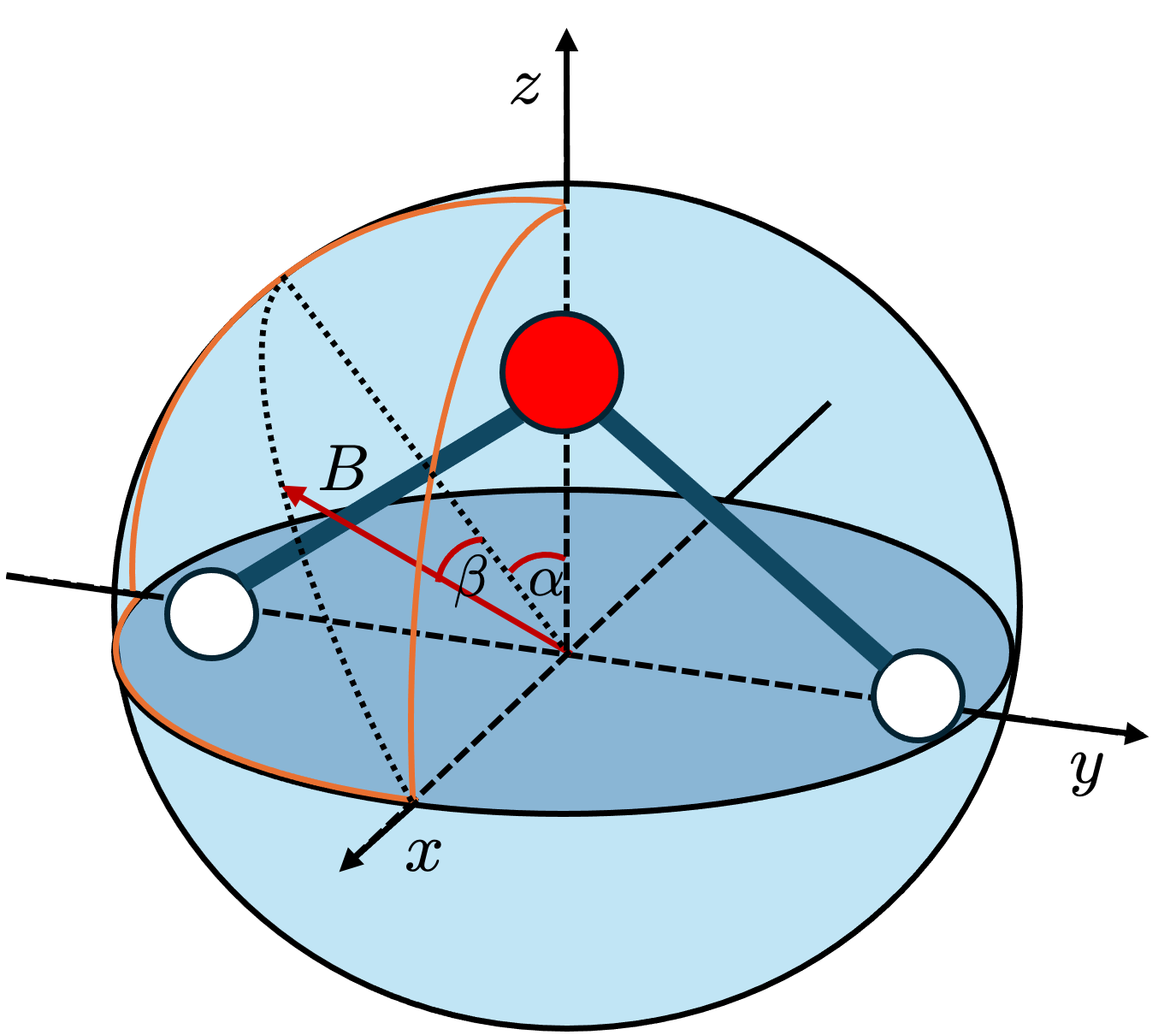}
    \caption{Water molecule in a magnetic field of B=0.5 B$_0$, whose orientation is allowed to vary corresponding to the polar coordinates $\alpha$ and $\beta$.}\label{fig:water_orientation}
\end{figure}

In this section, we focus on the water molecule in a magnetic field. The lowest singlet state of the system has been studied at the ff-CCSD level in ref.~\onlinecite{Thomas2021} within a strong magnetic field of of B=0.5 B$_0$ and the occurrence of complex energy eigenvalues in CC calculations was investigated. In general, except for special symmetries, the ff-CC energy can become complex valued, similar as for non-perturbative treatments of relativistic effects that include spin-orbit coupling. 
The orientation of the field was varied on the surface of the positive octant of the unit sphere (see Fig.~\ref{fig:water_orientation}) and is described by the two polar angles $\alpha,\beta$. It was shown that the imaginary part of the ground-state energy only vanishes for those orientations of the magnetic field which are aligned to one of the symmetry axes of the point group of the molecule in the field-free case. 
Here, apart from the ground state, we also investigate the first three excited singlet states of the water molecule. In the field-free case, these states are of $B_1$, $A_2$ and $A_1$ symmetry.
The aim is to compare the quality of the ff-CCSD results for both ground and excited states and the corresponding ff-UCC3 results. 
A clear advantage of ff-UCC3 as compared to ff-CCSD is that by construction all energies are real for all $\alpha$ and $\beta$.
The calculations have been performed with the uncontracted cc-pVTZ \cite{pritchard2019a,feller1996a,schuchardt2007a,dunning1989a, kendall1992a} basis set, using the geometry from ref.~\onlinecite{Thomas2021}. 

\begin{figure*}
    \centering
    \begin{subfigure}{.95\columnwidth}
    \includegraphics[width=1.\linewidth]{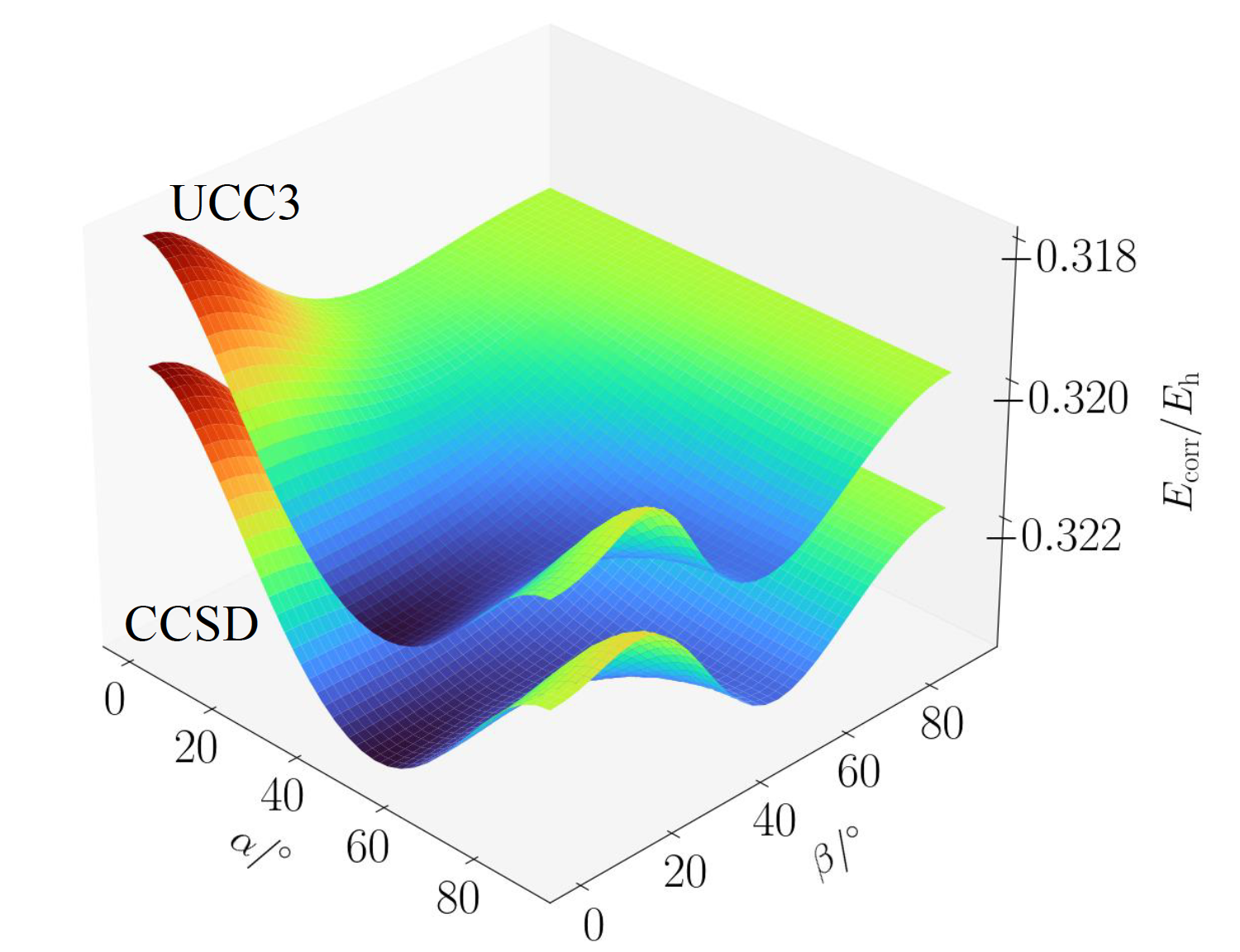}
    \caption{}
    \label{fig:gs}
    \end{subfigure}
    \begin{subfigure}{.95\columnwidth}
    \includegraphics[width=1.\linewidth]{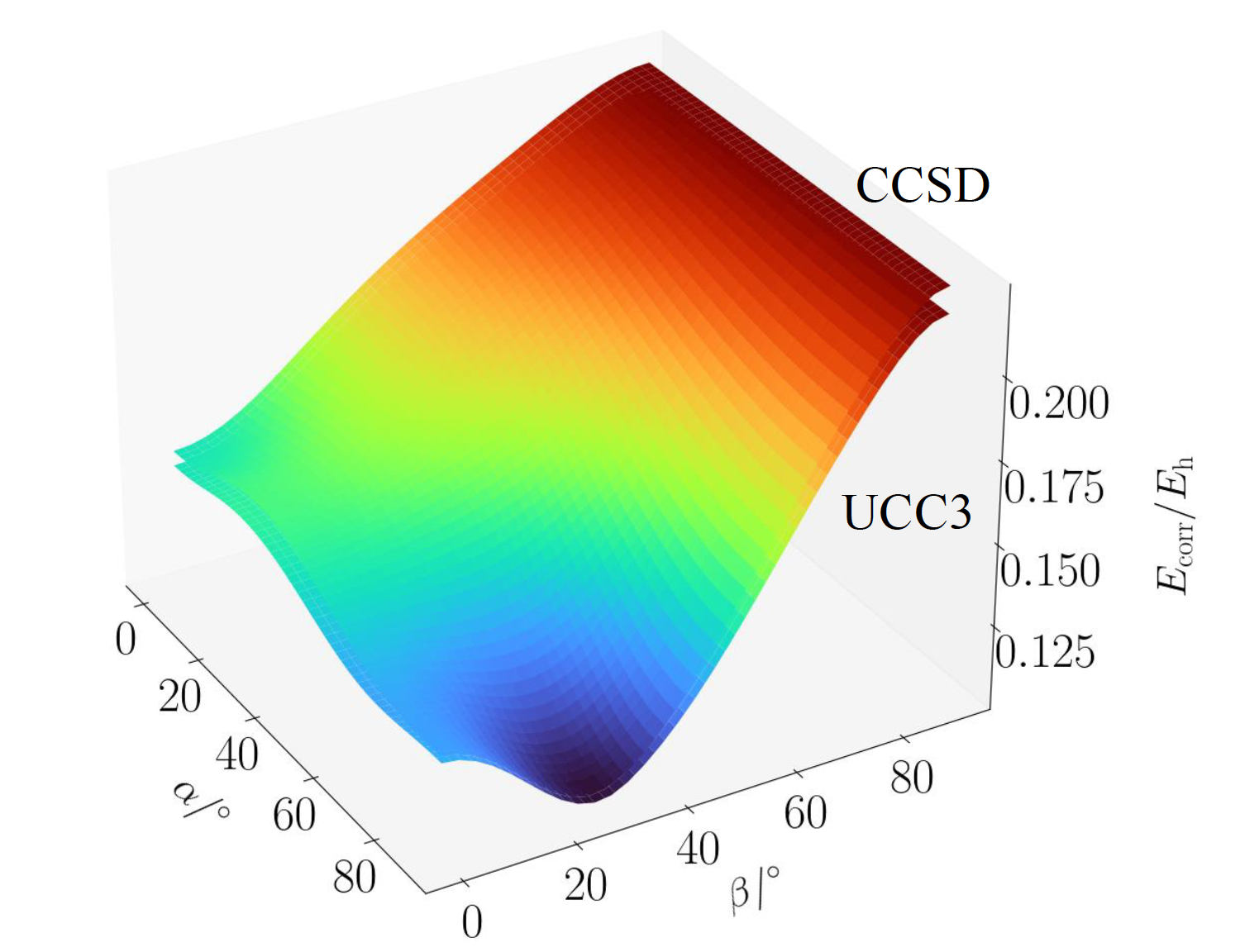}
    \caption{}
    \label{fig:eom_1}
    \end{subfigure}
    \begin{subfigure}{.95\columnwidth}
    \includegraphics[width=1.\linewidth]{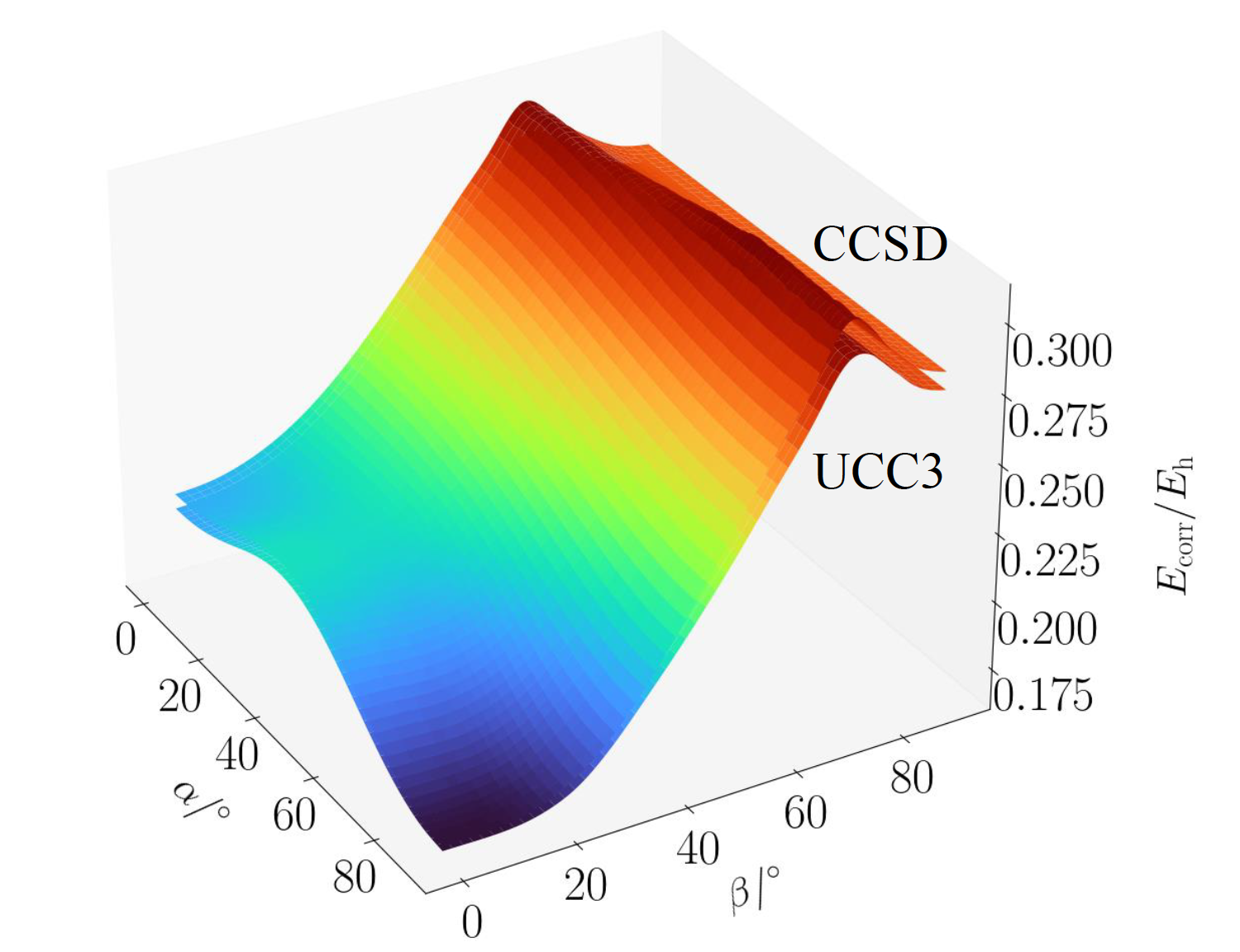}
    \caption{}
    \label{fig:eom_2}
    \end{subfigure}
    \begin{subfigure}{.95\columnwidth}
    \includegraphics[width=1.\linewidth]{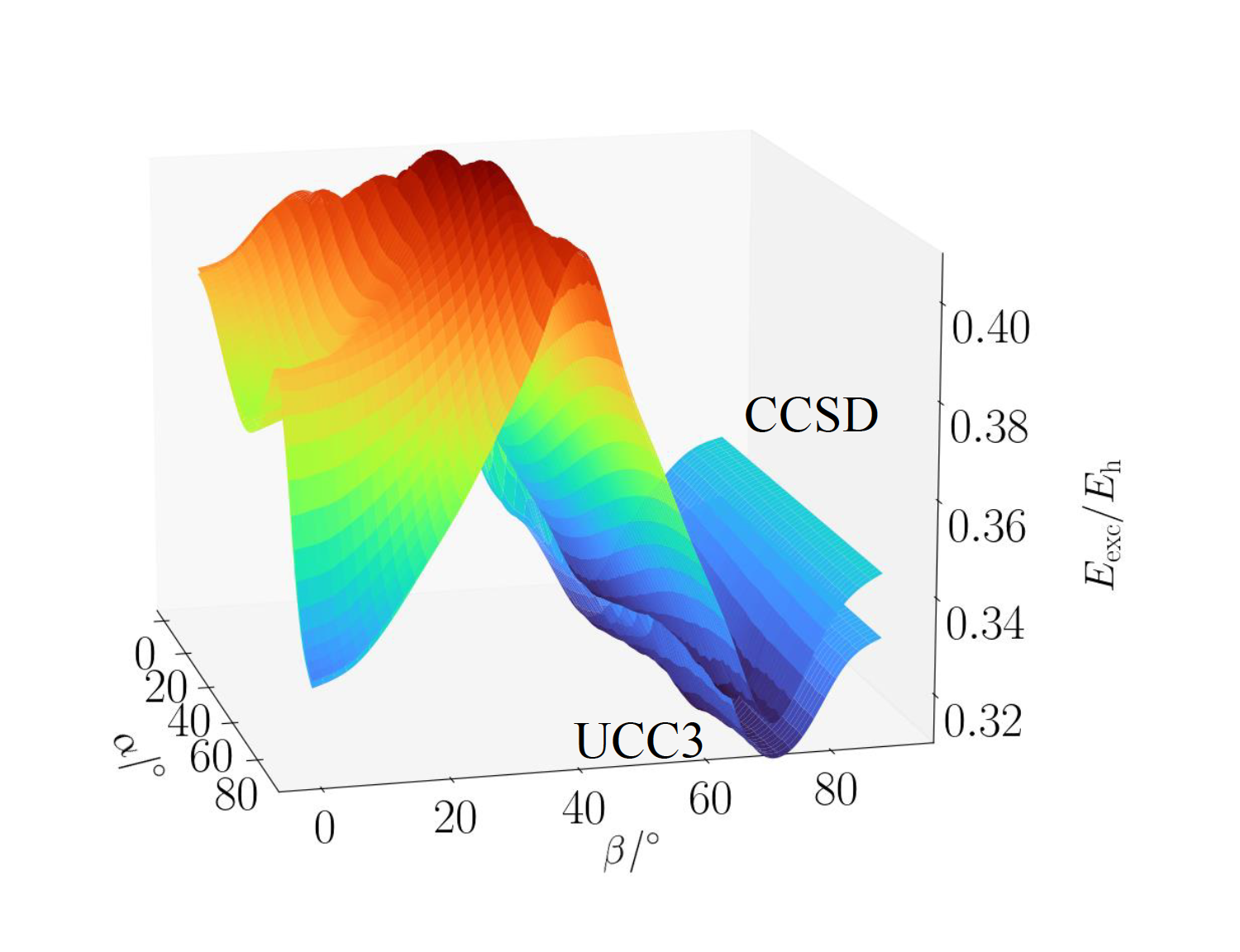}
    \caption{}
    \label{fig:eom_3}
    \end{subfigure}
    \caption{Correlation energy of the ground state (Fig.~\ref{fig:gs}) and excitation energies of the first three excited states (Figs.~\ref{fig:eom_1}-\ref{fig:eom_3}) of the water molecule in a magnetic field of B=0.5 B$_0$ as a function of its orientation, as pictured in Fig.~\ref{fig:water_orientation}, calculated at the CCSD and UCC3 level of theory, using the unc-cc-pVTZ basis set.}
    \label{fig:surf_en}
\end{figure*}

\begin{figure*}
    \centering
    \begin{subfigure}{.95\columnwidth}
    \includegraphics[width=1.\linewidth]{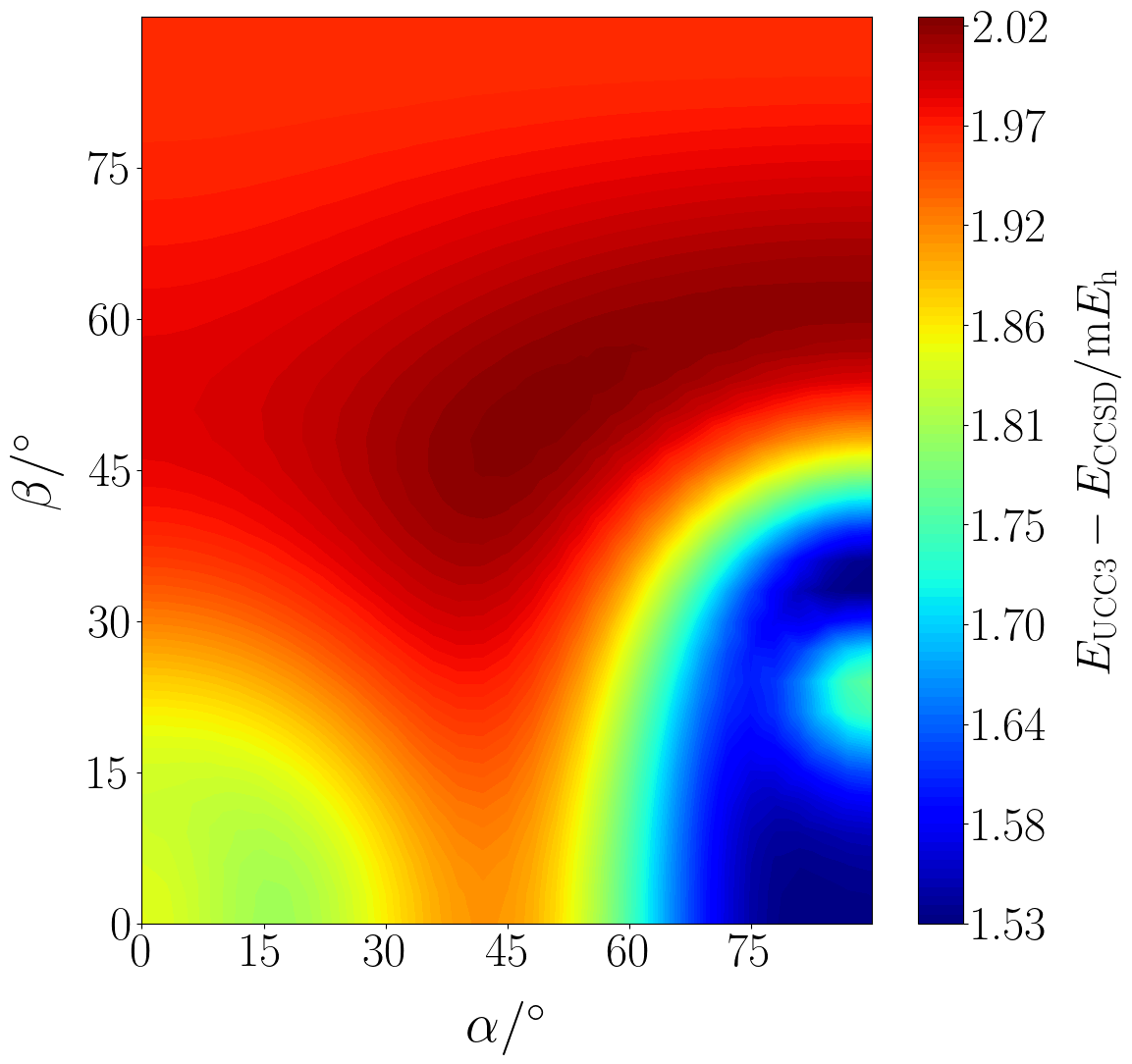}
    \caption{$\Delta E(\Psi_0)=E\ped{UCC3}(\Psi_0)-E\ped{CCSD}(\Psi_0)$: $\max\Delta E(\Psi_0)=2.028 \text{m}E\ped{h}$  ($48^\circ,51^\circ$),
     $\min\Delta E(\Psi_0)=1.531 \text{m}E\ped{h}$ ($81^\circ,0^\circ$).}
    \label{fig:diff_gs}
    \end{subfigure}
    \begin{subfigure}{.95\columnwidth}
    \includegraphics[width=1.\linewidth]{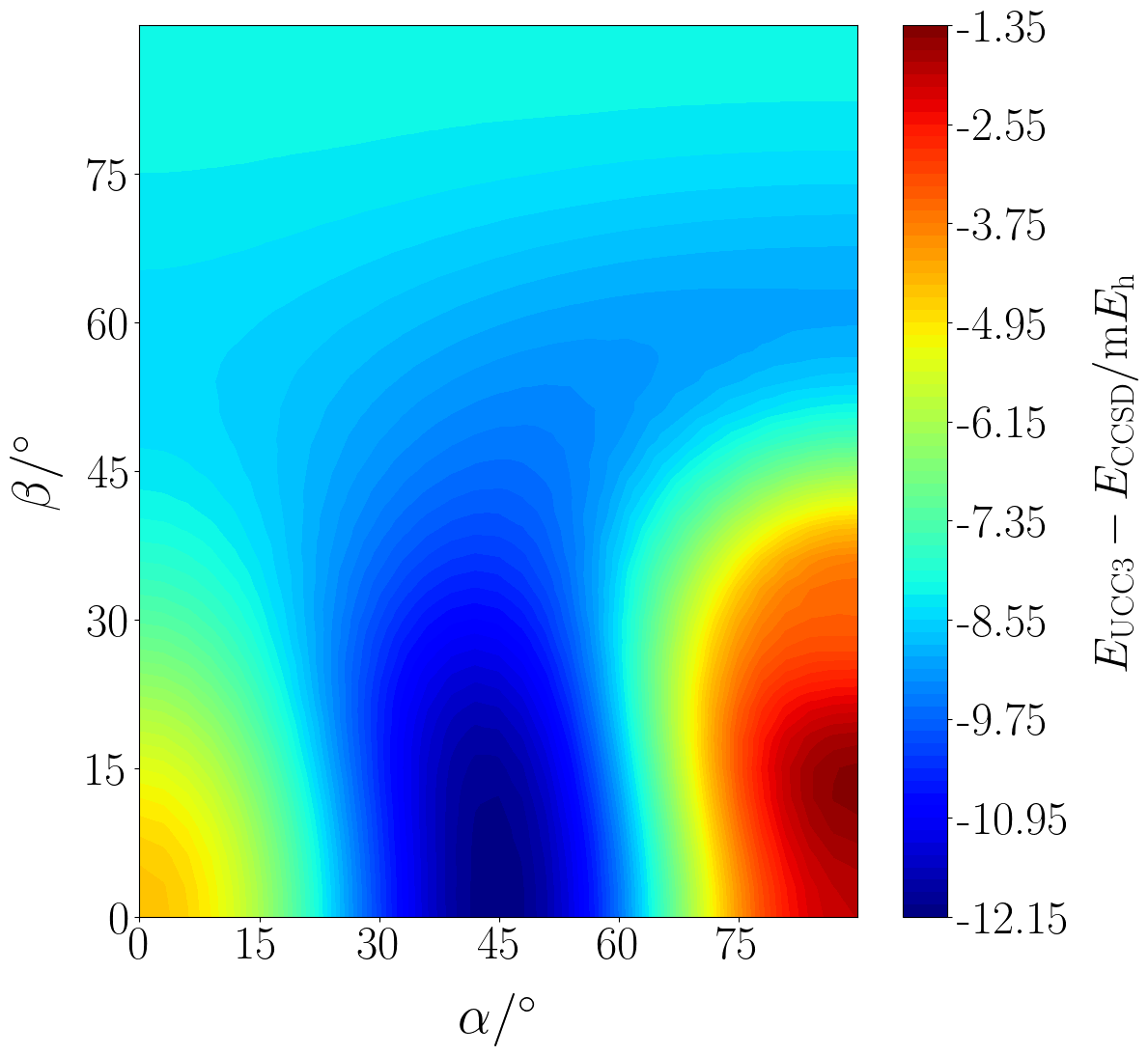}
     \caption{$\Delta E(\Psi_1)=E\ped{UCC3}(\Psi_1)-E\ped{CCSD}(\Psi_1)$: $\max\Delta E(\Psi_1)= -1.431\text{m}E\ped{h}$ ($90^\circ,12^\circ$), $\min\Delta E(\Psi_1)= -12.133\text{m}E\ped{h}$ ($45^\circ,0^\circ$).}
    \label{fig:diff1}
    \end{subfigure}
    \begin{subfigure}{.95\columnwidth}
    \includegraphics[width=1.\linewidth]{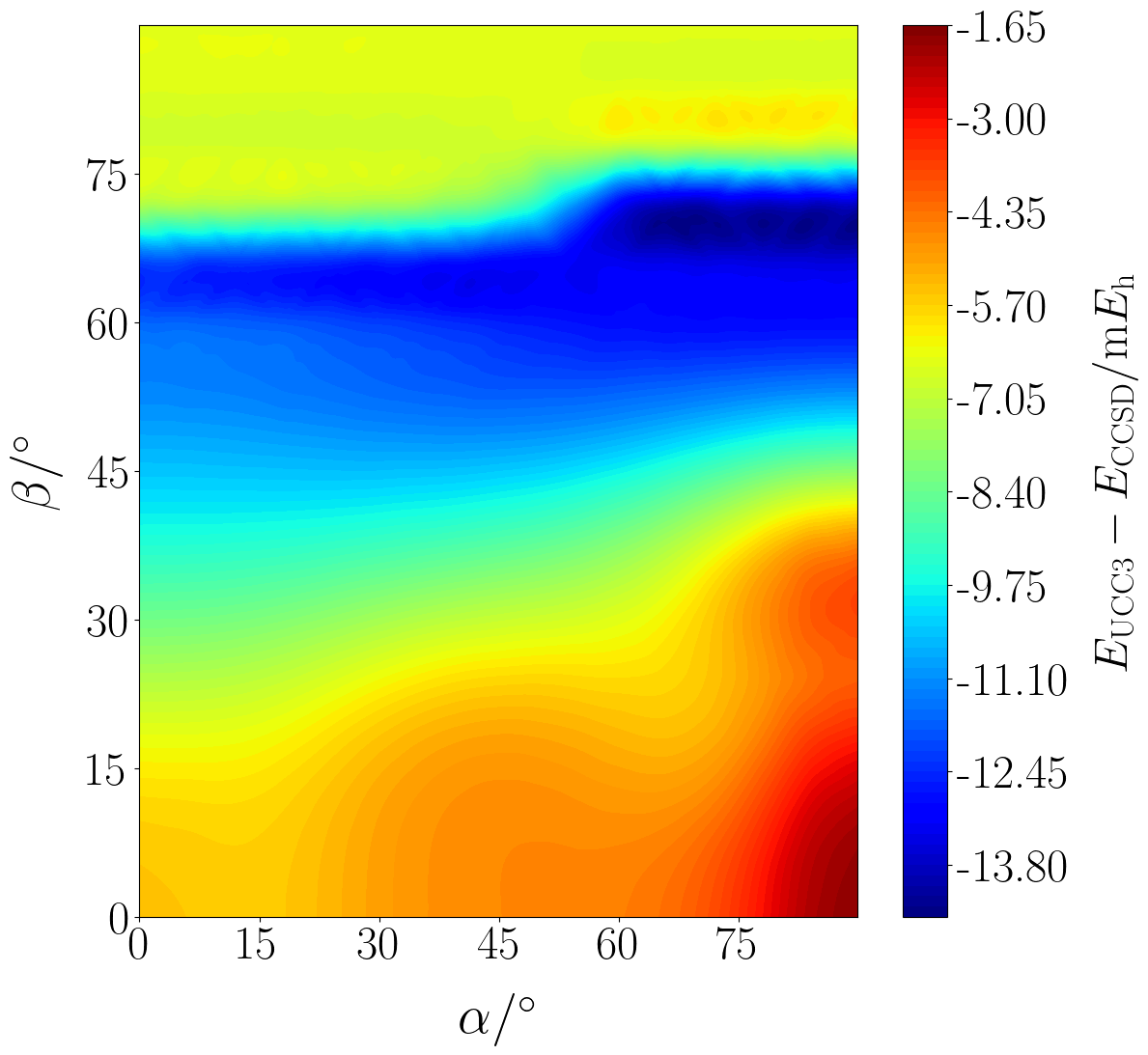}
     \caption{$\Delta E(\Psi_2)=E\ped{UCC3}(\Psi_2)-E\ped{CCSD}(\Psi_2)$: $\max\Delta E(\Psi_2)= -1.797\text{m}E\ped{h}$ ($90^\circ,0^\circ$), $\min\Delta E(\Psi_2)= -13.792\text{m}E\ped{h}$ ($90^\circ,72^\circ$).}
    \label{fig:diff2}
    \end{subfigure}
    \begin{subfigure}{.95\columnwidth}
    \includegraphics[width=1.\linewidth]{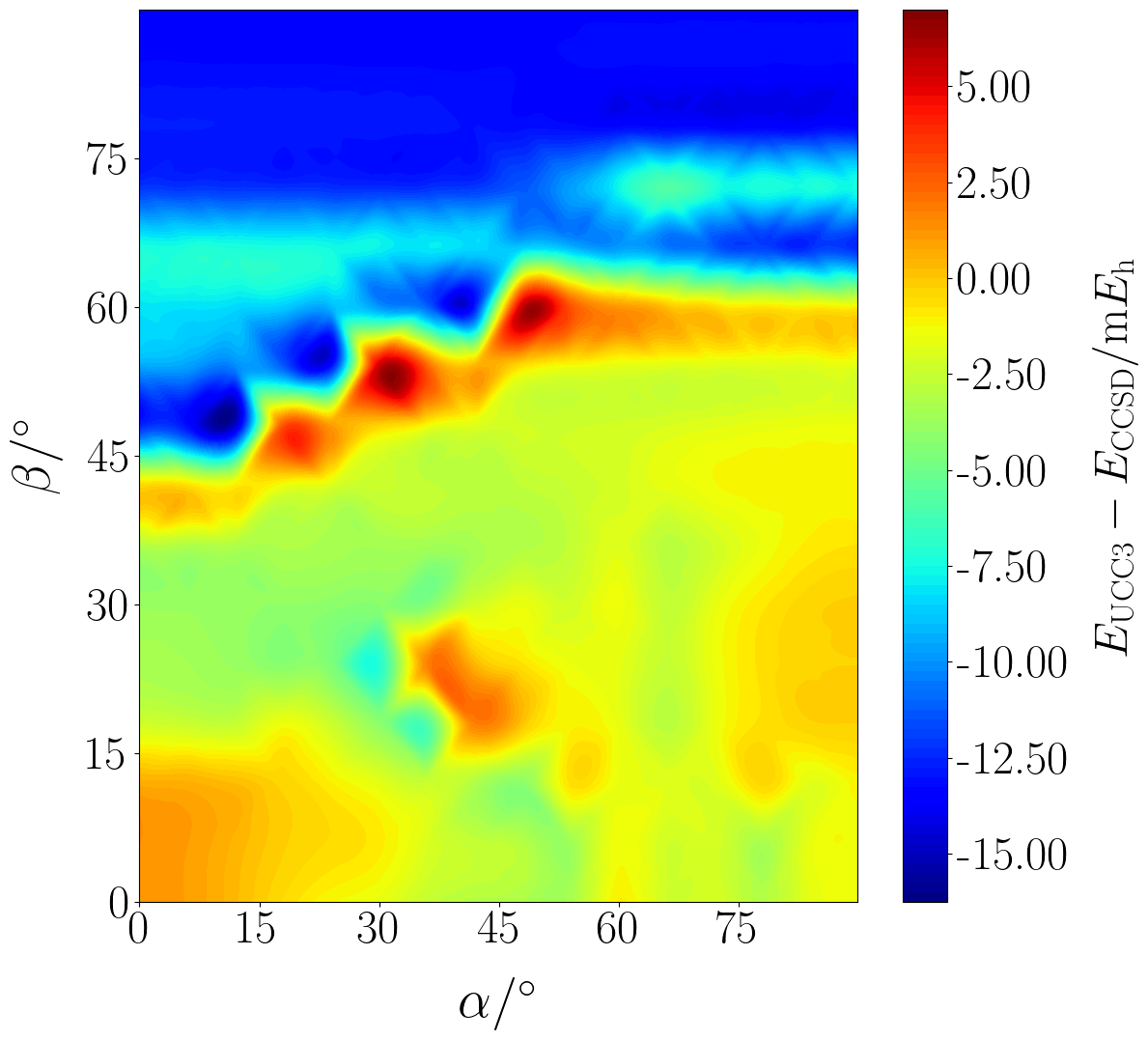}
     \caption{$\Delta E(\Psi_3)=E\ped{UCC3}(\Psi_3)-E\ped{CCSD}(\Psi_3)$: $\max\Delta E(\Psi_3)= 5.535\text{m}E\ped{h}$ ($48^\circ,60^\circ$), $\min\Delta E(\Psi_3)= -14.721\text{m}E\ped{h}$ ($12^\circ,48^\circ$).}
    \label{fig:diff3}
    \end{subfigure}
    \caption{Energy differences, calculated as $E\ped{UCC3}-E\ped{CCSD}$, for the ground (Fig.~\ref{fig:diff_gs}) and first three excited states (Figs.~\ref{fig:diff1}-\ref{fig:diff3}) of the water molecule in a magnetic field of B=0.5 B$_0$ as a function of its orientation, as pictured in Fig.~\ref{fig:water_orientation}.}\label{fig:diff}
\end{figure*}

\begin{figure*}
    \centering
    \begin{subfigure}{.95\columnwidth}
    \includegraphics[width=1.\linewidth]{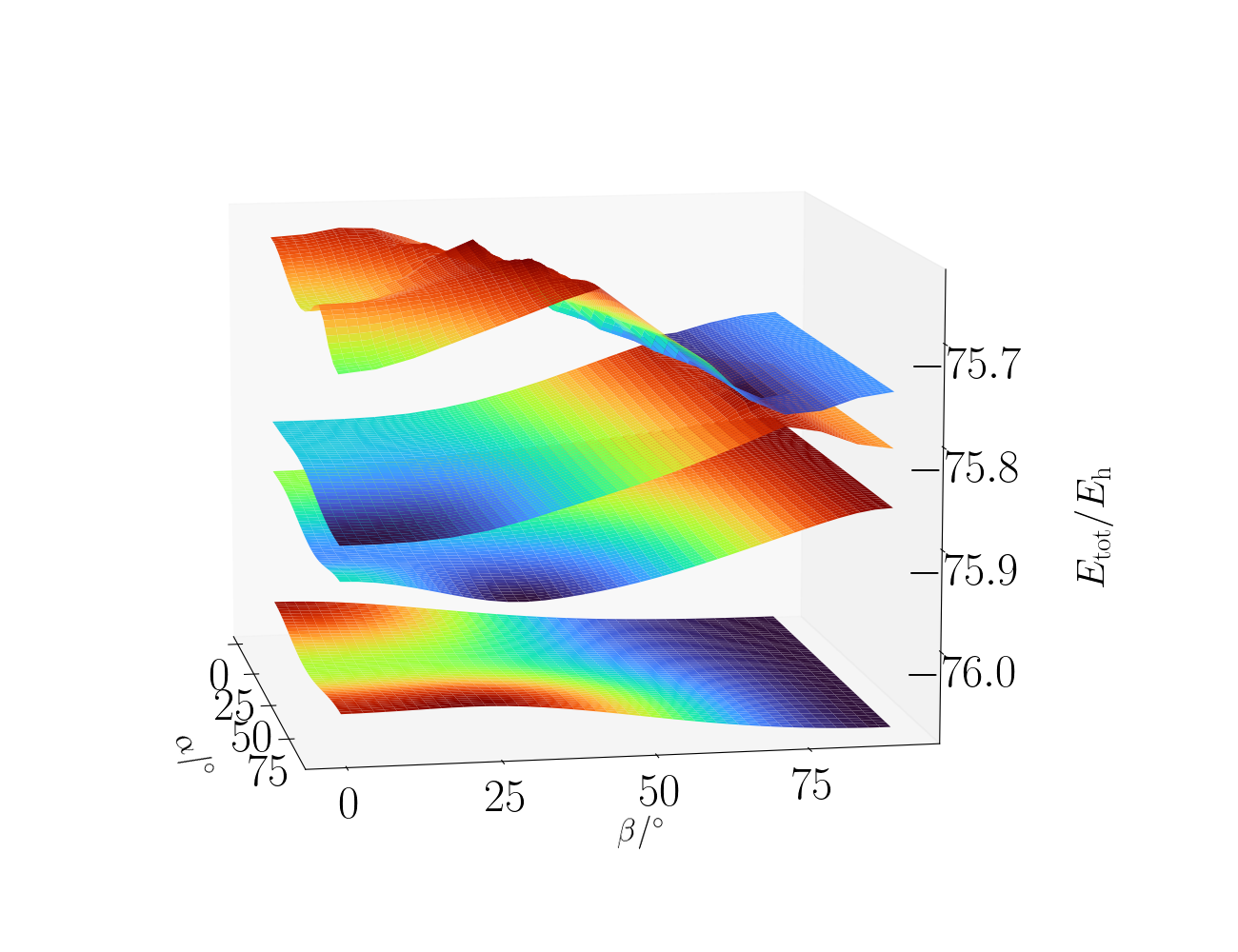}
    \caption{Four lowest-lying singlet states with UCC3.
    } 
    \label{fig:ucc3_tot}
    \end{subfigure}
    \begin{subfigure}{.95\columnwidth}
    \includegraphics[width=1.\linewidth]{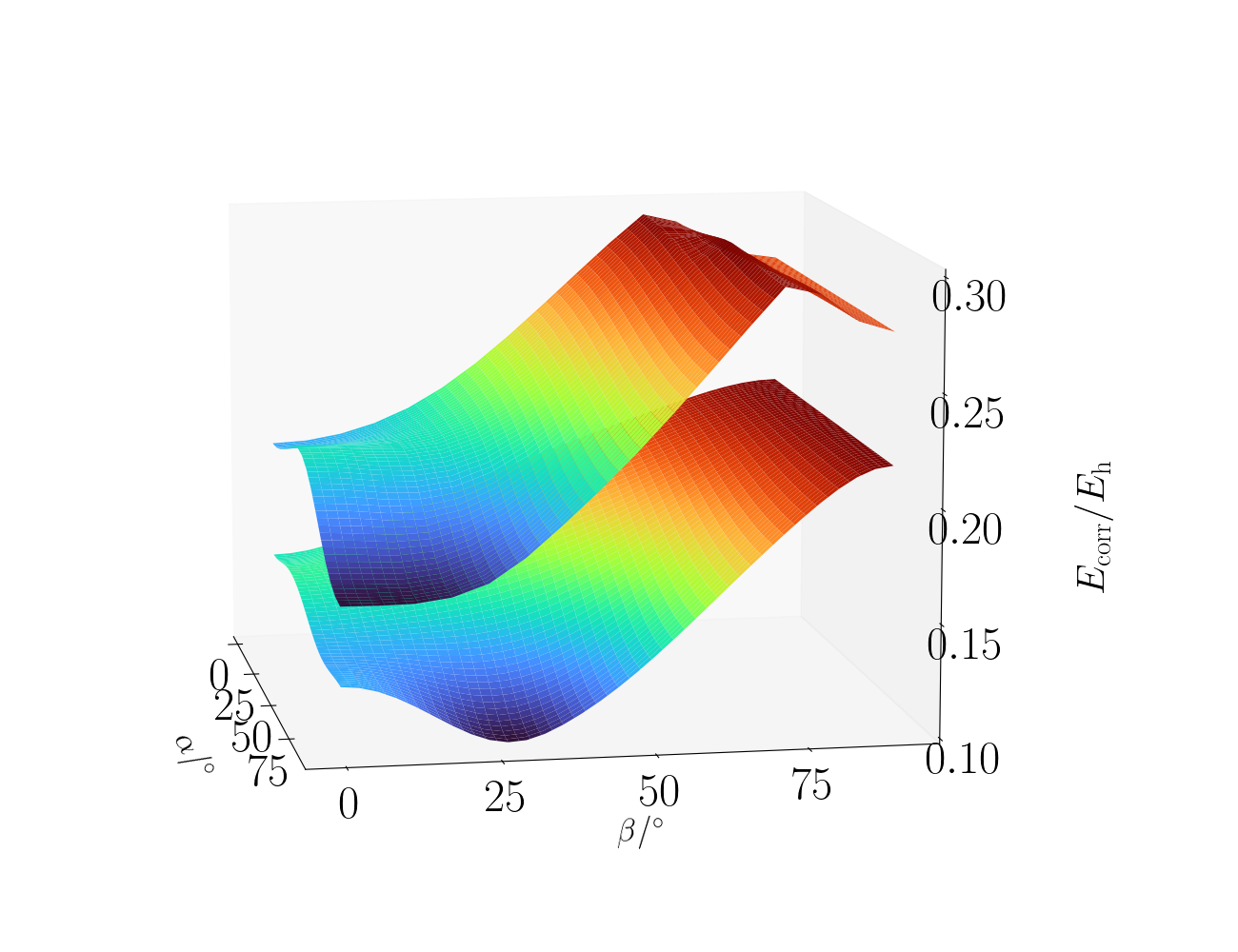}
    \caption{$\Psi_1$ and $\Psi_2$ with UCC3.
    }  
    \label{fig:ucc3_12}
    \end{subfigure}
    \begin{subfigure}{.95\columnwidth}
    \includegraphics[width=1.\linewidth]{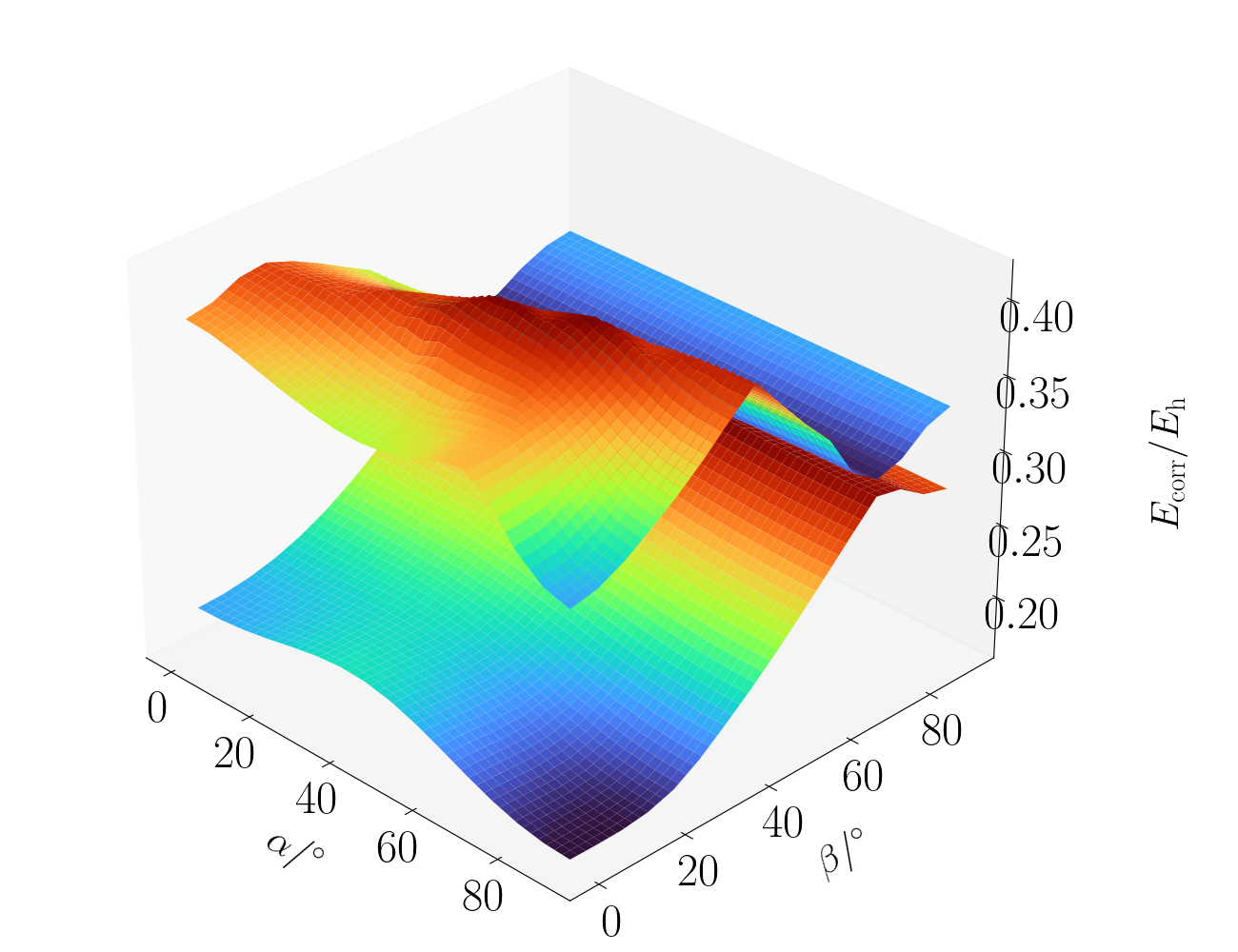}
    \caption{$\Psi_2$ and $\Psi_3$ with UCC3.
    }  
    \label{fig:ucc3_23}
    \end{subfigure}
    \caption{Real part of the energy surfaces of the ground and first three excited states of the water molecule in a magnetic field of B=0.5 B$_0$ as a function of its orientation, as shown in Fig.~\ref{fig:water_orientation}, calculated at the ff-UCC3 level of theory.}
    \label{fig:tot_h2o} 
\end{figure*}

\begin{figure*}
    \centering
    \begin{subfigure}{.95\columnwidth}
    \includegraphics[width=1.\linewidth]{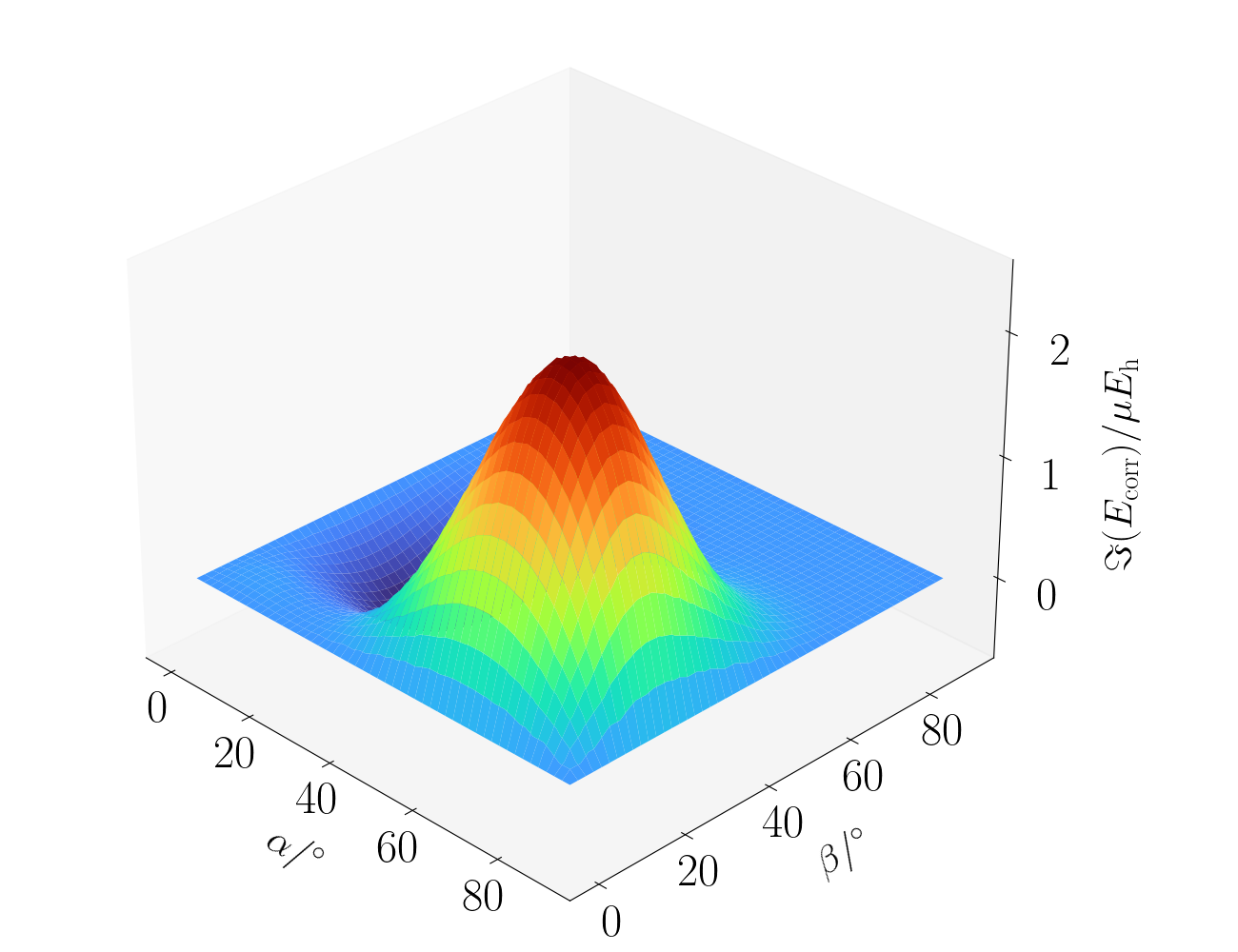}
    \caption{Imaginary ground-state energy: $\max(\Im{E\ped{CCSD}}(\Psi_0))=2.50\mu E\ped{h}$ ($66^\circ,24^\circ$).
        }
    \label{fig:gs_im}
    \end{subfigure}
    \begin{subfigure}{.95\columnwidth}
    \includegraphics[width=1.\linewidth]{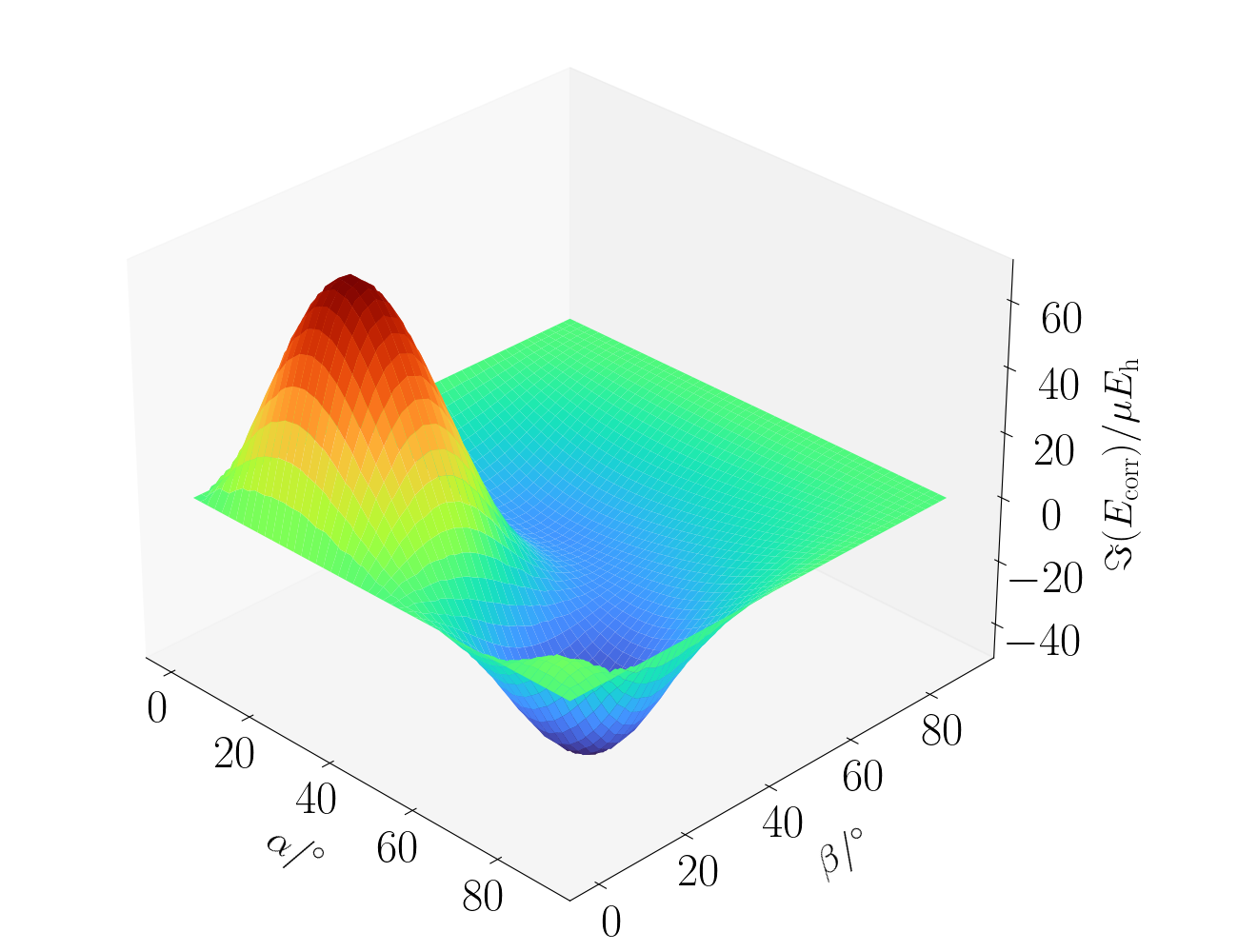}
    \caption{Imaginary energy of $\Psi_1$:  $\max(\Im{E\ped{CCSD}}(\Psi_1))= 69.87\mu E\ped{h}$ ($21^\circ,18^\circ$), $\min(\Im{E\ped{CCSD}}(\Psi_1))= -48.31\mu E\ped{h}$ ($69^\circ,24^\circ$).
    } 
    \label{fig:eom1_im}
    \end{subfigure}
    \begin{subfigure}{.95\columnwidth}
    \includegraphics[width=1.\linewidth]{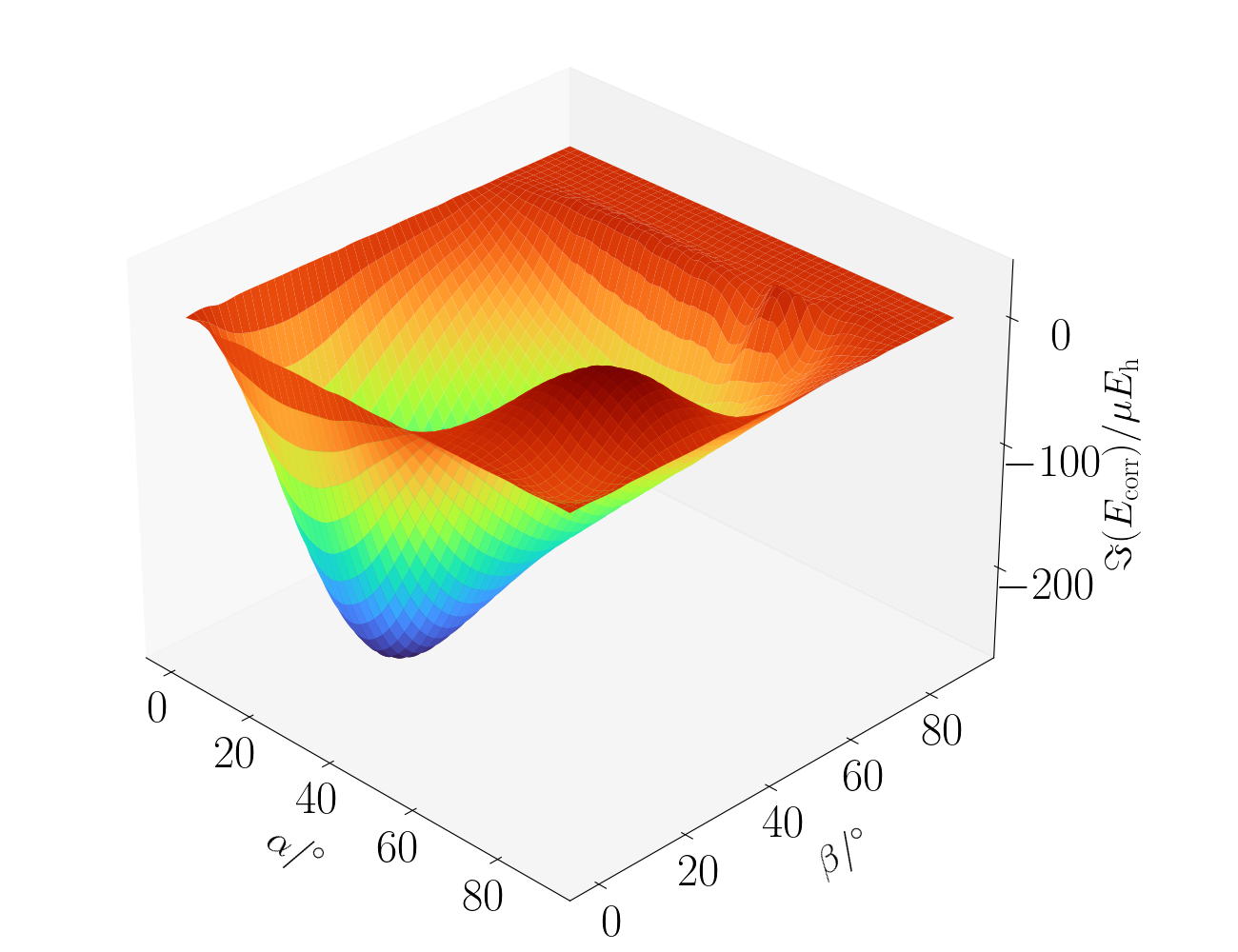}
    \caption{Imaginary energy of $\Psi_2$:  $\max(\Im{E\ped{CCSD}}(\Psi_2))= 38.25\mu E\ped{h}$ ($72^\circ,24^\circ$), $\min(\Im{E\ped{CCSD}}(\Psi_2))= -266.87\mu E\ped{h}$ ($24^\circ,24^\circ$).
    }  
    \label{fig:eom2_im}
    \end{subfigure}
    \begin{subfigure}{.95\columnwidth}
    \includegraphics[width=1.\linewidth]{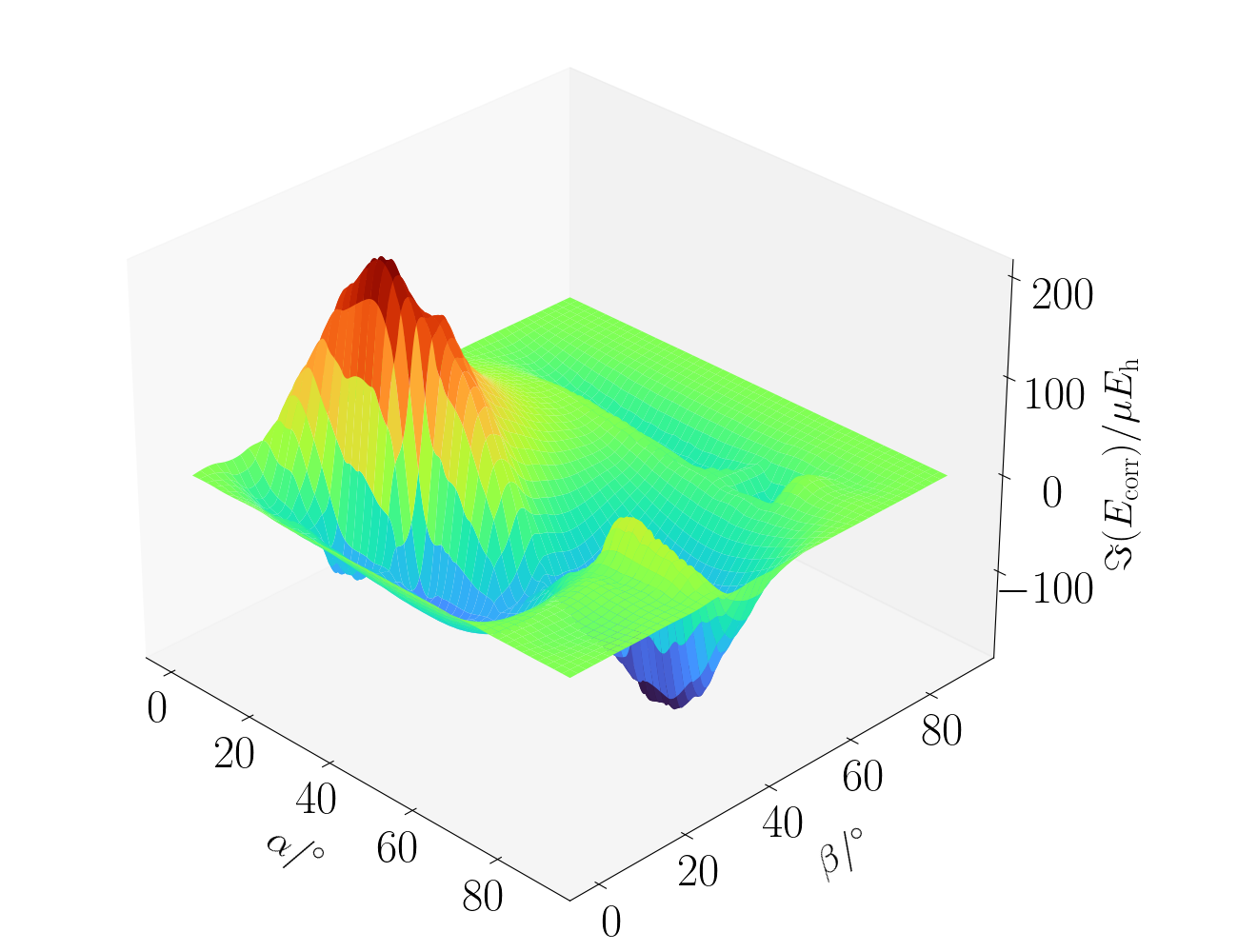}
    \caption{Imaginary energy of $\Psi_3$:  $\max(\Im{E\ped{CCSD}}(\Psi_3))= 182.48\mu E\ped{h}$ ($18^\circ,24^\circ$), $\min(\Im{E\ped{CCSD}}(\Psi_3))= -171.70\mu E\ped{h}$ ($72^\circ,42^\circ$).
    }  
    \label{fig:eom3_im}
    \end{subfigure}
    \caption{Imaginary part of the energy surfaces of the ground and first three excited states of the water molecule in a magnetic field of B=0.5 B$_0$ as a function of its orientation, as shown in Fig.~\ref{fig:water_orientation}, calculated at the ff-CCSD level of theory. The maximum and minimum values of the imaginary part of the ff-CCSD energy is given below each figure, where the positions of minima and maxima are indicated as coordinates $(\alpha,\beta)$.}
    \label{fig:im_h2o}
\end{figure*}

In Fig.~\ref{fig:surf_en}, the real parts of the ground- and excited-state ff-CCSD energies and the corresponding ff-UCC3 energies are plotted as a function of the coordinates $\alpha,\beta$. The underlying HF calculation is the same for both methods; therefore for the ground state only the correlation energy is plotted. In addition, in Fig.~\ref{fig:diff}, the differences $\Delta E=E\ped{UCC3}-E\ped{CCSD}$ between the ff-UCC3 and the ff-CCSD results are shown as a color map. 
Small differences are shown in blue, while larger differences evolve towards red. 

Fig.~\ref{fig:tot_h2o} presents the energy surfaces of all states concurrently to facilitate analysis of their interactions. Here we show the ff-UCC3 surfaces, while the corresponding (and very similar) ff-CCSD surfaces can be found in the \hyperlink{si}{supplementary material}.

From a qualitative point of view, the shape of the energy surfaces describing the states, obtained at the ff-CCSD and ff-UCC3 levels of theory, respectively, is the same for all investigated states (Fig.~\ref{fig:surf_en}). 
We note that for state $\Psi_3$ (see Fig. \ref{fig:diff3}), the two energy surfaces intersect around $\beta=60^\circ$, while for the other states no intersection is found.

In Figs.~\ref{fig:gs}-\ref{fig:eom_3} the surfaces of the real part of the energy values of the ground- and first three excited singlet states, $\Psi_0$, $\Psi_1$, $\Psi_2$ and $\Psi_3$ are shown. Since the states have the same symmetry ($C_1$) for a generic orientation of the magnetic field, avoided crossings can be observed. This is also visible in Fig.~\ref{fig:tot_h2o} where the potential energy surfaces of the states are shown together for ff-UCC3. 
The excited states $\Psi_2$ and $\Psi_3$ (Figs.~\ref{fig:eom_2} and \ref{fig:eom_3}) also exhibit an avoided crossing, visible at around $\beta=80^\circ$, where the two surfaces are very close to each other for both methods. For the third state, the crest at about $\beta=40^\circ$ hints at the mixing with higher-lying states not investigated here. \\
As mentioned above, the surfaces obtained with the ff-CCSD and ff-UCC3 methods qualitatively agree for all states. A more quantitative analysis is possible through the colour-map plots in Figs.~\ref{fig:diff1}-\ref{fig:diff3}. Overall, it is found that:
\begin{itemize}
\item
for the ground state, the energy difference  $\Delta E=E\ped{UCC3}-E\ped{CCSD}$ has only positive values: the ff-UCC3 energy is here always larger than the real part of the ff-CCSD energy. 
\item 
for $\Psi_1$ and $\Psi_2$, the difference $\Delta E=E\ped{UCC3}-E\ped{CCSD}$ has negative values for all polar angles, showing that for these two states the ff-CCSD surface lies above the ff-UCC3 surface.
\item 
for the state $\Psi_3$, the potential energy surfaces intersect in the vicinity of the crest. Hence, the energy difference goes from positive values on one side of the crest to negative values on the other side. 
\end{itemize}
For the maximum and minimum energy differences $ \Delta E^{\max}$ and  $\Delta E^{\min}$, respectively, we find: 
\begin{itemize}
\item 
For $\Psi_0$, the maximum energy difference is  $\Delta E^{\max}=2.02\;\text{m}E\ped{h}$, at about $\alpha=48^\circ$ and $\beta=51^\circ$, while the minimum energy difference of $\Delta E^{\min}=1.53\;\text{m}E\ped{h}$ is found at about $\alpha=81^\circ$ and $\beta=0^\circ$.  
\item For $\Psi_1$, the minimum energy difference value is $ \Delta E^{\min}= -12.13 \;\text{m}E\ped{h}$, at $\alpha=45^\circ$ and $\beta=0^\circ$. 
We note that in this region of the PES no avoided crossing or more complicated electronic structure is observed.  The maximum energy difference is at $\alpha=90^\circ$  and $\beta=12^\circ$, where $\Delta E^{\max}= -1.43 \;\text{m}E\ped{h}$. 
\item
For $\Psi_2$ and $\Psi_3$, large energy differences are found around the avoided crossing between these two states: we note in Figs.~\ref{fig:diff2}-\ref{fig:diff3} that blue regions (i.e., large values of $\Delta E$) are located around $\beta\approx 70^\circ$, where the avoided crossing is observed.
\item 
For $\Psi_{2}$, the minimum energy difference is $\Delta E^{\min}=-13.79\;\text{m}E\ped{h}$  at $\alpha=90^\circ$  and $\beta=0^\circ$, while the maximum energy difference is  $\Delta E^{\max}= -1.80\;\text{m}E\ped{h}$ at $\alpha=90^\circ$ and $\beta=72^\circ$. 
\item
For $\Psi_3$, the sign of the energy difference changes throughout the surface: the largest positive value of $\Delta E^{\max}=5.54\;\text{m}E\ped{h}$ is obtained at $\alpha=48^\circ$ and $\beta=60^\circ$, while the largest negative value (the largest absolute difference) of $\Delta E^{\min}=-14.72\;\text{m}E\ped{h}$ is observed at $\alpha=12^\circ$ and $\beta=48^\circ$.
\end{itemize}
 Overall, we note that the excitation energy differences between the two methods are between a few $\text{m}E\ped{h}$ and 15 $\text{m}E\ped{h}$. 
For a clearer characterization of the states, we additionally performed CCSDT calculations for the orientations showing the largest imaginary components in Fig.~\ref{fig:im_h2o}. The corresponding results are collected in table~\ref{tab:H2O_ccsdt}.
As expected, the imaginary parts of the CCSDT calculations are smaller than those obtained at the CCSD level of theory.  
We note that even at the CCSDT level the imaginary part for the excited states is about two orders of magnitude larger than for the ground state, showing the importance of a Hermitian approach especially for the excited-state calculations. Furthermore, also from the CCSDT calculations no double-excitation character was detected at these points in correspondence to the CCSD results.

The imaginary part of the CCSD energy is plotted in Fig.~\ref{fig:gs_im} (see also ref. \onlinecite{Thomas2021} for the ground state).
The maximum absolute value of the imaginary contribution to the energy is of about 2.5 $\mu E\ped{h}$, found at $\alpha=66^\circ$ and $\beta=24^\circ$. 

The magnitude of the imaginary part of the ground-state correlation energy may be said to be negligible, but this is no longer true for the excitation energies: for the state $\Psi_1$, Fig.~\ref{fig:eom1_im} shows that the imaginary part reaches positive values up to $ 69.87\;\mu E\ped{h}$ (at $\alpha=21^\circ$ and $\beta=18^\circ$) and negative values up to  $-48.31\;\mu E\ped{h}$ (at $\alpha=69^\circ$ and $\beta=24^\circ$).  
For excited states $\Psi_2$ and $\Psi_3$, the occurrence of complex eigenvalues becomes even more significant. In Fig.~\ref{fig:eom2_im}, a negative imaginary part of $-266\; \mu E\ped{h}$ is observed at $\alpha=24^\circ$ and $\beta=24^\circ$, while for the third excited state a maximum value of the imaginary part of  $182\;\mu E\ped{h}$ is found at $\alpha=18^\circ$ and $\beta=24^\circ$. The imaginary parts therefore reach a magnitude in the m$E\ped{h}$ regime, i.e. two orders of magnitude larger than previously observed for the ground state.  
Considering the behaviour of the real part of the energy (Fig.~\ref{fig:surf_en}) and the difference plot (Fig.~\ref{fig:diff}), no correlation between the difference in the energy values obtained with the two methods and the magnitude of the imaginary part given by ff-CCSD is found. 

The investigation shows that no physical interpretation could be derived for the minima or maxima of the imaginary parts of the CC energies. The obtained results all show a good agreement between UCC3 energies and the real part of CCSD results, validating the hypothesis that for this system the imaginary parts can indeed be neglected. These do not necessarily appear to be related to the avoided crossings that occur between the excited states.

\begin{table}[]
    \centering
    \begin{tabular}{|c|c|r|r|r|r|}
    \hline
         $\alpha,\beta$&State& $\Re E_{\text{CCSD}}$ &$\Im E_{\text{CCSD}}$&$\Re E_{\text{CCSDT}}$&$\Im E_{\text{CCSDT}}$  \\
         \hline
$66^\circ,24^\circ$ & $\Psi_0$    & -76.06 & 2.50    & -76.07       &-0.37     \\
$21^\circ,18^\circ$ & $\Psi_1$    & -75.90 & 69.64   & -75.91       & 11.38    \\
$69^\circ,24^\circ$ & $\Psi_1$    & -75.93 &-45.82   & -75.95       & -5.77    \\
$72^\circ,24^\circ$ & $\Psi_2$    & -75.87 & 40.63   & -75.88       & 8.30     \\
$24^\circ,24^\circ$ & $\Psi_2$    & -75.85 & -267.27 & -75.86       & -54.87   \\
$18^\circ,24^\circ$ & $\Psi_3$    & -75.66 & 182.08  & -75.91       & 8.72     \\
$72^\circ,42^\circ$ & $\Psi_3$    & -75.65 & -170.58 & -75.66       & -38.85   \\
         \hline
    \end{tabular}
    \caption{Total energies of the water molecule computed at the ff-CCSD and ff-CCSDT levels of theory, with the unc-cc-pVTZ basis set. The orientations were chosen for which the largest imaginary parts in the CCSD case have been found. The real part of the energy is given in Hartree, the imaginary part in $\mu H$. }
    \label{tab:H2O_ccsdt}
\end{table}

\subsection{Boric acid}\label{sec:Boh3}
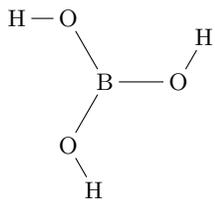
\begin{figure}
    \centering
        \centering
        \begin{tikzpicture}
          \node (fig) at(0,0){\chemfig{B([:0,1]-O(-[::60,0.7]H))([:120,1]-O(-[::60,0.7]H))([:-120,1]-O(-[::60,0.7]H))}};
        \end{tikzpicture}
        \caption{Boric acid B(OH)$_3$, exhibiting $C_{3h}$ symmetry.}
        \label{fig:BOH3}
\end{figure}
Systems with a complex Abelian point group symmetry have excited states that belong to pairs of complex conjugate IRREPs. These states are pairwise degenerate and are characterized by a complex wave function even in the absence of a magnetic field. Nevertheless, in a Hermitian framework these states can be calculated using real algebra by forming real linear combinations of the complex wave functions. These linear combinations no longer transform as the irreducible representations of the point group.
However, this real representation is no longer possible within the EOM-CC framework, where the non-Hermitian expression of the energy leads to the states belonging to the complex irreducible representations which have pairwise complex conjugate energy values and therefore are not truly degenerate. 
A real EOM-CC code therefore cannot compute these states. While it is possible to access these states using a complex EOM-CC code\cite{Kitsarasthes}, the use of complex algebra is more memory intensive and computationally expensive. 
Consequently, employing a real-valued program is advantageous in the field-free scenario. Accordingly, formalisms such as UCC theory, in which  energies are calculated via expectation values of Hermitian operators, are favored. 
We note that the standard CC framework effectively describes the closed-shell ground state, which corresponds to a real IRREP, while challenges arise primarily for the excited states.

In this study, boric acid is taken as an example. The molecule belongs to the complex Abelian point group $C_{3h}$ (see Fig.~\ref{fig:BOH3}). The point group $C_{3h}$ possesses two real IRREPs, $A^\prime$ and $A^{\prime\prime}$, and two pairs of complex-conjugate ones, $E_1^\prime, E_2^\prime$ and $E^{\prime\prime}_1,E^{\prime\prime}_2$. First, the field-free case is investigated; then the case of a perpendicular magnetic field is analyzed. We note that the latter orientation conserves the point-group symmetry of the system even though a magnetic field is applied. The magnetic field strength is varied up to 0.8 B$_0$, in steps of 0.5 B$_0$. The geometry used for all calculations was optimized at the field-free CCSD/unc-aug-cc-pVDZ\cite{pritchard2019a,feller1996a,schuchardt2007a,dunning1989a, kendall1992a} level of theory: $R\ped{BO}=2.6018\; a_0$, $R\ped{OH}=1.8181\; a_0$ and $\angle\text{BOH}=68.23^\circ$. The UCC3 energies of the ground state and the first excited state of each IRREP have been obtained with the \textsc{Qcumbre} program package, using the unc-aug-cc-pVDZ basis set. The CC3\cite{Kitsaras2024}, CCSD, and CISD results are taken from Ref. \onlinecite{Kitsarasthes}.

In table \ref{tab:boh3_0} the excitation energies of the lowest excited states of the IRREPs $A^{\prime}$, $A^{\prime\prime}$, $E^\prime$ and $ E^{\prime\prime}$, obtained at the CCSD, CC3, CISD, and UCC3 levels of theory are listed. As expected, the energies of the $A^\prime$ and $A^{\prime\prime}$ states are real for all methods, while for the complex IRREPs $E^\prime$ and $ E^{\prime\prime}$, the CC methods find pairs of complex-conjugate values. The UCC3 results, on the other hand, correctly predict real degenerate energies. The discrepancies between the CCSD and CC3 results are of the order of $0.001\;E\ped{h}$, while the differences between the CC3 and UCC3 values are of the order of $0.01\;E\ped{h}$. 
The better agreement between CCSD and CC3 can be attributed to the fact that they are different truncations of the same wave function ansatz. In table~\ref{tab:boh3_0}, the CISD results for the EOM-CC energies show large discrepancies with respect to the CC3 results of the order of $0.1\;E\ped{h}$. Among the two methods presented here that yield real energies, UCC3 is preferred to CISD because of its superior accuracy. In the following discussion, to account for the large discrepancies in correlation energies observed for CISD compared to the other methods, the CISD results have been shifted to coincide with the CCSD energies at $B=0$.  

\begin{figure*}[]
\centering
    \includegraphics[width=1.\linewidth]{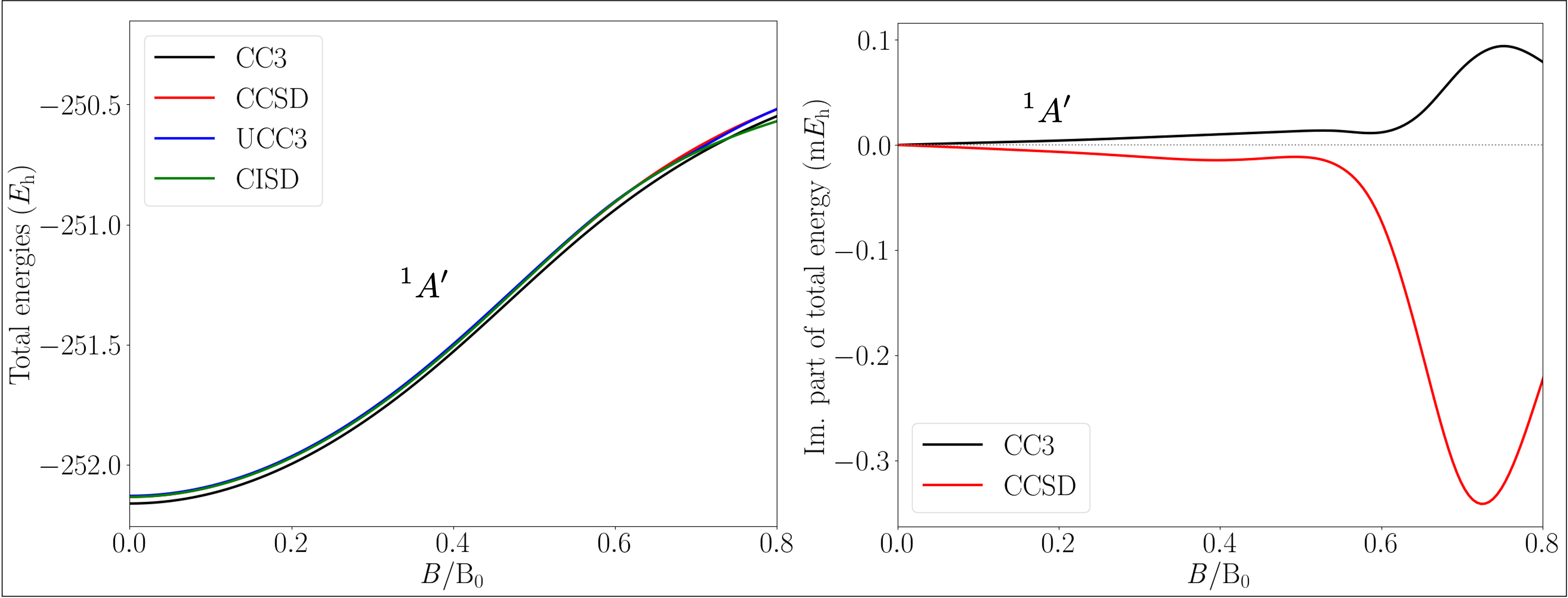}
    \caption{Total energy of the ground state of  boric acid B(OH)$_3$, in an external magnetic field, directed perpendicularly to the molecular plane. The field strength varies in the interval between 0 B$_0$-0.8 B$_0$. The left panel shows the comparison between the real parts of the energies computed at the CC3, CISD, CCSD, and UCC3 levels of theory. The right panel shows the non-vanishing imaginary parts of the CC energies.}
    \label{fig:BOH3_tot_en}
\end{figure*}

\begin{table*}
    \centering
    \begin{tabular}{|l|l|l|l|l|l|l|}
    \hline
    \multirow{2}{*}{Methods}& \multicolumn{4}{|c|}{Excitation energies/$E\ped{h}$}\\
      \cline{2-5}
     &$A^{\prime}$ & $A^{\prime\prime}$ &  $E^\prime$ &  $ E^{\prime\prime}$\\
    \hline
    CCSD& 0.356378 & 0.332552 &0.364306$\pm$0.000047\textit{i}&0.301831$\pm$0.000051\textit{i}\\
    CC3 & 0.354183& 0.331149 & 0.362247$\pm$0.000004\textit{i}&0.298958$\pm$0.000011\textit{i}\\
    UCC3 &  0.366560 &  0.338875 & 0.373787	& 0.308129\\
    CISD &0.255217 & 0.232414& 0.263483	& 0.201056\\
   \hline
        \end{tabular}
    \caption{Excitation energies  ($E\ped{h}$) of the lowest singlet states  $1A^{\prime}$, $1A^{\prime\prime}$, $1E^\prime$ and $1E^{\prime\prime}$, at the CCSD, CC3, UCC3, and CISD levels of theory, computed with the unc-aug-cc-pVDZ basis set.\cite{pritchard2019a,feller1996a,schuchardt2007a,dunning1989a, kendall1992a} For CCSD and CC3, the energies are pairs of complex-conjugate values.}
    \label{tab:boh3_0}
\end{table*}
Fig.~\ref{fig:BOH3_tot_en} shows the total energy of the ground state as a function of the magnetic field strength. In an increasing magnetic field, the energy increases, due to the action of the diamagnetic term in the Hamiltonian. The spin-Zeeman term does not influence the energy, as the ground state is a closed-shell singlet state. The left panel displays the results for the real part in the energy, with practically indistinguishable curves for the CCSD, shifted CISD, and UCC3 methods, while the inclusion of triple excitations shifts the CC3 energy to slightly lower values, on average about $0.03\;E_{h}$ below CCSD and UCC3.  However, for both the CC3 and CCSD methods, a non-vanishing imaginary part arises in a magnetic field (right panel of Fig.~\ref{fig:BOH3_tot_en}).  For the range below 0.3 B$_0$, the imaginary part of the ground state energy $\Im{E\ped{GS}}$ is of the order of $\approx 10^{-5}\;E\ped{h}$. Around 0.7 B$_0$, however, both CCSD and CC3 are affected by an increasing imaginary part of the total energy, up to a maximum of $\approx 0.34\;\text{m}E\ped{h}$ and $\approx 0.09\;\text{m}E\ped{h}$, respectively. When going from CCSD to CC3, the magnitude of the imaginary part decreases, as could be expected from the fact that in the limit of considering the full excitation operator in the CC parameterization, the FCI limit is reached and no imaginary components occur. The presence of complex energies does not seem to provide particular insight into the accuracy of the real part which does not change significantly.  

\begin{figure*}[]
\centering
    \begin{subfigure}{\textwidth}
    \includegraphics[width=1.\linewidth]{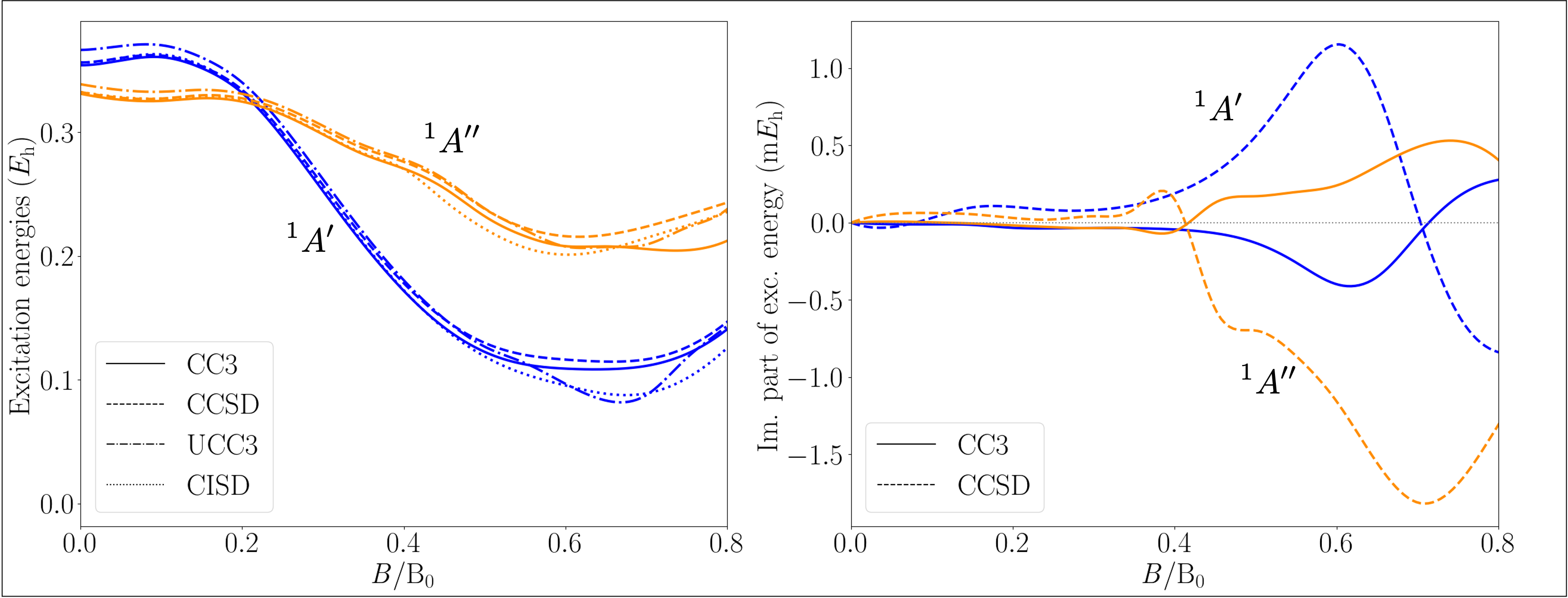}
    \caption{}
    \label{fig:BOH3_1}
    \end{subfigure}
    \begin{subfigure}{\textwidth}
    \includegraphics[width=1.\linewidth]{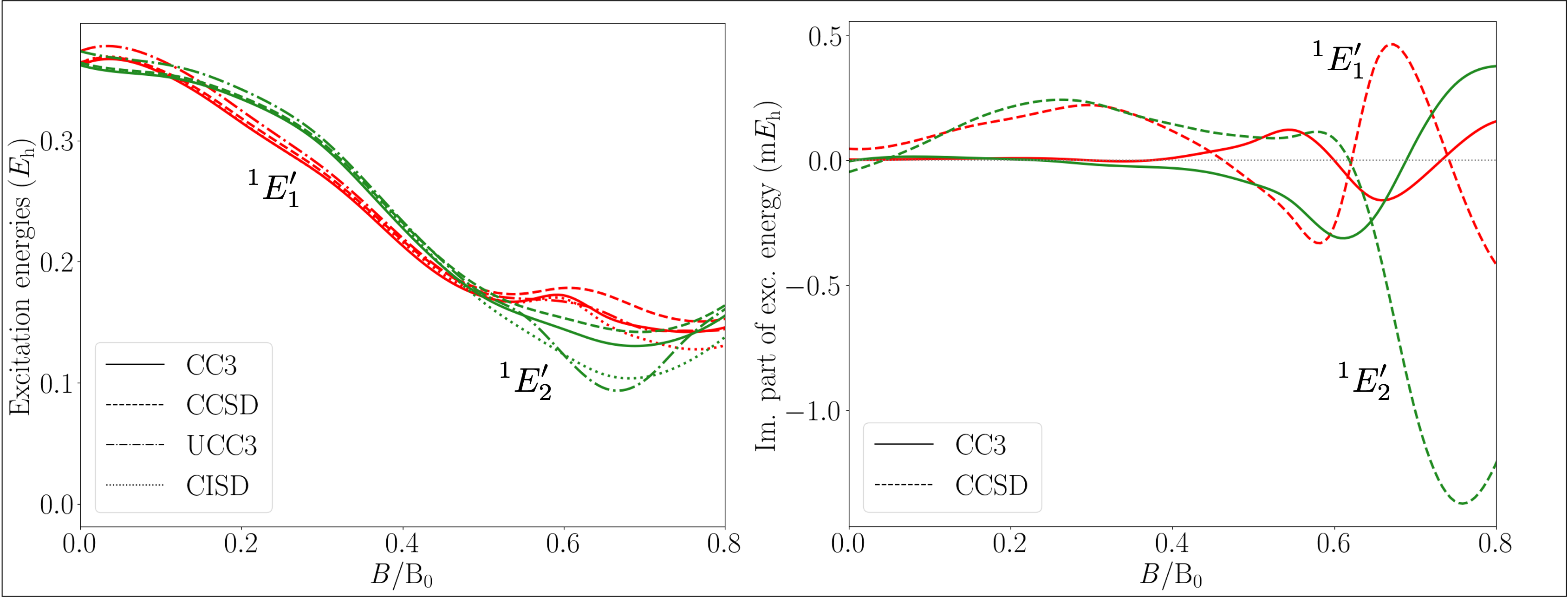}
    \caption{}
    \label{fig:BOH3_2}
    \end{subfigure}
    \begin{subfigure}{\textwidth}
    \includegraphics[width=1.\linewidth]{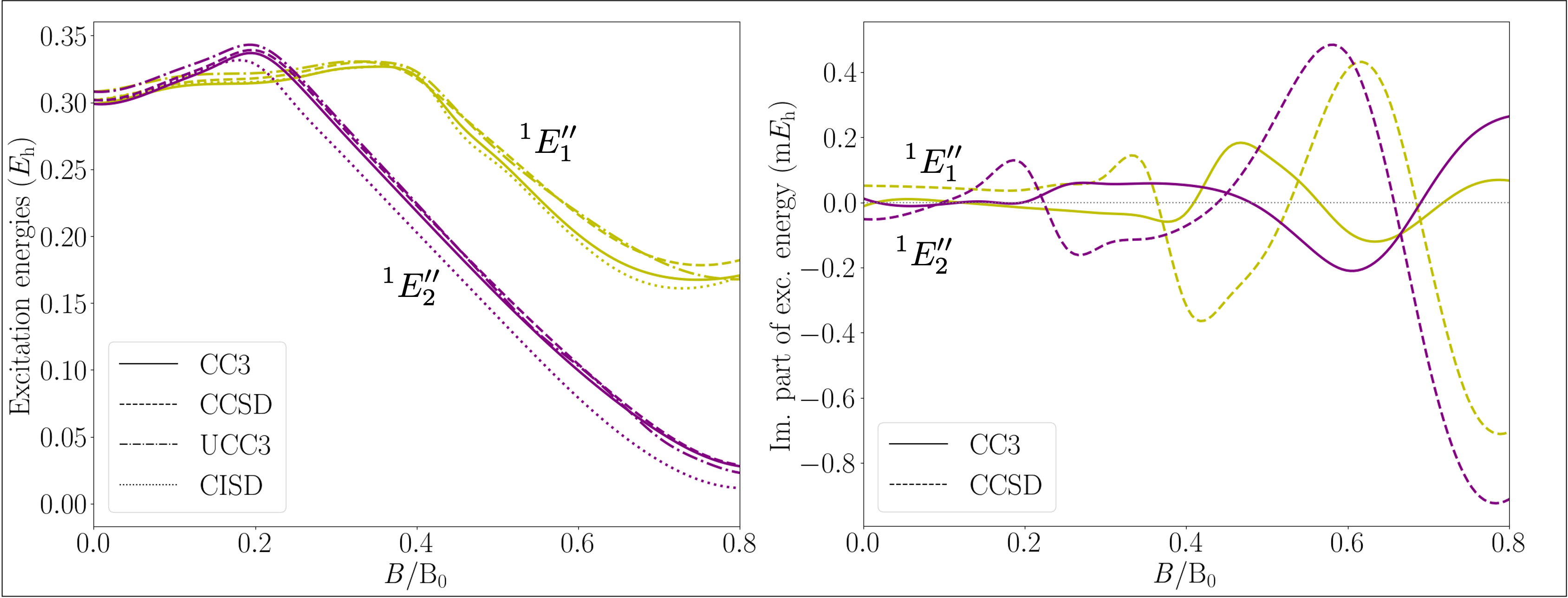}
    \caption{}
    \label{fig:BOH3_3}
    \end{subfigure}
    \caption{Excitation energies of low-lying singlet states of each IRREP for B(OH)$_3$, in an external magnetic field, directed perpendicularly to the molecular plane. The field strength varies in the interval 0 B$_0$-0.8 B$_0$. In the left column, the comparison between the real parts of the energies computed at the CC3, CCSD, CISD, and UCC3 levels of theory is shown. In the right column, the non-vanishing imaginary parts of the CC energies are shown.}
    \label{fig:BOH3_exc}
\end{figure*}

\begin{table}[]
    \centering
    \begin{tabular}{|c|r|r|r|r|}
    \hline
    State & $\Re E_{\text{CCSD,TZ}}$& $\Im E_{\text{CCSD,TZ}}$ & $\Re E_{\text{CCSD,QZ}}$ & $\Im E_{\text{CCSD,QZ}}$ \\
    \hline
$^1A^\prime$   &1.13$\cdot 10^{-01}$     & 9.99$\cdot 10^{-04}$  &     1.13$\cdot 10^{-01}$   &    9.57$\cdot 10^{-04}$\\
$^1E_1^\prime$ &1.69$\cdot 10^{-01}$     & 5.28$\cdot 10^{-04}$  &     1.67$\cdot 10^{-01}$   &    5.03$\cdot 10^{-04}$\\
$^1E_2^\prime$ &1.44$\cdot 10^{-01}$     &-2.17$\cdot 10^{-04}$  &     1.44$\cdot 10^{-01}$   &   -2.87$\cdot 10^{-04}$\\
$^1A^{\prime\prime}$   &2.12$\cdot 10^{-01}$     &-1.27$\cdot 10^{-03}$  &     2.10$\cdot 10^{-01}$   &   -1.25$\cdot 10^{-03}$\\
$^1E_1^{\prime\prime}$  & 1.91$\cdot 10^{-01}$     & 4.79$\cdot 10^{-04}$  &     1.91$\cdot 10^{-01}$   &    4.67$\cdot 10^{-04}$\\
$^1E_2^{\prime\prime}$  & 7.53$\cdot 10^{-02}$     & 2.84$\cdot 10^{-04}$  &     7.55$\cdot 10^{-02}$   &    2.68$\cdot 10^{-04}$\\
\hline
    \end{tabular}
    \caption{Excitation energies of boric acid in an external magnetic field of 0.65 B$_0$, computed at the CCSD/aug-cc-pVTZ and CCSD/aug-cc-pVQZ levels of theory. The energies are given in Hartree.}
    \label{tab:boh3_ccsd_tz_qz}
\end{table}

\begin{table*}[]
    \centering
    \begin{tabular}{|c|r|r|r|r|r|r|}
    \hline
    State & $\Re E_{\text{CCSD,DZ}}$ ($E_{\rm h}$)& $\Im E_{\text{CCSD,DZ}}$ (m$E_{\rm h}$)& $\Re E_{\text{CC3,DZ}}$ ($E_{\rm h}$)& $\Im E_{\text{CC3,DZ}}$ (m$E_{\rm h}$)&$\Re E_{\text{CCSDT,DZ}}$ ($E_{\rm h}$)& $\Im E_{\text{CCSDT,DZ}}$ (m$E_{\rm h}$) \\
    \hline
   $^1A^\prime$ & 1.15$\cdot10^{-1}$&  1.05 &  1.09$\cdot 10^{-1}$& 3.83$\cdot 10^{-1}$ &1.08$\cdot 10^{-1}$ & 4.48$\cdot 10^{-1}$ \\
$^1E_1^\prime$ &  1.73$\cdot10^{-1}$ & 5.75$\cdot10^{-1}$ & 1.86$\cdot 10^{-1}$&-8.13$\cdot 10^{-2}$ &1.82$\cdot 10^{-1}$ &-1.98$\cdot 10^{-1}$ \\
$^1E_2^\prime$ &  1.45$\cdot10^{-1}$ &-9.70$\cdot10^{-2}$ & 1.33$\cdot 10^{-1}$& 2.56$\cdot 10^{-1}$  &  1.31$\cdot 10^{-1}$ & 2.07$\cdot 10^{-1}$\\
$^1A^{\prime\prime}$ &  2.17$\cdot10^{-1}$ &-1.36 & 2.08$\cdot 10^{-1}$& -3.34$\cdot 10^{-1}$ & 2.06$\cdot 10^{-1}$ &-4.80$\cdot 10^{-1}$ \\
$^1E_1^{\prime\prime}$ & 1.96$\cdot10^{-1}$ & 5.19$\cdot10^{-1}$ &  1.81$\cdot 10^{-1}$& 1.45$\cdot 10^{-1}$ & 1.75$\cdot 10^{-1}$ & 1.50$\cdot 10^{-1}$\\
$^1E_2^{\prime\prime}$ &  7.84$\cdot10^{-2}$ & 2.79$\cdot10^{-1}$ & 7.51$\cdot 10^{-2}$& 1.77$\cdot 10^{-1}$  &  7.42$\cdot 10^{-2}$ & 1.56$\cdot 10^{-1}$ \\
\hline
    \end{tabular}
    \caption{Excitation energies of boric acid in an external magnetic field of 0.65 B$_0$, using an uncontracted aug-cc-pVDZ basis set, computed at the frozen-core CCSD, CC3, CCSDT levels of theory.}
    \label{tab:boh3_ccsdt}
\end{table*}

\begin{table*}[]
    \centering
    \begin{tabular}{|c|r|r|}
    \hline
    State & $\Re (\Delta E_{\text{CCSD,DZ}})$ ($E_{\rm h}$)&$\Re (\Delta E_{\text{UCC3,DZ}})$ ($E_{\rm h}$)\\
    \hline
$^1A^\prime$ &  6.78$\cdot 10^{-3}$& -2.47$\cdot 10^{-2}$\\
$^1E_1^\prime$ &   -9.14$\cdot 10^{-3}$   & -2.41$\cdot 10^{-2}$ \\
$^1E_2^\prime$ &     1.44$\cdot 10^{-2}$ & -3.53$\cdot 10^{-2}$ \\
$^1A^{\prime\prime}$ &  1.13$\cdot 10^{-2}$  & 1.50$\cdot 10^{-3}$\\
$^1E_1^{\prime\prime}$ &  2.07$\cdot 10^{-2}$  &  3.05$\cdot 10^{-2}$\\
$^1E_2^{\prime\prime}$ &   4.23$\cdot 10^{-3}$  & 2.43$\cdot 10^{-3}$\\
\hline
    \end{tabular}
    \caption{Differences in excitation energies of CCSD and UCC3 with respect to the CCSDT reference, of boric acid in an external magnetic field of 0.65 B$_0$, using an uncontracted aug-cc-pVDZ basis set. The differences are calculated as $\Re (\Delta E_{\text{CCSD,DZ}})= \Re E_{\text{CCSD,DZ}}-\Re E_{\text{CCSDT,DZ}}$ and $\Re (\Delta E_{\text{UCC3,DZ}})= \Re E_{\text{UCC3,DZ}}-\Re E_{\text{CCSDT,DZ}}$.}
    \label{tab:boh3_delta_ccsdt}
\end{table*}

Similar to the ground-state energy, Fig.~\ref{fig:BOH3_exc} shows the excitation energies of the first excited singlet states of each IRREP as function of the magnetic field strength. For each figure, the left panel compares the real part of the excitation energies, computed with the four methods. In Fig.~\ref{fig:BOH3_1}, the states $^1A^\prime$ and $^1A^{\prime\prime}$, in Fig.~\ref{fig:BOH3_2}, the states $^1E^\prime_1$ and $^1E^\prime_2$ and in Fig.~\ref{fig:BOH3_3}, the states $^1E^{\prime\prime}_1$ and $^1E^{\prime\prime}_2$ are shown. It is observed that the imaginary part of the CCSD energies is larger than the corresponding imaginary part of the CC3 energies. The extrema are observed at similar field strengths, but the role of minima and maxima is reversed.  The states $^1E^{\prime\prime}_1,^1E^{\prime\prime}_2$, are the energetically lowest excited states as they are characterized by the HOMO-LUMO transition. As explained before, the states belonging to the complex IRREPs, $E^\prime$ and $E^{\prime\prime}$, start off as a degenerate pair and are split by the magnetic field. For all states a decrease in the excitation energy when going to higher magnetic-field strengths is observed. The three  methods are in good agreement with each other for field strengths up to 0.5 B$_0$ for each of the inspected states, while qualitative differences are observed when going to higher field strengths. These differences are likely due to different avoided crossings with higher-lying states of the same symmetry.

In Fig.~\ref{fig:BOH3_1}, the energies of the states $^1A^\prime$ and $^1A^{\prime\prime}$ are displayed as a function of the magnetic-field strength. For field strengths larger than 0.2 B$_0$, the $^1A^\prime$ state has a lower energy than the $^1A^{\prime\prime}$ state. Major differences are observed in the range between 0.55 B$_0$ and 0.75 B$_0$. From the inspection of the amplitudes of the $^1A^{\prime\prime}$ state obtained with the CC3 method, a double-excitation character appears to be present. The double-excitation character is also found by the UCC3 method, while it is absent in the CCSD results. The energy lowering at about 0.7 B$_0$, found by CC3, is described differently by UCC3, CISD, and CCSD. The discrepancy with respect to CC3 might stem from the fact that the other methods, due to the limitation of the excitation space to singles and doubles, do not describe the double-excitation character well. For the $^1A^\prime$ state, the energy lowering is observed for both CISD and UCC3. However, in this region the UCC3 description of the state acquires a partial double-excitation character, which is absent in the CC3 results. Therefore, the shape of the $^1A^\prime$ curve differs from that obtained with the other methods which describe the $^1A^\prime$ state via a single excitation. The right panel of Fig.~\ref{fig:BOH3_1} shows the corresponding imaginary parts of the CC3 and CCSD results. For the states belonging to the real IRREPs, $^1A^\prime$ and $^1A^{\prime\prime}$, the excitation energies in the field-free case are real. The plotted imaginary values in the right panel of Fig.~\ref{fig:BOH3_1} therefore depart from 0 $E\ped{h}$ at B=0 B$_0$. For CCSD, the maximum of $\abs{\Im{E\ped{exc}}}$ of 1.8 m$E\ped{h}$ is reached by the $^1A^{\prime\prime}$ state, while the same state for CC3 has a maximum of $\abs{\Im{E\ped{exc}}}$ of 0.4 m$E\ped{h}$. The largest values of $\abs{\Im{E\ped{exc}}}$ for the CC3 results are found in correspondence to the largest double-excitation character. The CCSD curves show maxima at similar field strengths where CC3 shows minima and vice versa.\\ 
For the IRREP $E^\prime$ (Fig.~\ref{fig:BOH3_2}), the same features as in Fig.~\ref{fig:BOH3_1} can be observed for magnetic field strengths larger than 0.55 B$_0$. The  $^1E_1^\prime$ state acquires a double-excitation character, causing major differences in the results for the four different methods between 0.55 B$_0$ and 0.70 B$_0$. Again, the CC3 and UCC3 results exhibit a double-excitation character, while CCSD possesses single-excitation character. For the $^1E_2^\prime$ state, the CC3 and UCC3 results possess a double-excitation character between 0.60 B$_0$ and 0.75 B$_0$. From the right panel, it is observed that the two states have complex-conjugate energy values in the field-free case, starting symmetrically around the $x$-axis, as expected. In the finite field, the energies do no longer occur as pairs of complex-conjugate values and evolve independently. The imaginary part is no longer negligible at higher magnetic-field strengths, especially in the range 0.4 B$_0$-0.8 B$_0$. In particular, the maximum of $\abs{\Im{E\ped{exc}}}$ for the $^1E_1^{\prime}$ state is 0.53 m$E\ped{h}$ for CCSD and  0.16 m$E\ped{h}$  for CC3, while for the $^1E_2^{\prime}$ state it is 1.37 m$E\ped{h}$ for CCSD and 0.38 m$E\ped{h}$ for CC3. Similarly to the case shown in Fig.~\ref{fig:BOH3_2}, it can be noticed that the maxima of $\abs{\Im{E\ped{exc}}}$ for the CC3 energies correspond to the presence of a double-excitation character of the states, as seen from the contributions to the R amplitudes. In addition, we observe that the overall development of the imaginary component is quite different between CCSD and CC3. For CC3 the extrema are smaller than for CCSD. Often, the extrema of CCSD and CC3 have a different sign, respectively.  

Fig.~\ref{fig:BOH3_3} shows the energies of the two lowest-lying states belonging to the IRREP $E^{\prime\prime}$. Here, no major differences in the development of the energies as a function of the field strength are observed. Avoided crossings occur at 0.2 B$_0$ for the $^1E_2^{\prime\prime}$ state and 0.4 B$_0$ for the $^1E_1^{\prime\prime}$ state. Both states have a small double-excitation character (however not predominant over the single-excitation character), observed in the CC3 and UCC3 results which is absent for CCSD. The maximum of $\abs{\Im{E\ped{exc}}}$ for the $^1E_1^{\prime\prime}$ state is 0.71 m$E\ped{h}$ for CCSD and 0.16 m$E\ped{h}$ for CC3, while for the $^1E_2^{\prime\prime}$  state the maximum of $\abs{\Im{E\ped{exc}}}$ is 0.91 m$E\ped{h}$ for CCSD and 0.27 m$E\ped{h}$ for CC3. The extrema of $\Im{E\ped{exc}}$ for CC3 once more correspond to field strengths at which the largest double-excitation character is observed. The extrema of $\Im{E\ped{exc}}$ for the CCSD energies are found at similar field strengths, with discrepancies of at most 0.1 B$_0$.

Table~\ref{tab:boh3_ccsd_tz_qz} shows the excitation energies of boric acid in an external magnetic field of 0.65 B$_0$, calculated at the CCSD/aug-cc-pVTZ and CCSD/aug-cc-pVQZ levels of theory. This magnitude of the magnetic field was chosen because around this value all excited states are characterized by a non-vanishing imaginary part. From the listed values, it appears that increasing the basis set from aug-cc-pVTZ to aug-cc-pVQZ does not reduce the magnitude of the imaginary part, thus showing that it is not an artifact of the finite basis set. \\
In table~\ref{tab:boh3_ccsdt}, for a field strength of 0.65 B$_0$ using the aug-cc-pVDZ basis, calculations for full triples (CCSDT) have been performed as well. The calculations were performed using the frozen-core approximation. The table shows that the imaginary part of the CCSDT results is of the same order of magnitude as the imaginary part of the CCSD and CC3 results. The inspection of the excitation amplitudes shows that both CC3 and CCSDT detect a partial double-excitation character for the states $^1 E^{\prime}_1,^1 E^{\prime}_2,^1 A^{\prime\prime},^1 E^{\prime\prime}_1,^1 E^{\prime\prime}_2$, while CCSD describes these states only through single excitations.
\\
In table~\ref{tab:boh3_delta_ccsdt}, the discrepancies between the real parts of CCSD and UCC3 results with respect to full triples (CCSDT) are shown. These differences, calculated as $\Re (\Delta E_{\rm method})=\Re E_{\rm method}-\Re E_{\rm CCSDT}$, don't show a unique trend. For some states ($^1 A^{\prime},^1E_1^{\prime}, ^1E_2^\prime, ^1 E^{\prime\prime}$), $\Re (\Delta E_{\rm CCSD})$ is slightly smaller than $\Re (\Delta E_{\rm UCC3})$, while for the other states ($^1 A^{\prime\prime},^1E_2^{\prime\prime}$) $\Re (\Delta E_{\rm UCC3})$ is smaller than $\Re (\Delta E_{\rm CCSD})$. The discrepancies are in most cases of the same order of magnitude. Therefore, the accuracy of the two methods is comparable, as it was shown in section~\ref{sec:CH+_en}.

In summary, UCC3 constitutes an accurate method to calculate degenerate excited states of a system belonging to a complex Abelian point group, without having to resort to the use of a complex code. Here, the importance of having a Hermitian formalism becomes apparent, as EOM-CC cannot find these states in a real-valued framework.\\
The problems arising from the non-Hermiticity of the CC theory are also evident in the finite-field case: the large imaginary parts observed in the excitation energies at higher field strengths show a complicated behaviour as a function of the magnetic field strength. From this study, it seems that the largest values of the imaginary parts are found in correspondence to a partial double-excitation character in the description of the excited states. Concluding, the UCC3 approach seems advantageous at least in cases where complex Abelian point groups or imaginary components to the energy occur. 

\section{Conclusions and perspectives}\label{sec:concl}
In this paper, we have described and investigated a finite-field version of the UCC$n$ approach. This development was motivated by the limitations known for CC theory, arising from the non-Hermiticity of the theory.  These limitations are well-documented in the literature: complex energy eigenvalues are found for systems in a magnetic field\cite{Thomas2021} and for excited states in the proximity of conical intersections.\cite{Khn2007,kjonstad2019orbital} 

The adopted UCC ansatz maintains the advantages of an exponential parameterization of the wave function, and the Hermiticity of the energy expression guarantees real energies. It was shown in section ~\ref{sec:results} that UCC theory is an alternative to CC theory for the calculation of molecular energies and properties.

Following the formalism first proposed by Liu \textit{et al.},\cite{Liu2018} the method has been adapted to the finite magnetic-field case, implying the use of complex algebra. Through this adaptation, both ground- and excited states could be targeted, maintaining a structure of the equations similar to CC theory.  This work focuses on two approximations of UCC theory, determined by a perturbative truncation of the amplitude equations at second (UCC2) and  third order (UCC3). Both methods have been implemented in the QCUMBRE program package.\cite{QCUMBRE} The methylidyne  cation CH$^+$ was taken as an example to investigate the comparability of different CC and UCC truncations, where different orientations of the magnetic field were considered to analyze the different descriptions of avoided crossings between different states. In fact, this system was chosen because of its low-lying doubly-excited state, which for some orientations possesses avoided crossings with singly-excited ones. The comparable accuracy of the CCSD and UCC3 results has been outlined, while UCC2 proved unsuitable to treat states with significant double-excitation character.

The analysis of complex eigenvalues was performed for water and boric acid. 
For the water molecule in a magnetic field of different orientations, the qualitative description of the ground- and excited-state energies was found to agree between ff-CCSD and ff-UCC3. For ff-CCSD, the imaginary contributions to the energies were found to be significant, especially in the excited states. Also, occurrence of complex energies did not turn out to be a viable diagnostic criterion for the quality of the real part of the ff-CCSD results. \\
Boric acid invstigated  which is a case for which a real-valued quantum-chemical code cannot find the excited states of the IRREPs $E^\prime$ and $E^{\prime\prime}$ using a standard CC code. 
In the field-free case, a corresponding UCC calculation involves only real algebra. This is a clear advantage over standard CC theory, where the non-Hermiticity of the energy expression necessitates the use of complex algebra for the excited states of B(OH)$_3$, even without an external field.
B(OH)$_3$ has also been analyzed within a strong magnetic field. Here, UCC represents a solution to the problem of non-negligible imaginary parts in the excited-state energies. Similar to the analysis of water, it was found that the imaginary part conveys no clear diagnostic criterion. 
A large imaginary component did not signify a corresponding discrepancy between the ff-UCC3 and ff-CCSD results. The largest imaginary components could be found in correspondence to a partial double-excitation character in the description of the excited states. The comparison with UCC showed that the real part of the CC energies is often a good approximation. However, without a comparison the quality of the CC data remains unclear.

The discussed truncation scheme is not unique and different UCC formalisms may be explored. Most UCC methods are based on a perturbative truncation of the expansion. However, recent studies have investigated truncation schemes based on expansions up to a certain rank of commutators.\cite{liu2021unitary, Liu2022, Bauman2022, Hoffmann1988, Neuscamman2009,UCC_deprince} These methods seem to be advantageous for molecules for which the Møller-Plesset series does not show smooth convergence at low orders. For example, the qUCCSD approach is discussed to improve the accuracy of UCC3 for molecules with strong orbital relaxation and electron correlation.\cite{liu2021unitary, Liu2022} Exploration of such methods in the ff context would hence presumably lead to more accurate results for more molecules exhibiting strong electron-correlation effects.

This study focuses on the Hamiltonian in a magnetic field which leads to complex energy values in the CC framework. However, in ref.~\onlinecite{Thomas2021} also other conditions determining a complex part in the Hamiltonian have been analyzed extensively. Among these, the vicinity to conical intersections is documented in the literature to cause the CCSD results to become complex. 
In addition to other possible solutions to this problem,\cite{kjonstad2019orbital,Kjnstad2024,Stoll2025} the Hermitian formulation of the UCC energy clearly provides a natural way to eliminate unphysical results in this context.

The increasing interest in unitary formulations is also motivated by the direct application of these methods in the field of quantum computing.\cite{Romero2018, Lee2018, Evangelista2019, Anand2022} Therefore the different truncation schemes in quantum chemistry could be compared to the strategies used in the encoding of states in quantum computing, especially in the presence of an external magnetic field.\cite{Culpitt2023} 

The perspectives presented here highlight just one part of the broader research landscape opened up by the application of UCC methods in quantum chemistry. Systematic and in-depth investigations into the behavior, scalability, and accuracy of UCC across a wide spectrum of molecular systems and correlation regimes are essential to rigorously assess its computational utility and to delineate its range of applicability.

 \section{Supplementary material}
Supplementary material provides:
\begin{itemize}
\item a diagrammatic representation of all UCC equations
\item the CCSD potential energy surfaces for the excited states of water, in analogy to fig.~\ref{fig:tot_h2o}
\item tables with data supporting the plots in the results sections \ref{sec:CH+_en} and \ref{sec:Boh3}.
\end{itemize} 
 
\section{Acknowledgments}
The authors thank Prof. Jürgen Gauss (Johannes Gutenberg-Universität Mainz) for helpful discussions concerning unitary and traditional Coupled-Cluster theory; Prof. Lan Cheng (Johns Hopkins University) and Prof. Junzi Liu (University of Science and Technology Beijing) for helpful discussions concerning Unitary Coupled-Cluster diagrams and for providing reference data; Prof. Anna Krylov (University of Southern California) for bringing to our attention the issue of the killer condition.

Funding for this work was provided by the Deutsche Forschungsgemeinschaft via grant STO-1239/1-1.

\section{Conflicts of interest}
The authors have no conflict to disclose.

\section{Author declarations}
\subsection{Author contributions}
Laura Grazioli: conceptualization, investigation, methodology, software, validation, writing – original draft\\
Marios-Petros Kitsaras: investigation, software, writing - review\\
Stella Stopkowicz: conceptualization, funding acquisition, project administration, resources, supervision, writing – review and editing.

\subsection{Data availability}
The data that support the findings of this study are available within the article and its supplementary material.

\section{ORCID}

Laura Grazioli https://orcid.org/0000-0002-9404-8587

Marios-Petros Kitsaras https://orcid.org/0000-0002-9549-3674

Stella Stopkowicz https://orcid.org/0000-0002-0037-7962

\section{Appendix}
\begin{figure}
    \centering
    \includegraphics[width=\linewidth]{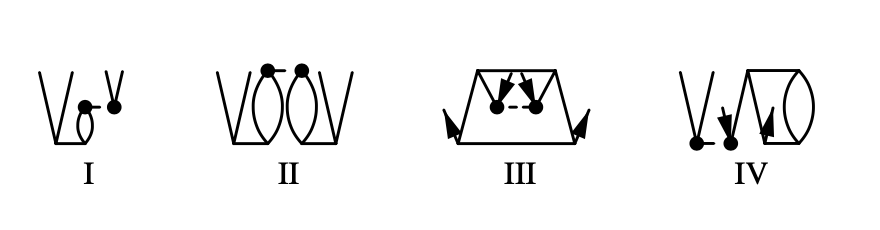}
    \caption{}
        \label{fig:diag_rules}
\end{figure}
In Fig. \ref{fig:diag_rules}, four examples of UCC diagrams are provided, in order to explain the diagrammatic rules  listed in sec.~\ref{sec:diag}.
\begin{itemize}
    \item I: In this term, the involved $\hat{V}$ operator is part of the so-called rest part $\hat{V}\ped{R}$, as it is not a pure excitation nor a pure de-excitation operator. This term arises from the single commutator and belongs to both $\frac{1}{2}[\hat{V}\ped{R},\hat{\sigma}]$ and $\frac{1}{2}[\hat{V},\hat{\sigma}]$. These two contributions add and the term has hence a prefactor of $\frac{1}{2}+\frac{1}{2}=1$ in front of the commutator. The same global prefactor is found for the single commutator in the BCH expansion, i.e., $1\cdot[\hat{V},\hat{\sigma}]$. The same rules as for the diagrammatic formalism of CC apply. The term is evaluated as $P(ij)P(ab)\braket{ak||ic}\sigma_{jk}^{bc}$.
    \item II: The potential operator here is part of $\hat{V}\ped{ND}$, as it only involves de-excitations within the given truncation level. Furthermore, the contraction with the $\hat{\sigma}$ operator results in a term belonging to $[\hat{V},\hat{\sigma}]\ped{R}$. According to Rule 2, this term is therefore part of the double commutators $\frac{1}{12}[[\hat{V}\ped{ND},\hat{\sigma}],\hat{\sigma}]$ and $\frac{1}{4}[[\hat{V},\hat{\sigma}]\ped{R},\hat{\sigma}]$. The global prefactor to this double commutator for this diagram is $\frac{1}{12}+\frac{1}{4}=\frac{1}{3}$. This prefactor differs from the prefactor of the BCH expansion, i.e., $\frac{1}{2}[[\hat{V},\hat{\sigma}],\hat{\sigma}]$. \\
    From the CC diagrammatic rules, a factor $\frac{1}{2}$ arises from the connection of the potential operator with two equivalent operators. This factor needs to be scaled according to rule 3 in sec.~\ref{sec:diag} obtaining $(\frac{1}{2}\cdot\frac{1}{3})\cdot 2!=\frac{1}{3}$. We hence find $\frac{1}{3}P(ij)P(ab)\braket{kl||cd}\sigma_{ik}^{ac}\sigma_{jl}^{bd}$.
    \item III: The potential operator is part of $\hat{V}\ped{ND}$ as it only involves excitations. Same as for example 
    II, the term acquires a global prefactor of $\frac{1}{3}$, given by the sum $\frac{1}{12}[[V\ped{ND},\tilde{\sigma}_2],\tilde{\sigma}_2]+\frac{1}{4}[[V,\tilde{\sigma}_2]\ped{R},\tilde{\sigma}_2]$. 
    From the CC rules, a prefactor $\frac{1}{4}$ is obtained, due to  two pairs of equivalent lines in the diagrams. Scaling this factor according to rule 3 in sec.~\ref{sec:diag} yields $(\frac{1}{4}\cdot\frac{1}{3})\cdot 2!=\frac{1}{6}$. Furthermore, the last rule of sec.~\ref{sec:diag} applies here as well, as the potential operator is connected only to one of the amplitudes, giving an additional factor of $\frac{1}{2}$. In total, one hence obtains $\frac{1}{12}\braket{cd||ij}\sigma_{kl}^{cd*}\sigma_{kl}^{ab}$ for this term.
    \item IV: This diagram involves a pure excitation part of $\hat{V}$. It contributes to $\frac{1}{12}[[\hat{V}\ped{ND},\hat{\sigma}],\hat{\sigma}]$ but due to the connection to $\sigma^\dagger$ also to $\frac{1}{4}[[\hat{V},\hat{\sigma}]\ped{R},\hat{\sigma}]$. Both have one pair of equivalent lines and the scaling of the prefactor leads to $\frac{1}{3}$. Furthermore, the potential operator is connected to only one of the amplitudes ($\sigma^\dagger$) thus requiring an additional factor $\frac{1}{2}$. The diagram therefore 
    is evaluated as $-P(ab)\frac{1}{6}\braket{ad||ij}\sigma_{kl}^{cd*}\sigma_{kl}^{cb}$.
\end{itemize}

\newpage

\bibliography{article}

\begin{thebibliography}{96}%
\makeatletter
\providecommand \@ifxundefined [1]{%
 \@ifx{#1\undefined}
}%
\providecommand \@ifnum [1]{%
 \ifnum #1\expandafter \@firstoftwo
 \else \expandafter \@secondoftwo
 \fi
}%
\providecommand \@ifx [1]{%
 \ifx #1\expandafter \@firstoftwo
 \else \expandafter \@secondoftwo
 \fi
}%
\providecommand \natexlab [1]{#1}%
\providecommand \enquote  [1]{``#1''}%
\providecommand \bibnamefont  [1]{#1}%
\providecommand \bibfnamefont [1]{#1}%
\providecommand \citenamefont [1]{#1}%
\providecommand \href@noop [0]{\@secondoftwo}%
\providecommand \href [0]{\begingroup \@sanitize@url \@href}%
\providecommand \@href[1]{\@@startlink{#1}\@@href}%
\providecommand \@@href[1]{\endgroup#1\@@endlink}%
\providecommand \@sanitize@url [0]{\catcode `\\12\catcode `\$12\catcode
  `\&12\catcode `\#12\catcode `\^12\catcode `\_12\catcode `\%12\relax}%
\providecommand \@@startlink[1]{}%
\providecommand \@@endlink[0]{}%
\providecommand \url  [0]{\begingroup\@sanitize@url \@url }%
\providecommand \@url [1]{\endgroup\@href {#1}{\urlprefix }}%
\providecommand \urlprefix  [0]{URL }%
\providecommand \Eprint [0]{\href }%
\providecommand \doibase [0]{http://dx.doi.org/}%
\providecommand \selectlanguage [0]{\@gobble}%
\providecommand \bibinfo  [0]{\@secondoftwo}%
\providecommand \bibfield  [0]{\@secondoftwo}%
\providecommand \translation [1]{[#1]}%
\providecommand \BibitemOpen [0]{}%
\providecommand \bibitemStop [0]{}%
\providecommand \bibitemNoStop [0]{.\EOS\space}%
\providecommand \EOS [0]{\spacefactor3000\relax}%
\providecommand \BibitemShut  [1]{\csname bibitem#1\endcsname}%
\let\auto@bib@innerbib\@empty
%</preamble>
\bibitem [{\citenamefont {Tyagi}\ and\ \citenamefont
  {Banerjee}(2017)}]{tyagi2017materials}%
  \BibitemOpen
  \bibfield  {author} {\bibinfo {author} {\bibfnamefont {A.}~\bibnamefont
  {Tyagi}}\ and\ \bibinfo {author} {\bibfnamefont {S.}~\bibnamefont
  {Banerjee}},\ }\href@noop {} {\emph {\bibinfo {title} {Materials under
  extreme conditions: recent trends and future prospects}}}\ (\bibinfo
  {publisher} {Elsevier},\ \bibinfo {year} {2017})\BibitemShut {NoStop}%
\bibitem [{\citenamefont {Goldman}(2019)}]{goldman2019computational}%
  \BibitemOpen
  \bibfield  {author} {\bibinfo {author} {\bibfnamefont {N.}~\bibnamefont
  {Goldman}},\ }\href@noop {} {\emph {\bibinfo {title} {Computational
  approaches for chemistry under extreme conditions}}},\ Vol.~\bibinfo {volume}
  {28}\ (\bibinfo  {publisher} {Springer},\ \bibinfo {year} {2019})\BibitemShut
  {NoStop}%
\bibitem [{\citenamefont {Hampe}\ and\ \citenamefont
  {Stopkowicz}(2017)}]{Hampe2017}%
  \BibitemOpen
  \bibfield  {author} {\bibinfo {author} {\bibfnamefont {F.}~\bibnamefont
  {Hampe}}\ and\ \bibinfo {author} {\bibfnamefont {S.}~\bibnamefont
  {Stopkowicz}},\ }\bibfield  {title} {\enquote {\bibinfo {title}
  {{Equation-of-motion coupled-cluster methods for atoms and molecules in
  strong magnetic fields}},}\ }\href {\doibase 10.1063/1.4979624} {\bibfield
  {journal} {\bibinfo  {journal} {J. Chem. Phys.}\ }\textbf {\bibinfo {volume}
  {146}},\ \bibinfo {pages} {154105} (\bibinfo {year} {2017})}\BibitemShut
  {NoStop}%
\bibitem [{\citenamefont {Stopkowicz}(2017)}]{Stopkowicz2017}%
  \BibitemOpen
  \bibfield  {author} {\bibinfo {author} {\bibfnamefont {S.}~\bibnamefont
  {Stopkowicz}},\ }\bibfield  {title} {\enquote {\bibinfo {title} {Perspective:
  Coupled cluster theory for atoms and molecules in strong magnetic fields},}\
  }\href {\doibase 10.1002/qua.25391} {\bibfield  {journal} {\bibinfo
  {journal} {Int. J. Quantum Chem.}\ }\textbf {\bibinfo {volume} {118}},\
  \bibinfo {pages} {e25391} (\bibinfo {year} {2017})}\BibitemShut {NoStop}%
\bibitem [{\citenamefont {Holzer}\ \emph {et~al.}(2019)\citenamefont {Holzer},
  \citenamefont {Teale}, \citenamefont {Hampe}, \citenamefont {Stopkowicz},
  \citenamefont {Helgaker},\ and\ \citenamefont {Klopper}}]{holzer2019gw}%
  \BibitemOpen
  \bibfield  {author} {\bibinfo {author} {\bibfnamefont {C.}~\bibnamefont
  {Holzer}}, \bibinfo {author} {\bibfnamefont {A.~M.}\ \bibnamefont {Teale}},
  \bibinfo {author} {\bibfnamefont {F.}~\bibnamefont {Hampe}}, \bibinfo
  {author} {\bibfnamefont {S.}~\bibnamefont {Stopkowicz}}, \bibinfo {author}
  {\bibfnamefont {T.}~\bibnamefont {Helgaker}}, \ and\ \bibinfo {author}
  {\bibfnamefont {W.}~\bibnamefont {Klopper}},\ }\bibfield  {title} {\enquote
  {\bibinfo {title} {{GW} quasiparticle energies of atoms in strong magnetic
  fields},}\ }\href {\doibase https://doi.org/10.1063/1.5093396} {\bibfield
  {journal} {\bibinfo  {journal} {J. Chem. Phys.}\ }\textbf {\bibinfo {volume}
  {150}},\ \bibinfo {pages} {214112} (\bibinfo {year} {2019})}\BibitemShut
  {NoStop}%
\bibitem [{\citenamefont {Hampe}\ and\ \citenamefont
  {Stopkowicz}(2019)}]{Hampe2019}%
  \BibitemOpen
  \bibfield  {author} {\bibinfo {author} {\bibfnamefont {F.}~\bibnamefont
  {Hampe}}\ and\ \bibinfo {author} {\bibfnamefont {S.}~\bibnamefont
  {Stopkowicz}},\ }\bibfield  {title} {\enquote {\bibinfo {title}
  {Transition-dipole moments for electronic excitations in strong magnetic
  fields using equation-of-motion and linear response coupled-cluster
  theory},}\ }\href {\doibase 10.1021/acs.jctc.9b00242} {\bibfield  {journal}
  {\bibinfo  {journal} {J. Chem. Theory Comput.}\ }\textbf {\bibinfo {volume}
  {15}},\ \bibinfo {pages} {4036--4043} (\bibinfo {year} {2019})}\BibitemShut
  {NoStop}%
\bibitem [{\citenamefont {Hampe}, \citenamefont {Gross},\ and\ \citenamefont
  {Stopkowicz}(2020)}]{Hampe2020}%
  \BibitemOpen
  \bibfield  {author} {\bibinfo {author} {\bibfnamefont {F.}~\bibnamefont
  {Hampe}}, \bibinfo {author} {\bibfnamefont {N.}~\bibnamefont {Gross}}, \ and\
  \bibinfo {author} {\bibfnamefont {S.}~\bibnamefont {Stopkowicz}},\ }\bibfield
   {title} {\enquote {\bibinfo {title} {{Full triples contribution in
  coupled-cluster and equation-of-motion coupled-cluster methods for atoms and
  molecules in strong magnetic fields}},}\ }\href {\doibase 10.1039/D0CP04169F}
  {\bibfield  {journal} {\bibinfo  {journal} {Phys. Chem. Chem. Phys.}\
  }\textbf {\bibinfo {volume} {22}},\ \bibinfo {pages} {23522--23529} (\bibinfo
  {year} {2020})}\BibitemShut {NoStop}%
\bibitem [{\citenamefont {Hollands}\ \emph {et~al.}(2023)\citenamefont
  {Hollands}, \citenamefont {Stopkowicz}, \citenamefont {Kitsaras},
  \citenamefont {Hampe}, \citenamefont {Blaschke},\ and\ \citenamefont
  {Hermes}}]{hollands2023dz}%
  \BibitemOpen
  \bibfield  {author} {\bibinfo {author} {\bibfnamefont {M.~A.}\ \bibnamefont
  {Hollands}}, \bibinfo {author} {\bibfnamefont {S.}~\bibnamefont
  {Stopkowicz}}, \bibinfo {author} {\bibfnamefont {M.-P.}\ \bibnamefont
  {Kitsaras}}, \bibinfo {author} {\bibfnamefont {F.}~\bibnamefont {Hampe}},
  \bibinfo {author} {\bibfnamefont {S.}~\bibnamefont {Blaschke}}, \ and\
  \bibinfo {author} {\bibfnamefont {J.}~\bibnamefont {Hermes}},\ }\bibfield
  {title} {\enquote {\bibinfo {title} {A {DZ} white dwarf with a 30 {MG}
  magnetic field},}\ }\href {\doibase https://doi.org/10.1093/mnras/stad143}
  {\bibfield  {journal} {\bibinfo  {journal} {Mon. Not. R. Astron. Soc.}\
  }\textbf {\bibinfo {volume} {520}},\ \bibinfo {pages} {3560--3575} (\bibinfo
  {year} {2023})}\BibitemShut {NoStop}%
\bibitem [{\citenamefont {Franzon}\ and\ \citenamefont
  {Schramm}(2015)}]{Franzon2015}%
  \BibitemOpen
  \bibfield  {author} {\bibinfo {author} {\bibfnamefont {B.}~\bibnamefont
  {Franzon}}\ and\ \bibinfo {author} {\bibfnamefont {S.}~\bibnamefont
  {Schramm}},\ }\bibfield  {title} {\enquote {\bibinfo {title} {Effects of
  strong magnetic fields and rotation on white dwarf structure},}\ }\href
  {https://doi.org/10.1103/physrevd.92.083006} {\bibfield  {journal} {\bibinfo
  {journal} {Phys. Rev. D}\ }\textbf {\bibinfo {volume} {92}},\ \bibinfo
  {pages} {083006} (\bibinfo {year} {2015})}\BibitemShut {NoStop}%
\bibitem [{\citenamefont {Franzon}\ and\ \citenamefont
  {Schramm}(2017)}]{Franzon2017}%
  \BibitemOpen
  \bibfield  {author} {\bibinfo {author} {\bibfnamefont {B.}~\bibnamefont
  {Franzon}}\ and\ \bibinfo {author} {\bibfnamefont {S.}~\bibnamefont
  {Schramm}},\ }\bibfield  {title} {\enquote {\bibinfo {title} {Effects of
  magnetic fields in white dwarfs},}\ }\href {\doibase
  10.1088/1742-6596/861/1/012015} {\bibfield  {journal} {\bibinfo  {journal}
  {J. Phys. Conf. Ser.}\ }\textbf {\bibinfo {volume} {861}},\ \bibinfo {pages}
  {012015} (\bibinfo {year} {2017})}\BibitemShut {NoStop}%
\bibitem [{\citenamefont {Bera}\ and\ \citenamefont
  {Bhattacharya}(2014)}]{Bera2014}%
  \BibitemOpen
  \bibfield  {author} {\bibinfo {author} {\bibfnamefont {P.}~\bibnamefont
  {Bera}}\ and\ \bibinfo {author} {\bibfnamefont {D.}~\bibnamefont
  {Bhattacharya}},\ }\bibfield  {title} {\enquote {\bibinfo {title}
  {Mass{\textendash}radius relation of strongly magnetized white dwarfs: nearly
  independent of {L}andau quantization},}\ }\href {\doibase
  10.1093/mnras/stu2014} {\bibfield  {journal} {\bibinfo  {journal} {Mon. Not.
  R. Astron. Soc.}\ }\textbf {\bibinfo {volume} {445}},\ \bibinfo {pages}
  {3951--3958} (\bibinfo {year} {2014})}\BibitemShut {NoStop}%
\bibitem [{\citenamefont {Boshkayev}\ \emph {et~al.}(2014)\citenamefont
  {Boshkayev}, \citenamefont {Rueda}, \citenamefont {Ruffini},\ and\
  \citenamefont {Siutsou}}]{Boshkayev2014}%
  \BibitemOpen
  \bibfield  {author} {\bibinfo {author} {\bibfnamefont {K.}~\bibnamefont
  {Boshkayev}}, \bibinfo {author} {\bibfnamefont {J.~A.}\ \bibnamefont
  {Rueda}}, \bibinfo {author} {\bibfnamefont {R.}~\bibnamefont {Ruffini}}, \
  and\ \bibinfo {author} {\bibfnamefont {I.}~\bibnamefont {Siutsou}},\
  }\bibfield  {title} {\enquote {\bibinfo {title} {General relativistic white
  dwarfs and their astrophysical implications},}\ }\href {\doibase
  10.3938/jkps.65.855} {\bibfield  {journal} {\bibinfo  {journal} {J. Korean
  Chem. Soc.}\ }\textbf {\bibinfo {volume} {65}},\ \bibinfo {pages} {855--860}
  (\bibinfo {year} {2014})}\BibitemShut {NoStop}%
\bibitem [{\citenamefont {Terada}\ \emph {et~al.}(2008)\citenamefont {Terada},
  \citenamefont {Hayashi}, \citenamefont {Ishida}, \citenamefont {Mukai},
  \citenamefont {Dotani}, \citenamefont {Okada}, \citenamefont {Nakamura},
  \citenamefont {Naik}, \citenamefont {Bamba},\ and\ \citenamefont
  {Makishima}}]{Terada2008}%
  \BibitemOpen
  \bibfield  {author} {\bibinfo {author} {\bibfnamefont {Y.}~\bibnamefont
  {Terada}}, \bibinfo {author} {\bibfnamefont {T.}~\bibnamefont {Hayashi}},
  \bibinfo {author} {\bibfnamefont {M.}~\bibnamefont {Ishida}}, \bibinfo
  {author} {\bibfnamefont {K.}~\bibnamefont {Mukai}}, \bibinfo {author}
  {\bibfnamefont {T.}~\bibnamefont {Dotani}}, \bibinfo {author} {\bibfnamefont
  {S.}~\bibnamefont {Okada}}, \bibinfo {author} {\bibfnamefont
  {R.}~\bibnamefont {Nakamura}}, \bibinfo {author} {\bibfnamefont
  {S.}~\bibnamefont {Naik}}, \bibinfo {author} {\bibfnamefont {A.}~\bibnamefont
  {Bamba}}, \ and\ \bibinfo {author} {\bibfnamefont {K.}~\bibnamefont
  {Makishima}},\ }\bibfield  {title} {\enquote {\bibinfo {title} {Suzaku
  discovery of hard {X}-ray pulsations from a rotating magnetized white dwarf,
  {AEAquarii}},}\ }\href {\doibase 10.1093/pasj/60.2.387} {\bibfield  {journal}
  {\bibinfo  {journal} {Publ. Astron. Soc. Jpn.}\ }\textbf {\bibinfo {volume}
  {60}},\ \bibinfo {pages} {387--397} (\bibinfo {year} {2008})}\BibitemShut
  {NoStop}%
\bibitem [{\citenamefont {Reimers}\ \emph {et~al.}(1996)\citenamefont
  {Reimers}, \citenamefont {Jordan}, \citenamefont {Koester}, \citenamefont
  {Bade}, \citenamefont {Koehler},\ and\ \citenamefont
  {Wisotzki}}]{Reimers1996}%
  \BibitemOpen
  \bibfield  {author} {\bibinfo {author} {\bibfnamefont {D.}~\bibnamefont
  {Reimers}}, \bibinfo {author} {\bibfnamefont {S.}~\bibnamefont {Jordan}},
  \bibinfo {author} {\bibfnamefont {D.}~\bibnamefont {Koester}}, \bibinfo
  {author} {\bibfnamefont {N.}~\bibnamefont {Bade}}, \bibinfo {author}
  {\bibfnamefont {T.}~\bibnamefont {Koehler}}, \ and\ \bibinfo {author}
  {\bibfnamefont {L.}~\bibnamefont {Wisotzki}},\ }\bibfield  {title} {\enquote
  {\bibinfo {title} {Discovery of four white dwarfs with strong magnetic fields
  by the {H}amburg/{ESO} survey},}\ }\href
  {https://arxiv.org/abs/astro-ph/9604104} {\  (\bibinfo {year}
  {1996})}\BibitemShut {NoStop}%
\bibitem [{\citenamefont {Koester}\ and\ \citenamefont
  {Chanmugam}(1990)}]{Koester1990}%
  \BibitemOpen
  \bibfield  {author} {\bibinfo {author} {\bibfnamefont {D.}~\bibnamefont
  {Koester}}\ and\ \bibinfo {author} {\bibfnamefont {G.}~\bibnamefont
  {Chanmugam}},\ }\bibfield  {title} {\enquote {\bibinfo {title} {{Physics of
  white dwarf stars}},}\ }\href {\doibase 10.1088/0034-4885/53/7/001}
  {\bibfield  {journal} {\bibinfo  {journal} {Reports Prog. Phys.}\ }\textbf
  {\bibinfo {volume} {53}},\ \bibinfo {pages} {837--915} (\bibinfo {year}
  {1990})}\BibitemShut {NoStop}%
\bibitem [{\citenamefont {Greenstein}(1984)}]{Greenstein1984}%
  \BibitemOpen
  \bibfield  {author} {\bibinfo {author} {\bibfnamefont {J.~L.}\ \bibnamefont
  {Greenstein}},\ }\bibfield  {title} {\enquote {\bibinfo {title} {{The
  identification of hydrogen in GRW +70 deg 8247}},}\ }\href {\doibase
  10.1086/184282} {\bibfield  {journal} {\bibinfo  {journal} {Astrophys. J.}\
  }\textbf {\bibinfo {volume} {281}},\ \bibinfo {pages} {L47--L50} (\bibinfo
  {year} {1984})}\BibitemShut {NoStop}%
\bibitem [{\citenamefont {Henry}\ and\ \citenamefont
  {Oconnell}(1985)}]{Henry1985}%
  \BibitemOpen
  \bibfield  {author} {\bibinfo {author} {\bibfnamefont {R.~J.~W.}\
  \bibnamefont {Henry}}\ and\ \bibinfo {author} {\bibfnamefont {R.~F.}\
  \bibnamefont {Oconnell}},\ }\bibfield  {title} {\enquote {\bibinfo {title}
  {{Hydrogen spectrum in magnetic white dwarfs - H-alpha, H-beta and H-gamma
  transitions}},}\ }\href {\doibase 10.1086/131540} {\bibfield  {journal}
  {\bibinfo  {journal} {Publ. Astron. Soc. Pacific}\ }\textbf {\bibinfo
  {volume} {97}},\ \bibinfo {pages} {333} (\bibinfo {year} {1985})}\BibitemShut
  {NoStop}%
\bibitem [{\citenamefont {Dufour}\ \emph {et~al.}(2007)\citenamefont {Dufour},
  \citenamefont {Liebert}, \citenamefont {Fontaine},\ and\ \citenamefont
  {Behara}}]{Dufour2007}%
  \BibitemOpen
  \bibfield  {author} {\bibinfo {author} {\bibfnamefont {P.}~\bibnamefont
  {Dufour}}, \bibinfo {author} {\bibfnamefont {J.}~\bibnamefont {Liebert}},
  \bibinfo {author} {\bibfnamefont {G.}~\bibnamefont {Fontaine}}, \ and\
  \bibinfo {author} {\bibfnamefont {N.}~\bibnamefont {Behara}},\ }\bibfield
  {title} {\enquote {\bibinfo {title} {{White dwarf stars with carbon
  atmospheres}},}\ }\href {\doibase 10.1038/nature06318} {\bibfield  {journal}
  {\bibinfo  {journal} {Nature}\ }\textbf {\bibinfo {volume} {450}},\ \bibinfo
  {pages} {522--524} (\bibinfo {year} {2007})}\BibitemShut {NoStop}%
\bibitem [{\citenamefont {Berdyugina}, \citenamefont {Berdyugin},\ and\
  \citenamefont {Piirola}(2007)}]{Berdyugina2007}%
  \BibitemOpen
  \bibfield  {author} {\bibinfo {author} {\bibfnamefont {S.~V.}\ \bibnamefont
  {Berdyugina}}, \bibinfo {author} {\bibfnamefont {A.~V.}\ \bibnamefont
  {Berdyugin}}, \ and\ \bibinfo {author} {\bibfnamefont {V.}~\bibnamefont
  {Piirola}},\ }\bibfield  {title} {\enquote {\bibinfo {title} {Molecular
  magnetic dichroism in spectra of white dwarfs},}\ }\href {\doibase
  10.1103/PhysRevLett.99.091101} {\bibfield  {journal} {\bibinfo  {journal}
  {Phys. Rev. Lett.}\ }\textbf {\bibinfo {volume} {99}},\ \bibinfo {pages}
  {091101} (\bibinfo {year} {2007})}\BibitemShut {NoStop}%
\bibitem [{\citenamefont {Jordan}(2008)}]{Jordan2008}%
  \BibitemOpen
  \bibfield  {author} {\bibinfo {author} {\bibfnamefont {S.}~\bibnamefont
  {Jordan}},\ }\bibfield  {title} {\enquote {\bibinfo {title} {Magnetic fields
  in white dwarfs and their direct progenitors},}\ }\href {\doibase
  10.1017/S1743921309030749} {\bibfield  {journal} {\bibinfo  {journal} {Proc.
  Int. Astron. Union}\ }\textbf {\bibinfo {volume} {4}},\ \bibinfo {pages}
  {369--378} (\bibinfo {year} {2008})}\BibitemShut {NoStop}%
\bibitem [{\citenamefont {Xu}\ \emph {et~al.}(2013)\citenamefont {Xu},
  \citenamefont {Jura}, \citenamefont {Koester}, \citenamefont {Klein},\ and\
  \citenamefont {Zuckerman}}]{Xu2013}%
  \BibitemOpen
  \bibfield  {author} {\bibinfo {author} {\bibfnamefont {S.}~\bibnamefont
  {Xu}}, \bibinfo {author} {\bibfnamefont {M.}~\bibnamefont {Jura}}, \bibinfo
  {author} {\bibfnamefont {D.}~\bibnamefont {Koester}}, \bibinfo {author}
  {\bibfnamefont {B.}~\bibnamefont {Klein}}, \ and\ \bibinfo {author}
  {\bibfnamefont {B.}~\bibnamefont {Zuckerman}},\ }\bibfield  {title} {\enquote
  {\bibinfo {title} {Discovery of molecular hydrogen in white dwarf
  atmospheres},}\ }\href {\doibase 10.1088/2041-8205/766/2/L18} {\bibfield
  {journal} {\bibinfo  {journal} {Astrophys. J.}\ }\textbf {\bibinfo {volume}
  {766}},\ \bibinfo {pages} {L18} (\bibinfo {year} {2013})}\BibitemShut
  {NoStop}%
\bibitem [{\citenamefont {Landstreet}\ \emph {et~al.}(2012)\citenamefont
  {Landstreet}, \citenamefont {Bagnulo}, \citenamefont {Valyavin},
  \citenamefont {Fossati}, \citenamefont {Jordan}, \citenamefont {Monin},\ and\
  \citenamefont {Wade}}]{Landstreet2012}%
  \BibitemOpen
  \bibfield  {author} {\bibinfo {author} {\bibfnamefont {J.~D.}\ \bibnamefont
  {Landstreet}}, \bibinfo {author} {\bibfnamefont {S.}~\bibnamefont {Bagnulo}},
  \bibinfo {author} {\bibfnamefont {G.~G.}\ \bibnamefont {Valyavin}}, \bibinfo
  {author} {\bibfnamefont {L.}~\bibnamefont {Fossati}}, \bibinfo {author}
  {\bibfnamefont {S.}~\bibnamefont {Jordan}}, \bibinfo {author} {\bibfnamefont
  {D.}~\bibnamefont {Monin}}, \ and\ \bibinfo {author} {\bibfnamefont {G.~A.}\
  \bibnamefont {Wade}},\ }\bibfield  {title} {\enquote {\bibinfo {title} {{On
  the incidence of weak magnetic fields in DA white dwarfs}},}\ }\href
  {\doibase 10.1051/0004-6361/201219829} {\bibfield  {journal} {\bibinfo
  {journal} {Astron. Astrophys.}\ }\textbf {\bibinfo {volume} {545}},\ \bibinfo
  {pages} {A30} (\bibinfo {year} {2012})}\BibitemShut {NoStop}%
\bibitem [{\citenamefont {Bagnulo}\ and\ \citenamefont
  {Landstreet}(2020)}]{Bagnulo2020}%
  \BibitemOpen
  \bibfield  {author} {\bibinfo {author} {\bibfnamefont {S.}~\bibnamefont
  {Bagnulo}}\ and\ \bibinfo {author} {\bibfnamefont {J.~D.}\ \bibnamefont
  {Landstreet}},\ }\bibfield  {title} {\enquote {\bibinfo {title} {Discovery of
  six new strongly magnetic white dwarfs in the 20 pc local population},}\
  }\href {\doibase 10.1051/0004-6361/202038565} {\bibfield  {journal} {\bibinfo
   {journal} {Astron. Astrophys.}\ }\textbf {\bibinfo {volume} {643}},\
  \bibinfo {pages} {A134} (\bibinfo {year} {2020})}\BibitemShut {NoStop}%
\bibitem [{\citenamefont {Rosner}\ \emph {et~al.}(1984)\citenamefont {Rosner},
  \citenamefont {Wunner}, \citenamefont {Herold},\ and\ \citenamefont
  {Ruder}}]{Rosner1984}%
  \BibitemOpen
  \bibfield  {author} {\bibinfo {author} {\bibfnamefont {W.}~\bibnamefont
  {Rosner}}, \bibinfo {author} {\bibfnamefont {G.}~\bibnamefont {Wunner}},
  \bibinfo {author} {\bibfnamefont {H.}~\bibnamefont {Herold}}, \ and\ \bibinfo
  {author} {\bibfnamefont {H.}~\bibnamefont {Ruder}},\ }\bibfield  {title}
  {\enquote {\bibinfo {title} {{Hydrogen atoms in arbitrary magnetic fields. I.
  Energy levels and wavefunctions}},}\ }\href {\doibase
  10.1088/0022-3700/17/1/010} {\bibfield  {journal} {\bibinfo  {journal} {J.
  Phys. B At. Mol. Phys.}\ }\textbf {\bibinfo {volume} {17}},\ \bibinfo {pages}
  {29--52} (\bibinfo {year} {1984})}\BibitemShut {NoStop}%
\bibitem [{\citenamefont {Schmelcher}, \citenamefont {Cederbaum},\ and\
  \citenamefont {Meyer}(1988{\natexlab{a}})}]{Schmelcher1988}%
  \BibitemOpen
  \bibfield  {author} {\bibinfo {author} {\bibfnamefont {P.}~\bibnamefont
  {Schmelcher}}, \bibinfo {author} {\bibfnamefont {L.~S.}\ \bibnamefont
  {Cederbaum}}, \ and\ \bibinfo {author} {\bibfnamefont {H.~D.}\ \bibnamefont
  {Meyer}},\ }\bibfield  {title} {\enquote {\bibinfo {title} {{Electronic and
  nuclear motion and their couplings in the presence of a magnetic field}},}\
  }\href {\doibase 10.1103/PhysRevA.38.6066} {\bibfield  {journal} {\bibinfo
  {journal} {Phys. Rev. A}\ }\textbf {\bibinfo {volume} {38}},\ \bibinfo
  {pages} {6066--6079} (\bibinfo {year} {1988}{\natexlab{a}})}\BibitemShut
  {NoStop}%
\bibitem [{\citenamefont {Schmelcher}, \citenamefont {Cederbaum},\ and\
  \citenamefont {Meyer}(1988{\natexlab{b}})}]{Schmelcher1988a}%
  \BibitemOpen
  \bibfield  {author} {\bibinfo {author} {\bibfnamefont {P.}~\bibnamefont
  {Schmelcher}}, \bibinfo {author} {\bibfnamefont {L.~S.}\ \bibnamefont
  {Cederbaum}}, \ and\ \bibinfo {author} {\bibfnamefont {H.~D.}\ \bibnamefont
  {Meyer}},\ }\bibfield  {title} {\enquote {\bibinfo {title} {{On the validity
  of the Born-Oppenhei- mer approximation in magnetic fields}},}\ }\href
  {\doibase 10.1088/0953-4075/21/15/005} {\bibfield  {journal} {\bibinfo
  {journal} {J. Phys. B At. Mol. Opt. Phys.}\ }\textbf {\bibinfo {volume}
  {21}},\ \bibinfo {pages} {L445--L450} (\bibinfo {year}
  {1988}{\natexlab{b}})}\BibitemShut {NoStop}%
\bibitem [{\citenamefont {Schmelcher}\ and\ \citenamefont
  {Cederbaum}(1988)}]{Schmelcher1988b}%
  \BibitemOpen
  \bibfield  {author} {\bibinfo {author} {\bibfnamefont {P.}~\bibnamefont
  {Schmelcher}}\ and\ \bibinfo {author} {\bibfnamefont {L.~S.}\ \bibnamefont
  {Cederbaum}},\ }\bibfield  {title} {\enquote {\bibinfo {title} {{Molecules in
  strong magnetic fields: Properties of atomic orbitals}},}\ }\href {\doibase
  10.1103/PhysRevA.37.672} {\bibfield  {journal} {\bibinfo  {journal} {Phys.
  Rev. A}\ }\textbf {\bibinfo {volume} {37}},\ \bibinfo {pages} {672--681}
  (\bibinfo {year} {1988})}\BibitemShut {NoStop}%
\bibitem [{\citenamefont {Schmelcher}\ and\ \citenamefont
  {Cederbaum}(1991)}]{Schmelcher1991}%
  \BibitemOpen
  \bibfield  {author} {\bibinfo {author} {\bibfnamefont {P.}~\bibnamefont
  {Schmelcher}}\ and\ \bibinfo {author} {\bibfnamefont {L.~S.}\ \bibnamefont
  {Cederbaum}},\ }\bibfield  {title} {\enquote {\bibinfo {title} {{On molecules
  and ions in strong magnetic fields}},}\ }\href {\doibase
  10.1002/QUA.560400836} {\bibfield  {journal} {\bibinfo  {journal} {Int. J.
  Quantum Chem.}\ }\textbf {\bibinfo {volume} {40}},\ \bibinfo {pages}
  {371--385} (\bibinfo {year} {1991})}\BibitemShut {NoStop}%
\bibitem [{\citenamefont {{Jordan}}\ \emph {et~al.}(1998)\citenamefont
  {{Jordan}}, \citenamefont {{Schmelcher}}, \citenamefont {{Becken}},\ and\
  \citenamefont {{Schweizer}}}]{Jordan1998}%
  \BibitemOpen
  \bibfield  {author} {\bibinfo {author} {\bibfnamefont {S.}~\bibnamefont
  {{Jordan}}}, \bibinfo {author} {\bibfnamefont {P.}~\bibnamefont
  {{Schmelcher}}}, \bibinfo {author} {\bibfnamefont {W.}~\bibnamefont
  {{Becken}}}, \ and\ \bibinfo {author} {\bibfnamefont {W.}~\bibnamefont
  {{Schweizer}}},\ }\bibfield  {title} {\enquote {\bibinfo {title} {Evidence
  for helium in the magnetic white dwarf {GD} 229},}\ }\href@noop {} {\bibfield
   {journal} {\bibinfo  {journal} {Astron. Astrophys.}\ }\textbf {\bibinfo
  {volume} {336}},\ \bibinfo {pages} {L33--L36} (\bibinfo {year}
  {1998})}\BibitemShut {NoStop}%
\bibitem [{\citenamefont {Becken}, \citenamefont {Schmelcher},\ and\
  \citenamefont {Diakonos}(1999)}]{Becken1999}%
  \BibitemOpen
  \bibfield  {author} {\bibinfo {author} {\bibfnamefont {W.}~\bibnamefont
  {Becken}}, \bibinfo {author} {\bibfnamefont {P.}~\bibnamefont {Schmelcher}},
  \ and\ \bibinfo {author} {\bibfnamefont {F.~K.}\ \bibnamefont {Diakonos}},\
  }\bibfield  {title} {\enquote {\bibinfo {title} {The helium atom in a strong
  magnetic field},}\ }\href {\doibase 10.1088/0953-4075/32/6/018} {\bibfield
  {journal} {\bibinfo  {journal} {J. Phys. B At. Mol. Opt. Phys.}\ }\textbf
  {\bibinfo {volume} {32}},\ \bibinfo {pages} {1557–1584} (\bibinfo {year}
  {1999})}\BibitemShut {NoStop}%
\bibitem [{\citenamefont {Al-Hujaj}\ and\ \citenamefont
  {Schmelcher}(2004{\natexlab{a}})}]{AlHujaj2004_Li}%
  \BibitemOpen
  \bibfield  {author} {\bibinfo {author} {\bibfnamefont {O.-A.}\ \bibnamefont
  {Al-Hujaj}}\ and\ \bibinfo {author} {\bibfnamefont {P.}~\bibnamefont
  {Schmelcher}},\ }\bibfield  {title} {\enquote {\bibinfo {title} {Lithium in
  strong magnetic fields},}\ }\href {\doibase 10.1103/physreva.70.033411}
  {\bibfield  {journal} {\bibinfo  {journal} {Phys. Rev. A}\ }\textbf {\bibinfo
  {volume} {70}},\ \bibinfo {pages} {033411} (\bibinfo {year}
  {2004}{\natexlab{a}})}\BibitemShut {NoStop}%
\bibitem [{\citenamefont {Al-Hujaj}\ and\ \citenamefont
  {Schmelcher}(2004{\natexlab{b}})}]{AlHujaj2004_Be}%
  \BibitemOpen
  \bibfield  {author} {\bibinfo {author} {\bibfnamefont {O.-A.}\ \bibnamefont
  {Al-Hujaj}}\ and\ \bibinfo {author} {\bibfnamefont {P.}~\bibnamefont
  {Schmelcher}},\ }\bibfield  {title} {\enquote {\bibinfo {title} {Beryllium in
  strong magnetic fields},}\ }\href {\doibase 10.1103/physreva.70.023411}
  {\bibfield  {journal} {\bibinfo  {journal} {Phys. Rev. A}\ }\textbf {\bibinfo
  {volume} {70}},\ \bibinfo {pages} {023411} (\bibinfo {year}
  {2004}{\natexlab{b}})}\BibitemShut {NoStop}%
\bibitem [{\citenamefont {London}(1937)}]{London37}%
  \BibitemOpen
  \bibfield  {author} {\bibinfo {author} {\bibfnamefont {F.}~\bibnamefont
  {London}},\ }\bibfield  {title} {\enquote {\bibinfo {title} {Théorie
  quantique des courants interatomiques dans les combinaisons aromatiques},}\
  }\href {\doibase 10.1051/jphysrad:01937008010039700} {\bibfield  {journal}
  {\bibinfo  {journal} {J. Phys. Radium}\ }\textbf {\bibinfo {volume} {8}},\
  \bibinfo {pages} {397--409} (\bibinfo {year} {1937})}\BibitemShut {NoStop}%
\bibitem [{\citenamefont {Hameka}(1958)}]{Hameka58}%
  \BibitemOpen
  \bibfield  {author} {\bibinfo {author} {\bibfnamefont {H.~F.}\ \bibnamefont
  {Hameka}},\ }\bibfield  {title} {\enquote {\bibinfo {title} {On the nuclear
  magnetic shielding in the hydrogen molecule},}\ }\href {\doibase
  10.1080/00268975800100261} {\bibfield  {journal} {\bibinfo  {journal} {Mol.
  Phys.}\ }\textbf {\bibinfo {volume} {1}},\ \bibinfo {pages} {203--215}
  (\bibinfo {year} {1958})}\BibitemShut {NoStop}%
\bibitem [{\citenamefont {Ditchfield}(1972)}]{Ditchfield72}%
  \BibitemOpen
  \bibfield  {author} {\bibinfo {author} {\bibfnamefont {R.}~\bibnamefont
  {Ditchfield}},\ }\bibfield  {title} {\enquote {\bibinfo {title} {Molecular
  orbital theory of magnetic shielding and magnetic susceptibility},}\ }\href
  {\doibase 10.1063/1.1677088} {\bibfield  {journal} {\bibinfo  {journal} {J.
  Chem. Phys.}\ }\textbf {\bibinfo {volume} {56}},\ \bibinfo {pages}
  {5688--5691} (\bibinfo {year} {1972})}\BibitemShut {NoStop}%
\bibitem [{\citenamefont {Wolinski}, \citenamefont {Hinton},\ and\
  \citenamefont {Pulay}(1990)}]{Wolinski90}%
  \BibitemOpen
  \bibfield  {author} {\bibinfo {author} {\bibfnamefont {K.}~\bibnamefont
  {Wolinski}}, \bibinfo {author} {\bibfnamefont {J.~F.}\ \bibnamefont
  {Hinton}}, \ and\ \bibinfo {author} {\bibfnamefont {P.}~\bibnamefont
  {Pulay}},\ }\bibfield  {title} {\enquote {\bibinfo {title} {Efficient
  implementation of the gauge-independent atomic orbital method for {NMR}
  chemical shift calculations},}\ }\href {\doibase 10.1021/ja00179a005}
  {\bibfield  {journal} {\bibinfo  {journal} {J. Am. Chem. Soc.}\ }\textbf
  {\bibinfo {volume} {112}},\ \bibinfo {pages} {8251--8260} (\bibinfo {year}
  {1990})}\BibitemShut {NoStop}%
\bibitem [{\citenamefont {Soncini}\ and\ \citenamefont
  {Fowler}(2004)}]{Soncini2004}%
  \BibitemOpen
  \bibfield  {author} {\bibinfo {author} {\bibfnamefont {A.}~\bibnamefont
  {Soncini}}\ and\ \bibinfo {author} {\bibfnamefont {P.}~\bibnamefont
  {Fowler}},\ }\bibfield  {title} {\enquote {\bibinfo {title} {{Non-linear ring
  currents: effect of strong magnetic fields on $\pi$-electron circulation}},}\
  }\href {\doibase 10.1016/j.cplett.2004.10.110} {\bibfield  {journal}
  {\bibinfo  {journal} {Chem. Phys. Lett.}\ }\textbf {\bibinfo {volume}
  {400}},\ \bibinfo {pages} {213--220} (\bibinfo {year} {2004})}\BibitemShut
  {NoStop}%
\bibitem [{\citenamefont {Tellgren}, \citenamefont {Soncini},\ and\
  \citenamefont {Helgaker}(2008)}]{Tellgren2008}%
  \BibitemOpen
  \bibfield  {author} {\bibinfo {author} {\bibfnamefont {E.~I.}\ \bibnamefont
  {Tellgren}}, \bibinfo {author} {\bibfnamefont {A.}~\bibnamefont {Soncini}}, \
  and\ \bibinfo {author} {\bibfnamefont {T.}~\bibnamefont {Helgaker}},\
  }\bibfield  {title} {\enquote {\bibinfo {title} {{Nonperturbative ab initio
  calculations in strong magnetic fields using London orbitals}},}\ }\href
  {\doibase 10.1063/1.2996525} {\bibfield  {journal} {\bibinfo  {journal} {J.
  Chem. Phys.}\ }\textbf {\bibinfo {volume} {129}},\ \bibinfo {pages} {154114}
  (\bibinfo {year} {2008})}\BibitemShut {NoStop}%
\bibitem [{\citenamefont {Lange}\ \emph {et~al.}(2012)\citenamefont {Lange},
  \citenamefont {Tellgren}, \citenamefont {Hoffmann},\ and\ \citenamefont
  {Helgaker}}]{Lange2012}%
  \BibitemOpen
  \bibfield  {author} {\bibinfo {author} {\bibfnamefont {K.~K.}\ \bibnamefont
  {Lange}}, \bibinfo {author} {\bibfnamefont {E.~I.}\ \bibnamefont {Tellgren}},
  \bibinfo {author} {\bibfnamefont {M.~R.}\ \bibnamefont {Hoffmann}}, \ and\
  \bibinfo {author} {\bibfnamefont {T.}~\bibnamefont {Helgaker}},\ }\bibfield
  {title} {\enquote {\bibinfo {title} {A paramagnetic bonding mechanism for
  diatomics in strong magnetic fields},}\ }\href {\doibase
  10.1126/science.1219703} {\bibfield  {journal} {\bibinfo  {journal}
  {Science}\ }\textbf {\bibinfo {volume} {337}},\ \bibinfo {pages} {327–331}
  (\bibinfo {year} {2012})}\BibitemShut {NoStop}%
\bibitem [{\citenamefont {Stopkowicz}\ \emph {et~al.}(2015)\citenamefont
  {Stopkowicz}, \citenamefont {Gauss}, \citenamefont {Lange}, \citenamefont
  {Tellgren},\ and\ \citenamefont {Helgaker}}]{Stopkowicz2015}%
  \BibitemOpen
  \bibfield  {author} {\bibinfo {author} {\bibfnamefont {S.}~\bibnamefont
  {Stopkowicz}}, \bibinfo {author} {\bibfnamefont {J.}~\bibnamefont {Gauss}},
  \bibinfo {author} {\bibfnamefont {K.~K.}\ \bibnamefont {Lange}}, \bibinfo
  {author} {\bibfnamefont {E.~I.}\ \bibnamefont {Tellgren}}, \ and\ \bibinfo
  {author} {\bibfnamefont {T.}~\bibnamefont {Helgaker}},\ }\bibfield  {title}
  {\enquote {\bibinfo {title} {{Coupled-cluster theory for atoms and molecules
  in strong magnetic fields}},}\ }\href {\doibase 10.1063/1.4928056} {\bibfield
   {journal} {\bibinfo  {journal} {J. Chem. Phys.}\ }\textbf {\bibinfo {volume}
  {143}},\ \bibinfo {pages} {074110} (\bibinfo {year} {2015})}\BibitemShut
  {NoStop}%
\bibitem [{\citenamefont {Tellgren}\ \emph {et~al.}(2014)\citenamefont
  {Tellgren}, \citenamefont {Teale}, \citenamefont {Furness}, \citenamefont
  {Lange}, \citenamefont {Ekstr{\"{o}}m},\ and\ \citenamefont
  {Helgaker}}]{Tellgren2014}%
  \BibitemOpen
  \bibfield  {author} {\bibinfo {author} {\bibfnamefont {E.~I.}\ \bibnamefont
  {Tellgren}}, \bibinfo {author} {\bibfnamefont {A.~M.}\ \bibnamefont {Teale}},
  \bibinfo {author} {\bibfnamefont {J.~W.}\ \bibnamefont {Furness}}, \bibinfo
  {author} {\bibfnamefont {K.~K.}\ \bibnamefont {Lange}}, \bibinfo {author}
  {\bibfnamefont {U.}~\bibnamefont {Ekstr{\"{o}}m}}, \ and\ \bibinfo {author}
  {\bibfnamefont {T.}~\bibnamefont {Helgaker}},\ }\bibfield  {title} {\enquote
  {\bibinfo {title} {{Non-perturbative calculation of molecular magnetic
  properties within current-density functional theory}},}\ }\href {\doibase
  10.1063/1.4861427} {\bibfield  {journal} {\bibinfo  {journal} {J. Chem.
  Phys.}\ }\textbf {\bibinfo {volume} {140}},\ \bibinfo {pages} {034101}
  (\bibinfo {year} {2014})}\BibitemShut {NoStop}%
\bibitem [{\citenamefont {Furness}\ \emph {et~al.}(2015)\citenamefont
  {Furness}, \citenamefont {Verbeke}, \citenamefont {Tellgren}, \citenamefont
  {Stopkowicz}, \citenamefont {Ekstr{\"{o}}m}, \citenamefont {Helgaker},\ and\
  \citenamefont {Teale}}]{Furness2015}%
  \BibitemOpen
  \bibfield  {author} {\bibinfo {author} {\bibfnamefont {J.~W.}\ \bibnamefont
  {Furness}}, \bibinfo {author} {\bibfnamefont {J.}~\bibnamefont {Verbeke}},
  \bibinfo {author} {\bibfnamefont {E.~I.}\ \bibnamefont {Tellgren}}, \bibinfo
  {author} {\bibfnamefont {S.}~\bibnamefont {Stopkowicz}}, \bibinfo {author}
  {\bibfnamefont {U.}~\bibnamefont {Ekstr{\"{o}}m}}, \bibinfo {author}
  {\bibfnamefont {T.}~\bibnamefont {Helgaker}}, \ and\ \bibinfo {author}
  {\bibfnamefont {A.~M.}\ \bibnamefont {Teale}},\ }\bibfield  {title} {\enquote
  {\bibinfo {title} {Current density functional theory using meta-generalized
  gradient exchange-correlation functionals},}\ }\href {\doibase
  10.1021/acs.jctc.5b00535} {\bibfield  {journal} {\bibinfo  {journal} {J.
  Chem. Theory Comput.}\ }\textbf {\bibinfo {volume} {11}},\ \bibinfo {pages}
  {4169--4181} (\bibinfo {year} {2015})}\BibitemShut {NoStop}%
\bibitem [{\citenamefont {Čížek}\ and\ \citenamefont
  {Paldus}(1980)}]{cizek1980coupled}%
  \BibitemOpen
  \bibfield  {author} {\bibinfo {author} {\bibfnamefont {J.}~\bibnamefont
  {Čížek}}\ and\ \bibinfo {author} {\bibfnamefont {J.}~\bibnamefont
  {Paldus}},\ }\bibfield  {title} {\enquote {\bibinfo {title} {Coupled cluster
  approach},}\ }\href {\doibase 10.1088/0031-8949/21/3-4/006} {\bibfield
  {journal} {\bibinfo  {journal} {Phys. Scr.}\ }\textbf {\bibinfo {volume}
  {21}},\ \bibinfo {pages} {251} (\bibinfo {year} {1980})}\BibitemShut
  {NoStop}%
\bibitem [{\citenamefont {Čížek}(1991)}]{vcivzek1991origins}%
  \BibitemOpen
  \bibfield  {author} {\bibinfo {author} {\bibfnamefont {J.}~\bibnamefont
  {Čížek}},\ }\bibfield  {title} {\enquote {\bibinfo {title} {Origins of
  coupled cluster technique for atoms and molecules},}\ }\href {\doibase
  https://doi.org/10.1007/BF01119616} {\bibfield  {journal} {\bibinfo
  {journal} {Theor. chim. acta}\ }\textbf {\bibinfo {volume} {80}},\ \bibinfo
  {pages} {91--94} (\bibinfo {year} {1991})}\BibitemShut {NoStop}%
\bibitem [{\citenamefont {Bartlett}\ and\ \citenamefont
  {Musia{\l}}(2007)}]{bartlett2007coupled}%
  \BibitemOpen
  \bibfield  {author} {\bibinfo {author} {\bibfnamefont {R.~J.}\ \bibnamefont
  {Bartlett}}\ and\ \bibinfo {author} {\bibfnamefont {M.}~\bibnamefont
  {Musia{\l}}},\ }\bibfield  {title} {\enquote {\bibinfo {title}
  {Coupled-cluster theory in quantum chemistry},}\ }\href {\doibase
  https://doi.org/10.1103/RevModPhys.79.291} {\bibfield  {journal} {\bibinfo
  {journal} {Rev. Mod. Phys.}\ }\textbf {\bibinfo {volume} {79}},\ \bibinfo
  {pages} {291} (\bibinfo {year} {2007})}\BibitemShut {NoStop}%
\bibitem [{\citenamefont {Purvis}\ and\ \citenamefont
  {Bartlett}(1982)}]{Purvis1982}%
  \BibitemOpen
  \bibfield  {author} {\bibinfo {author} {\bibfnamefont {G.~D.}\ \bibnamefont
  {Purvis}}\ and\ \bibinfo {author} {\bibfnamefont {R.~J.}\ \bibnamefont
  {Bartlett}},\ }\bibfield  {title} {\enquote {\bibinfo {title} {A full
  coupled-cluster singles and doubles model: The inclusion of disconnected
  triples},}\ }\href {\doibase 10.1063/1.443164} {\bibfield  {journal}
  {\bibinfo  {journal} {J. Chem. Phys.}\ }\textbf {\bibinfo {volume} {76}},\
  \bibinfo {pages} {1910–1918} (\bibinfo {year} {1982})}\BibitemShut
  {NoStop}%
\bibitem [{\citenamefont {Lee}, \citenamefont {Kucharski},\ and\ \citenamefont
  {Bartlett}(1984)}]{Lee1984}%
  \BibitemOpen
  \bibfield  {author} {\bibinfo {author} {\bibfnamefont {Y.~S.}\ \bibnamefont
  {Lee}}, \bibinfo {author} {\bibfnamefont {S.~A.}\ \bibnamefont {Kucharski}},
  \ and\ \bibinfo {author} {\bibfnamefont {R.~J.}\ \bibnamefont {Bartlett}},\
  }\bibfield  {title} {\enquote {\bibinfo {title} {A coupled cluster approach
  with triple excitations},}\ }\href {\doibase 10.1063/1.447591} {\bibfield
  {journal} {\bibinfo  {journal} {J. Chem. Phys.}\ }\textbf {\bibinfo {volume}
  {81}},\ \bibinfo {pages} {5906–5912} (\bibinfo {year} {1984})}\BibitemShut
  {NoStop}%
\bibitem [{\citenamefont {Christiansen}, \citenamefont {Koch},\ and\
  \citenamefont {J{\o}rgensen}(1995)}]{christiansen1995second}%
  \BibitemOpen
  \bibfield  {author} {\bibinfo {author} {\bibfnamefont {O.}~\bibnamefont
  {Christiansen}}, \bibinfo {author} {\bibfnamefont {H.}~\bibnamefont {Koch}},
  \ and\ \bibinfo {author} {\bibfnamefont {P.}~\bibnamefont {J{\o}rgensen}},\
  }\bibfield  {title} {\enquote {\bibinfo {title} {The second-order approximate
  coupled cluster singles and doubles model {CC2}},}\ }\href {\doibase
  https://doi.org/10.1016/0009-2614(95)00841-Q} {\bibfield  {journal} {\bibinfo
   {journal} {Chem. Phys. Lett.}\ }\textbf {\bibinfo {volume} {243}},\ \bibinfo
  {pages} {409--418} (\bibinfo {year} {1995})}\BibitemShut {NoStop}%
\bibitem [{\citenamefont {Koch}\ \emph {et~al.}(1997)\citenamefont {Koch},
  \citenamefont {Christiansen}, \citenamefont {Jørgensen}, \citenamefont
  {Sanchez~de Mer{\'a}s},\ and\ \citenamefont {Helgaker}}]{koch1997cc3}%
  \BibitemOpen
  \bibfield  {author} {\bibinfo {author} {\bibfnamefont {H.}~\bibnamefont
  {Koch}}, \bibinfo {author} {\bibfnamefont {O.}~\bibnamefont {Christiansen}},
  \bibinfo {author} {\bibfnamefont {P.}~\bibnamefont {Jørgensen}}, \bibinfo
  {author} {\bibfnamefont {A.~M.}\ \bibnamefont {Sanchez~de Mer{\'a}s}}, \ and\
  \bibinfo {author} {\bibfnamefont {T.}~\bibnamefont {Helgaker}},\ }\bibfield
  {title} {\enquote {\bibinfo {title} {The {CC3} model: {A}n iterative coupled
  cluster approach including connected triples},}\ }\href {\doibase
  https://doi.org/10.1063/1.473322} {\bibfield  {journal} {\bibinfo  {journal}
  {J. Chem. Phys.}\ }\textbf {\bibinfo {volume} {106}},\ \bibinfo {pages}
  {1808--1818} (\bibinfo {year} {1997})}\BibitemShut {NoStop}%
\bibitem [{\citenamefont {Stanton}(1997)}]{stanton1997ccsd}%
  \BibitemOpen
  \bibfield  {author} {\bibinfo {author} {\bibfnamefont {J.~F.}\ \bibnamefont
  {Stanton}},\ }\bibfield  {title} {\enquote {\bibinfo {title} {Why {CCSD (T)}
  works: a different perspective},}\ }\href {\doibase
  https://doi.org/10.1016/S0009-2614(97)01144-5} {\bibfield  {journal}
  {\bibinfo  {journal} {Chem. Phys. Lett.}\ }\textbf {\bibinfo {volume}
  {281}},\ \bibinfo {pages} {130--134} (\bibinfo {year} {1997})}\BibitemShut
  {NoStop}%
\bibitem [{\citenamefont {Stanton}\ and\ \citenamefont
  {Bartlett}(1993)}]{Stanton1993_eom}%
  \BibitemOpen
  \bibfield  {author} {\bibinfo {author} {\bibfnamefont {J.~F.}\ \bibnamefont
  {Stanton}}\ and\ \bibinfo {author} {\bibfnamefont {R.~J.}\ \bibnamefont
  {Bartlett}},\ }\bibfield  {title} {\enquote {\bibinfo {title} {The equation
  of motion coupled-cluster method. {A} systematic biorthogonal approach to
  molecular excitation energies, transition probabilities, and excited state
  properties},}\ }\href {\doibase 10.1063/1.464746} {\bibfield  {journal}
  {\bibinfo  {journal} {J. Chem. Phys.}\ }\textbf {\bibinfo {volume} {98}},\
  \bibinfo {pages} {7029–7039} (\bibinfo {year} {1993})}\BibitemShut
  {NoStop}%
\bibitem [{\citenamefont {Kitsaras}, \citenamefont {Grazioli},\ and\
  \citenamefont {Stopkowicz}(2024)}]{Kitsaras2024}%
  \BibitemOpen
  \bibfield  {author} {\bibinfo {author} {\bibfnamefont {M.-P.}\ \bibnamefont
  {Kitsaras}}, \bibinfo {author} {\bibfnamefont {L.}~\bibnamefont {Grazioli}},
  \ and\ \bibinfo {author} {\bibfnamefont {S.}~\bibnamefont {Stopkowicz}},\
  }\bibfield  {title} {\enquote {\bibinfo {title} {The approximate
  coupled-cluster methods {CC2} and {CC3} in a finite magnetic field},}\ }\href
  {http://dx.doi.org/10.1063/5.0189350} {\bibfield  {journal} {\bibinfo
  {journal} {J. Chem. Phys.}\ }\textbf {\bibinfo {volume} {160}},\ \bibinfo
  {pages} {094112} (\bibinfo {year} {2024})}\BibitemShut {NoStop}%
\bibitem [{\citenamefont {K\"{o}hn}\ and\ \citenamefont
  {Tajti}(2007)}]{Khn2007}%
  \BibitemOpen
  \bibfield  {author} {\bibinfo {author} {\bibfnamefont {A.}~\bibnamefont
  {K\"{o}hn}}\ and\ \bibinfo {author} {\bibfnamefont {A.}~\bibnamefont
  {Tajti}},\ }\bibfield  {title} {\enquote {\bibinfo {title} {Can
  coupled-cluster theory treat conical intersections?}}\ }\href
  {http://dx.doi.org/10.1063/1.2755681} {\bibfield  {journal} {\bibinfo
  {journal} {J. Chem. Phys.}\ }\textbf {\bibinfo {volume} {127}},\ \bibinfo
  {pages} {044105} (\bibinfo {year} {2007})}\BibitemShut {NoStop}%
\bibitem [{\citenamefont {Thomas}\ \emph {et~al.}(2021)\citenamefont {Thomas},
  \citenamefont {Hampe}, \citenamefont {Stopkowicz},\ and\ \citenamefont
  {Gauss}}]{Thomas2021}%
  \BibitemOpen
  \bibfield  {author} {\bibinfo {author} {\bibfnamefont {S.}~\bibnamefont
  {Thomas}}, \bibinfo {author} {\bibfnamefont {F.}~\bibnamefont {Hampe}},
  \bibinfo {author} {\bibfnamefont {S.}~\bibnamefont {Stopkowicz}}, \ and\
  \bibinfo {author} {\bibfnamefont {J.}~\bibnamefont {Gauss}},\ }\bibfield
  {title} {\enquote {\bibinfo {title} {Complex ground-state and excitation
  energies in coupled-cluster theory},}\ }\href
  {http://dx.doi.org/10.1080/00268976.2021.1968056} {\bibfield  {journal}
  {\bibinfo  {journal} {Mol. Phys.}\ }\textbf {\bibinfo {volume} {119}},\
  \bibinfo {pages} {e1968056} (\bibinfo {year} {2021})}\BibitemShut {NoStop}%
\bibitem [{\citenamefont {Bravaya}\ \emph {et~al.}(2013)\citenamefont
  {Bravaya}, \citenamefont {Zuev}, \citenamefont {Epifanovsky},\ and\
  \citenamefont {Krylov}}]{Bravaya2013}%
  \BibitemOpen
  \bibfield  {author} {\bibinfo {author} {\bibfnamefont {K.~B.}\ \bibnamefont
  {Bravaya}}, \bibinfo {author} {\bibfnamefont {D.}~\bibnamefont {Zuev}},
  \bibinfo {author} {\bibfnamefont {E.}~\bibnamefont {Epifanovsky}}, \ and\
  \bibinfo {author} {\bibfnamefont {A.~I.}\ \bibnamefont {Krylov}},\ }\bibfield
   {title} {\enquote {\bibinfo {title} {Complex-scaled equation-of-motion
  coupled-cluster method with single and double substitutions for autoionizing
  excited states: Theory, implementation, and examples},}\ }\href
  {http://dx.doi.org/10.1063/1.4795750} {\bibfield  {journal} {\bibinfo
  {journal} {J. Chem. Phys.}\ }\textbf {\bibinfo {volume} {138}},\ \bibinfo
  {pages} {124106} (\bibinfo {year} {2013})}\BibitemShut {NoStop}%
\bibitem [{\citenamefont {Jagau}\ \emph {et~al.}(2015)\citenamefont {Jagau},
  \citenamefont {Zuev}, \citenamefont {Bravaya}, \citenamefont {Epifanovsky},\
  and\ \citenamefont {Krylov}}]{Jagau2015}%
  \BibitemOpen
  \bibfield  {author} {\bibinfo {author} {\bibfnamefont {T.-C.}\ \bibnamefont
  {Jagau}}, \bibinfo {author} {\bibfnamefont {D.}~\bibnamefont {Zuev}},
  \bibinfo {author} {\bibfnamefont {K.~B.}\ \bibnamefont {Bravaya}}, \bibinfo
  {author} {\bibfnamefont {E.}~\bibnamefont {Epifanovsky}}, \ and\ \bibinfo
  {author} {\bibfnamefont {A.~I.}\ \bibnamefont {Krylov}},\ }\bibfield  {title}
  {\enquote {\bibinfo {title} {Correction to {\textquotedblleft}{A} fresh look
  at resonances and complex absorbing potentials: Density matrix-based
  approach{\textquotedblright}},}\ }\href {\doibase
  10.1021/acs.jpclett.5b02017} {\bibfield  {journal} {\bibinfo  {journal} {J.
  Phys. Chem. Lett.}\ }\textbf {\bibinfo {volume} {6}},\ \bibinfo {pages}
  {3866--3866} (\bibinfo {year} {2015})}\BibitemShut {NoStop}%
\bibitem [{\citenamefont {Jagau}, \citenamefont {Bravaya},\ and\ \citenamefont
  {Krylov}(2017)}]{Jagau2017}%
  \BibitemOpen
  \bibfield  {author} {\bibinfo {author} {\bibfnamefont {T.-C.}\ \bibnamefont
  {Jagau}}, \bibinfo {author} {\bibfnamefont {K.~B.}\ \bibnamefont {Bravaya}},
  \ and\ \bibinfo {author} {\bibfnamefont {A.~I.}\ \bibnamefont {Krylov}},\
  }\bibfield  {title} {\enquote {\bibinfo {title} {Extending quantum chemistry
  of bound states to electronic resonances},}\ }\href {\doibase
  10.1146/annurev-physchem-052516-050622} {\bibfield  {journal} {\bibinfo
  {journal} {Annu. Rev. Phys. Chem.}\ }\textbf {\bibinfo {volume} {68}},\
  \bibinfo {pages} {525–553} (\bibinfo {year} {2017})}\BibitemShut {NoStop}%
\bibitem [{\citenamefont {Benda}\ and\ \citenamefont
  {Jagau}(2018)}]{Benda2018}%
  \BibitemOpen
  \bibfield  {author} {\bibinfo {author} {\bibfnamefont {Z.}~\bibnamefont
  {Benda}}\ and\ \bibinfo {author} {\bibfnamefont {T.-C.}\ \bibnamefont
  {Jagau}},\ }\bibfield  {title} {\enquote {\bibinfo {title} {Locating
  exceptional points on multidimensional complex-valued potential energy
  surfaces},}\ }\href {\doibase 10.1021/acs.jpclett.8b03228} {\bibfield
  {journal} {\bibinfo  {journal} {J. Phys. Chem. Lett.}\ }\textbf {\bibinfo
  {volume} {9}},\ \bibinfo {pages} {6978–6984} (\bibinfo {year}
  {2018})}\BibitemShut {NoStop}%
\bibitem [{\citenamefont {Kutzelnigg}(1977)}]{kutzelnigg1977pair}%
  \BibitemOpen
  \bibfield  {author} {\bibinfo {author} {\bibfnamefont {W.}~\bibnamefont
  {Kutzelnigg}},\ }\bibfield  {title} {\enquote {\bibinfo {title} {Pair
  correlation theories},}\ }in\ \href {\doibase
  https://doi.org/10.1007/978-1-4757-0887-5_5} {\emph {\bibinfo {booktitle}
  {Methods of electronic structure theory}}},\ \bibinfo {editor} {edited by\
  \bibinfo {editor} {\bibfnamefont {H.~F.}\ \bibnamefont {Schaefer~III}}}\
  (\bibinfo  {publisher} {Springer},\ \bibinfo {address} {Boston, MA},\
  \bibinfo {year} {1977})\ pp.\ \bibinfo {pages} {129--188}\BibitemShut
  {NoStop}%
\bibitem [{\citenamefont {Bartlett}, \citenamefont {Kucharski},\ and\
  \citenamefont {Noga}(1989)}]{Bartlett1989}%
  \BibitemOpen
  \bibfield  {author} {\bibinfo {author} {\bibfnamefont {R.~J.}\ \bibnamefont
  {Bartlett}}, \bibinfo {author} {\bibfnamefont {S.~A.}\ \bibnamefont
  {Kucharski}}, \ and\ \bibinfo {author} {\bibfnamefont {J.}~\bibnamefont
  {Noga}},\ }\bibfield  {title} {\enquote {\bibinfo {title} {{Alternative
  coupled-cluster ans\"{a}tze II. The unitary coupled-cluster method}},}\
  }\href {\doibase 10.1016/s0009-2614(89)87372-5} {\bibfield  {journal}
  {\bibinfo  {journal} {Chem. Phys. Lett.}\ }\textbf {\bibinfo {volume}
  {155}},\ \bibinfo {pages} {133–140} (\bibinfo {year} {1989})}\BibitemShut
  {NoStop}%
\bibitem [{\citenamefont {Romero}\ \emph {et~al.}(2018)\citenamefont {Romero},
  \citenamefont {Babbush}, \citenamefont {McClean}, \citenamefont {Hempel},
  \citenamefont {Love},\ and\ \citenamefont {Aspuru-Guzik}}]{Romero2018}%
  \BibitemOpen
  \bibfield  {author} {\bibinfo {author} {\bibfnamefont {J.}~\bibnamefont
  {Romero}}, \bibinfo {author} {\bibfnamefont {R.}~\bibnamefont {Babbush}},
  \bibinfo {author} {\bibfnamefont {J.~R.}\ \bibnamefont {McClean}}, \bibinfo
  {author} {\bibfnamefont {C.}~\bibnamefont {Hempel}}, \bibinfo {author}
  {\bibfnamefont {P.~J.}\ \bibnamefont {Love}}, \ and\ \bibinfo {author}
  {\bibfnamefont {A.}~\bibnamefont {Aspuru-Guzik}},\ }\bibfield  {title}
  {\enquote {\bibinfo {title} {Strategies for quantum computing molecular
  energies using the unitary coupled cluster ansatz},}\ }\href {\doibase
  10.1088/2058-9565/aad3e4} {\bibfield  {journal} {\bibinfo  {journal} {Quantum
  Sci. Technol.}\ }\textbf {\bibinfo {volume} {4}},\ \bibinfo {pages} {014008}
  (\bibinfo {year} {2018})}\BibitemShut {NoStop}%
\bibitem [{\citenamefont {Lee}\ \emph {et~al.}(2018)\citenamefont {Lee},
  \citenamefont {Huggins}, \citenamefont {Head-Gordon},\ and\ \citenamefont
  {Whaley}}]{Lee2018}%
  \BibitemOpen
  \bibfield  {author} {\bibinfo {author} {\bibfnamefont {J.}~\bibnamefont
  {Lee}}, \bibinfo {author} {\bibfnamefont {W.~J.}\ \bibnamefont {Huggins}},
  \bibinfo {author} {\bibfnamefont {M.}~\bibnamefont {Head-Gordon}}, \ and\
  \bibinfo {author} {\bibfnamefont {K.~B.}\ \bibnamefont {Whaley}},\ }\bibfield
   {title} {\enquote {\bibinfo {title} {Generalized unitary coupled cluster
  wave functions for quantum computation},}\ }\href {\doibase
  10.1021/acs.jctc.8b01004} {\bibfield  {journal} {\bibinfo  {journal} {J.
  Chem. Theory Comput.}\ }\textbf {\bibinfo {volume} {15}},\ \bibinfo {pages}
  {311–324} (\bibinfo {year} {2018})}\BibitemShut {NoStop}%
\bibitem [{\citenamefont {Evangelista}, \citenamefont {Chan},\ and\
  \citenamefont {Scuseria}(2019)}]{Evangelista2019}%
  \BibitemOpen
  \bibfield  {author} {\bibinfo {author} {\bibfnamefont {F.~A.}\ \bibnamefont
  {Evangelista}}, \bibinfo {author} {\bibfnamefont {G.~K.-L.}\ \bibnamefont
  {Chan}}, \ and\ \bibinfo {author} {\bibfnamefont {G.~E.}\ \bibnamefont
  {Scuseria}},\ }\bibfield  {title} {\enquote {\bibinfo {title} {Exact
  parameterization of fermionic wave functions via unitary coupled cluster
  theory},}\ }\href {http://dx.doi.org/10.1063/1.5133059} {\bibfield  {journal}
  {\bibinfo  {journal} {J. Chem. Phys.}\ }\textbf {\bibinfo {volume} {151}},\
  \bibinfo {pages} {244112} (\bibinfo {year} {2019})}\BibitemShut {NoStop}%
\bibitem [{\citenamefont {Anand}\ \emph {et~al.}(2022)\citenamefont {Anand},
  \citenamefont {Schleich}, \citenamefont {Alperin-Lea}, \citenamefont
  {Jensen}, \citenamefont {Sim}, \citenamefont {Díaz-Tinoco}, \citenamefont
  {Kottmann}, \citenamefont {Degroote}, \citenamefont {Izmaylov},\ and\
  \citenamefont {Aspuru-Guzik}}]{Anand2022}%
  \BibitemOpen
  \bibfield  {author} {\bibinfo {author} {\bibfnamefont {A.}~\bibnamefont
  {Anand}}, \bibinfo {author} {\bibfnamefont {P.}~\bibnamefont {Schleich}},
  \bibinfo {author} {\bibfnamefont {S.}~\bibnamefont {Alperin-Lea}}, \bibinfo
  {author} {\bibfnamefont {P.~W.~K.}\ \bibnamefont {Jensen}}, \bibinfo {author}
  {\bibfnamefont {S.}~\bibnamefont {Sim}}, \bibinfo {author} {\bibfnamefont
  {M.}~\bibnamefont {Díaz-Tinoco}}, \bibinfo {author} {\bibfnamefont {J.~S.}\
  \bibnamefont {Kottmann}}, \bibinfo {author} {\bibfnamefont {M.}~\bibnamefont
  {Degroote}}, \bibinfo {author} {\bibfnamefont {A.~F.}\ \bibnamefont
  {Izmaylov}}, \ and\ \bibinfo {author} {\bibfnamefont {A.}~\bibnamefont
  {Aspuru-Guzik}},\ }\bibfield  {title} {\enquote {\bibinfo {title} {A quantum
  computing view on unitary coupled cluster theory},}\ }\href {\doibase
  10.1039/d1cs00932j} {\bibfield  {journal} {\bibinfo  {journal} {Chem. Soc.
  Rev.}\ }\textbf {\bibinfo {volume} {51}},\ \bibinfo {pages} {1659–1684}
  (\bibinfo {year} {2022})}\BibitemShut {NoStop}%
\bibitem [{\citenamefont {Culpitt}, \citenamefont {Tellgren},\ and\
  \citenamefont {Pavošević}(2023)}]{Culpitt2023}%
  \BibitemOpen
  \bibfield  {author} {\bibinfo {author} {\bibfnamefont {T.}~\bibnamefont
  {Culpitt}}, \bibinfo {author} {\bibfnamefont {E.~I.}\ \bibnamefont
  {Tellgren}}, \ and\ \bibinfo {author} {\bibfnamefont {F.}~\bibnamefont
  {Pavošević}},\ }\bibfield  {title} {\enquote {\bibinfo {title} {Unitary
  coupled-cluster for quantum computation of molecular properties in a strong
  magnetic field},}\ }\href {\doibase 10.1063/5.0177417} {\bibfield  {journal}
  {\bibinfo  {journal} {J. Chem. Phys.}\ }\textbf {\bibinfo {volume} {159}}
  (\bibinfo {year} {2023}),\ 10.1063/5.0177417}\BibitemShut {NoStop}%
\bibitem [{\citenamefont {Taube}\ and\ \citenamefont
  {Bartlett}(2006)}]{Taube2006}%
  \BibitemOpen
  \bibfield  {author} {\bibinfo {author} {\bibfnamefont {A.~G.}\ \bibnamefont
  {Taube}}\ and\ \bibinfo {author} {\bibfnamefont {R.~J.}\ \bibnamefont
  {Bartlett}},\ }\bibfield  {title} {\enquote {\bibinfo {title} {New
  perspectives on unitary coupled‐cluster theory},}\ }\href {\doibase
  10.1002/qua.21198} {\bibfield  {journal} {\bibinfo  {journal} {Int. J.
  Quantum Chem.}\ }\textbf {\bibinfo {volume} {106}},\ \bibinfo {pages}
  {3393–3401} (\bibinfo {year} {2006})}\BibitemShut {NoStop}%
\bibitem [{\citenamefont {Liu}\ and\ \citenamefont
  {Cheng}(2021)}]{liu2021unitary}%
  \BibitemOpen
  \bibfield  {author} {\bibinfo {author} {\bibfnamefont {J.}~\bibnamefont
  {Liu}}\ and\ \bibinfo {author} {\bibfnamefont {L.}~\bibnamefont {Cheng}},\
  }\bibfield  {title} {\enquote {\bibinfo {title} {{Unitary coupled-cluster
  based self-consistent polarization propagator theory: A quadratic unitary
  coupled-cluster singles and doubles scheme}},}\ }\href {\doibase
  https://doi.org/10.1063/5.0062090} {\bibfield  {journal} {\bibinfo  {journal}
  {J. Chem. Phys.}\ }\textbf {\bibinfo {volume} {155}},\ \bibinfo {pages}
  {174102} (\bibinfo {year} {2021})}\BibitemShut {NoStop}%
\bibitem [{\citenamefont {Liu}, \citenamefont {Matthews},\ and\ \citenamefont
  {Cheng}(2022)}]{Liu2022}%
  \BibitemOpen
  \bibfield  {author} {\bibinfo {author} {\bibfnamefont {J.}~\bibnamefont
  {Liu}}, \bibinfo {author} {\bibfnamefont {D.~A.}\ \bibnamefont {Matthews}}, \
  and\ \bibinfo {author} {\bibfnamefont {L.}~\bibnamefont {Cheng}},\ }\bibfield
   {title} {\enquote {\bibinfo {title} {Quadratic unitary coupled-cluster
  singles and doubles scheme: Efficient implementation, benchmark study, and
  formulation of an extended version},}\ }\href {\doibase
  10.1021/acs.jctc.1c01210} {\bibfield  {journal} {\bibinfo  {journal} {J.
  Chem. Theory Comput.}\ }\textbf {\bibinfo {volume} {18}},\ \bibinfo {pages}
  {2281–2291} (\bibinfo {year} {2022})}\BibitemShut {NoStop}%
\bibitem [{\citenamefont {Bauman}\ and\ \citenamefont
  {Kowalski}(2022)}]{Bauman2022}%
  \BibitemOpen
  \bibfield  {author} {\bibinfo {author} {\bibfnamefont {N.~P.}\ \bibnamefont
  {Bauman}}\ and\ \bibinfo {author} {\bibfnamefont {K.}~\bibnamefont
  {Kowalski}},\ }\bibfield  {title} {\enquote {\bibinfo {title} {Coupled
  cluster downfolding methods: The effect of double commutator terms on the
  accuracy of ground-state energies},}\ }\href
  {http://dx.doi.org/10.1063/5.0076260} {\bibfield  {journal} {\bibinfo
  {journal} {J. Chem. Phys.}\ }\textbf {\bibinfo {volume} {156}},\ \bibinfo
  {pages} {094106} (\bibinfo {year} {2022})}\BibitemShut {NoStop}%
\bibitem [{\citenamefont {Neuscamman}, \citenamefont {Yanai},\ and\
  \citenamefont {Chan}(2009)}]{Neuscamman2009}%
  \BibitemOpen
  \bibfield  {author} {\bibinfo {author} {\bibfnamefont {E.}~\bibnamefont
  {Neuscamman}}, \bibinfo {author} {\bibfnamefont {T.}~\bibnamefont {Yanai}}, \
  and\ \bibinfo {author} {\bibfnamefont {G.~K.-L.}\ \bibnamefont {Chan}},\
  }\bibfield  {title} {\enquote {\bibinfo {title} {Quadratic canonical
  transformation theory and higher order density matrices},}\ }\href
  {http://dx.doi.org/10.1063/1.3086932} {\bibfield  {journal} {\bibinfo
  {journal} {J. Chem. Phys.}\ }\textbf {\bibinfo {volume} {130}},\ \bibinfo
  {pages} {124102} (\bibinfo {year} {2009})}\BibitemShut {NoStop}%
\bibitem [{\citenamefont {Liu}\ \emph {et~al.}(2018)\citenamefont {Liu},
  \citenamefont {Asthana}, \citenamefont {Cheng},\ and\ \citenamefont
  {Mukherjee}}]{Liu2018}%
  \BibitemOpen
  \bibfield  {author} {\bibinfo {author} {\bibfnamefont {J.}~\bibnamefont
  {Liu}}, \bibinfo {author} {\bibfnamefont {A.}~\bibnamefont {Asthana}},
  \bibinfo {author} {\bibfnamefont {L.}~\bibnamefont {Cheng}}, \ and\ \bibinfo
  {author} {\bibfnamefont {D.}~\bibnamefont {Mukherjee}},\ }\bibfield  {title}
  {\enquote {\bibinfo {title} {{Unitary coupled-cluster based self-consistent
  polarization propagator theory: A third-order formulation and pilot
  applications}},}\ }\href {https://doi.org/10.1063/1.5030344} {\bibfield
  {journal} {\bibinfo  {journal} {J. Chem. Phys.}\ }\textbf {\bibinfo {volume}
  {148}},\ \bibinfo {pages} {244110} (\bibinfo {year} {2018})}\BibitemShut
  {NoStop}%
\bibitem [{\citenamefont {Phillips}\ \emph {et~al.}(2025)\citenamefont
  {Phillips}, \citenamefont {Koulias}, \citenamefont {Yuwono},\ and\
  \citenamefont {DePrince}}]{UCC_deprince}%
  \BibitemOpen
  \bibfield  {author} {\bibinfo {author} {\bibfnamefont {J.~T.}\ \bibnamefont
  {Phillips}}, \bibinfo {author} {\bibfnamefont {L.~N.}\ \bibnamefont
  {Koulias}}, \bibinfo {author} {\bibfnamefont {S.~H.}\ \bibnamefont {Yuwono}},
  \ and\ \bibinfo {author} {\bibfnamefont {A.~E.}\ \bibnamefont {DePrince}},\
  }\bibfield  {title} {\enquote {\bibinfo {title} {Comparing perturbative and
  commutator-rank-based truncation schemes in unitary coupled-cluster
  theory},}\ }\href {\doibase https://arxiv.org/abs/2503.00617} {\  (\bibinfo
  {year} {2025}),\ https://arxiv.org/abs/2503.00617}\BibitemShut {NoStop}%
\bibitem [{\citenamefont {Asthana}\ \emph {et~al.}(2023)\citenamefont
  {Asthana}, \citenamefont {Kumar}, \citenamefont {Abraham}, \citenamefont
  {Grimsley}, \citenamefont {Zhang}, \citenamefont {Cincio}, \citenamefont
  {Tretiak}, \citenamefont {Dub}, \citenamefont {Economou}, \citenamefont
  {Barnes} \emph {et~al.}}]{asthana2023quantum}%
  \BibitemOpen
  \bibfield  {author} {\bibinfo {author} {\bibfnamefont {A.}~\bibnamefont
  {Asthana}}, \bibinfo {author} {\bibfnamefont {A.}~\bibnamefont {Kumar}},
  \bibinfo {author} {\bibfnamefont {V.}~\bibnamefont {Abraham}}, \bibinfo
  {author} {\bibfnamefont {H.}~\bibnamefont {Grimsley}}, \bibinfo {author}
  {\bibfnamefont {Y.}~\bibnamefont {Zhang}}, \bibinfo {author} {\bibfnamefont
  {L.}~\bibnamefont {Cincio}}, \bibinfo {author} {\bibfnamefont
  {S.}~\bibnamefont {Tretiak}}, \bibinfo {author} {\bibfnamefont {P.~A.}\
  \bibnamefont {Dub}}, \bibinfo {author} {\bibfnamefont {S.~E.}\ \bibnamefont
  {Economou}}, \bibinfo {author} {\bibfnamefont {E.}~\bibnamefont {Barnes}},
  \emph {et~al.},\ }\bibfield  {title} {\enquote {\bibinfo {title} {Quantum
  self-consistent equation-of-motion method for computing molecular excitation
  energies, ionization potentials, and electron affinities on a quantum
  computer},}\ }\href {\doibase https://doi.org/10.1039/D2SC05371C} {\bibfield
  {journal} {\bibinfo  {journal} {Chem. Sci.}\ }\textbf {\bibinfo {volume}
  {14}},\ \bibinfo {pages} {2405--2418} (\bibinfo {year} {2023})}\BibitemShut
  {NoStop}%
\bibitem [{\citenamefont {Hodecker}\ \emph {et~al.}(2020)\citenamefont
  {Hodecker}, \citenamefont {Thielen}, \citenamefont {Liu}, \citenamefont
  {Rehn},\ and\ \citenamefont {Dreuw}}]{Hodecker2020}%
  \BibitemOpen
  \bibfield  {author} {\bibinfo {author} {\bibfnamefont {M.}~\bibnamefont
  {Hodecker}}, \bibinfo {author} {\bibfnamefont {S.~M.}\ \bibnamefont
  {Thielen}}, \bibinfo {author} {\bibfnamefont {J.}~\bibnamefont {Liu}},
  \bibinfo {author} {\bibfnamefont {D.~R.}\ \bibnamefont {Rehn}}, \ and\
  \bibinfo {author} {\bibfnamefont {A.}~\bibnamefont {Dreuw}},\ }\bibfield
  {title} {\enquote {\bibinfo {title} {Third-order {U}nitary {C}oupled
  {C}luster ({UCC}3) for excited electronic states: Efficient implementation
  and benchmarking},}\ }\href {https://doi.org/10.1021/acs.jctc.0c00335}
  {\bibfield  {journal} {\bibinfo  {journal} {J. Chem. Theory Comput.}\
  }\textbf {\bibinfo {volume} {16}},\ \bibinfo {pages} {3654--3663} (\bibinfo
  {year} {2020})}\BibitemShut {NoStop}%
\bibitem [{\citenamefont {Thielen}\ \emph {et~al.}(2021)\citenamefont
  {Thielen}, \citenamefont {Hodecker}, \citenamefont {Piazolo}, \citenamefont
  {Rehn},\ and\ \citenamefont {Dreuw}}]{Thielen2021}%
  \BibitemOpen
  \bibfield  {author} {\bibinfo {author} {\bibfnamefont {S.~M.}\ \bibnamefont
  {Thielen}}, \bibinfo {author} {\bibfnamefont {M.}~\bibnamefont {Hodecker}},
  \bibinfo {author} {\bibfnamefont {J.}~\bibnamefont {Piazolo}}, \bibinfo
  {author} {\bibfnamefont {D.~R.}\ \bibnamefont {Rehn}}, \ and\ \bibinfo
  {author} {\bibfnamefont {A.}~\bibnamefont {Dreuw}},\ }\bibfield  {title}
  {\enquote {\bibinfo {title} {Unitary coupled-cluster approach for the
  calculation of core-excited states and {X}-ray absorption spectra},}\ }\href
  {\doibase 10.1063/5.0047134} {\bibfield  {journal} {\bibinfo  {journal} {J.
  Chem. Phys.}\ }\textbf {\bibinfo {volume} {154}},\ \bibinfo {pages} {154108}
  (\bibinfo {year} {2021})}\BibitemShut {NoStop}%
\bibitem [{\citenamefont {Hodecker}\ \emph {et~al.}(2022)\citenamefont
  {Hodecker}, \citenamefont {Dempwolff}, \citenamefont {Schirmer},\ and\
  \citenamefont {Dreuw}}]{Hodecker2022}%
  \BibitemOpen
  \bibfield  {author} {\bibinfo {author} {\bibfnamefont {M.}~\bibnamefont
  {Hodecker}}, \bibinfo {author} {\bibfnamefont {A.~L.}\ \bibnamefont
  {Dempwolff}}, \bibinfo {author} {\bibfnamefont {J.}~\bibnamefont {Schirmer}},
  \ and\ \bibinfo {author} {\bibfnamefont {A.}~\bibnamefont {Dreuw}},\
  }\bibfield  {title} {\enquote {\bibinfo {title} {Theoretical analysis and
  comparison of unitary coupled-cluster and algebraic-diagrammatic construction
  methods for ionization},}\ }\href {http://dx.doi.org/10.1063/5.0070967}
  {\bibfield  {journal} {\bibinfo  {journal} {J. Chem. Phys.}\ }\textbf
  {\bibinfo {volume} {156}},\ \bibinfo {pages} {074104} (\bibinfo {year}
  {2022})}\BibitemShut {NoStop}%
\bibitem [{\citenamefont {Dreuw}, \citenamefont {Papapostolou},\ and\
  \citenamefont {Dempwolff}(2023)}]{Dreuw2023}%
  \BibitemOpen
  \bibfield  {author} {\bibinfo {author} {\bibfnamefont {A.}~\bibnamefont
  {Dreuw}}, \bibinfo {author} {\bibfnamefont {A.}~\bibnamefont {Papapostolou}},
  \ and\ \bibinfo {author} {\bibfnamefont {A.~L.}\ \bibnamefont {Dempwolff}},\
  }\bibfield  {title} {\enquote {\bibinfo {title} {Algebraic {D}iagrammatic
  {C}onstruction schemes employing the intermediate state formalism: Theory,
  capabilities, and interpretation},}\ }\href {\doibase
  10.1021/acs.jpca.3c02761} {\bibfield  {journal} {\bibinfo  {journal} {J.
  Phys. Chem. A}\ }\textbf {\bibinfo {volume} {127}},\ \bibinfo {pages}
  {6635--6646} (\bibinfo {year} {2023})}\BibitemShut {NoStop}%
\bibitem [{\citenamefont {Kitsaras}(2023)}]{Kitsarasthes}%
  \BibitemOpen
  \bibfield  {author} {\bibinfo {author} {\bibfnamefont {M.-P.}\ \bibnamefont
  {Kitsaras}},\ }\emph {\bibinfo {title} {Finite magnetic-field coupled-cluster
  methods: efficiency and utilities}},\ \href {\doibase
  10.25358/OPENSCIENCE-9599} {Ph.D. thesis},\ \bibinfo  {school} {Johannes
  Gutenberg-Universit\"{a}t Mainz} (\bibinfo {year} {2023})\BibitemShut
  {NoStop}%
\bibitem [{\citenamefont {Kim}\ and\ \citenamefont
  {Krylov}(2023)}]{kim2023two}%
  \BibitemOpen
  \bibfield  {author} {\bibinfo {author} {\bibfnamefont {Y.}~\bibnamefont
  {Kim}}\ and\ \bibinfo {author} {\bibfnamefont {A.~I.}\ \bibnamefont
  {Krylov}},\ }\bibfield  {title} {\enquote {\bibinfo {title} {Two algorithms
  for excited-state quantum solvers: Theory and application to {EOM-UCCSD}},}\
  }\href {http://dx.doi.org/10.1021/acs.jpca.3c02480} {\bibfield  {journal}
  {\bibinfo  {journal} {J. Phys. Chem. A}\ }\textbf {\bibinfo {volume} {127}},\
  \bibinfo {pages} {6552–6566} (\bibinfo {year} {2023})}\BibitemShut
  {NoStop}%
\bibitem [{\citenamefont {Evangelista}(2011)}]{Evangelista2011}%
  \BibitemOpen
  \bibfield  {author} {\bibinfo {author} {\bibfnamefont {F.~A.}\ \bibnamefont
  {Evangelista}},\ }\bibfield  {title} {\enquote {\bibinfo {title} {Alternative
  single-reference coupled cluster approaches for multireference problems: The
  simpler, the better},}\ }\href {http://dx.doi.org/10.1063/1.3598471}
  {\bibfield  {journal} {\bibinfo  {journal} {J. Chem. Phys.}\ }\textbf
  {\bibinfo {volume} {134}},\ \bibinfo {pages} {224102} (\bibinfo {year}
  {2011})}\BibitemShut {NoStop}%
\bibitem [{\citenamefont {Shavitt}\ and\ \citenamefont
  {Bartlett}(2009)}]{Shavitt2009}%
  \BibitemOpen
  \bibfield  {author} {\bibinfo {author} {\bibfnamefont {I.}~\bibnamefont
  {Shavitt}}\ and\ \bibinfo {author} {\bibfnamefont {R.~J.}\ \bibnamefont
  {Bartlett}},\ }\href {\doibase 10.1017/cbo9780511596834} {\emph {\bibinfo
  {title} {Many-Body Methods in Chemistry and Physics: MBPT and Coupled-Cluster
  Theory}}}\ (\bibinfo  {publisher} {Cambridge University Press},\ \bibinfo
  {address} {Cambridge},\ \bibinfo {year} {2009})\BibitemShut {NoStop}%
\bibitem [{\citenamefont {Hampe}\ \emph {et~al.}()\citenamefont {Hampe},
  \citenamefont {Stopkowicz}, \citenamefont {Groß}, \citenamefont {Kitsaras},
  \citenamefont {Grazioli}, \citenamefont {Blaschke}, \citenamefont {Monzel},
  \citenamefont {Yerg{\"u}n},\ and\ \citenamefont {Röper}}]{QCUMBRE}%
  \BibitemOpen
  \bibfield  {author} {\bibinfo {author} {\bibfnamefont {F.}~\bibnamefont
  {Hampe}}, \bibinfo {author} {\bibfnamefont {S.}~\bibnamefont {Stopkowicz}},
  \bibinfo {author} {\bibfnamefont {N.}~\bibnamefont {Groß}}, \bibinfo
  {author} {\bibfnamefont {M.-P.}\ \bibnamefont {Kitsaras}}, \bibinfo {author}
  {\bibfnamefont {L.}~\bibnamefont {Grazioli}}, \bibinfo {author}
  {\bibfnamefont {S.}~\bibnamefont {Blaschke}}, \bibinfo {author}
  {\bibfnamefont {L.}~\bibnamefont {Monzel}}, \bibinfo {author} {\bibfnamefont
  {P.}~\bibnamefont {Yerg{\"u}n}}, \ and\ \bibinfo {author} {\bibfnamefont
  {C.-M.}\ \bibnamefont {Röper}},\ }\href@noop {} {\enquote {\bibinfo {title}
  {{QCUMBRE}, quantum chemical utility enabling magnetic-field dependent
  investigations benefitting from rigorous electron-correlation treatment},}\
  }\bibinfo {note} {{F}or the current version, see
  https://www.qcumbre.org}\BibitemShut {NoStop}%
\bibitem [{\citenamefont {Gauss}\ \emph {et~al.}(2023)\citenamefont {Gauss},
  \citenamefont {Lipparini}, \citenamefont {Burger}, \citenamefont
  {S.Blaschke}, \citenamefont {Kitsaras},\ and\ \citenamefont
  {Stopkowicz}}]{mint}%
  \BibitemOpen
  \bibfield  {author} {\bibinfo {author} {\bibfnamefont {J.}~\bibnamefont
  {Gauss}}, \bibinfo {author} {\bibfnamefont {F.}~\bibnamefont {Lipparini}},
  \bibinfo {author} {\bibfnamefont {S.}~\bibnamefont {Burger}}, \bibinfo
  {author} {\bibnamefont {S.Blaschke}}, \bibinfo {author} {\bibfnamefont
  {M.-P.}\ \bibnamefont {Kitsaras}}, \ and\ \bibinfo {author} {\bibfnamefont
  {S.}~\bibnamefont {Stopkowicz}},\ }\href@noop {} {\enquote {\bibinfo {title}
  {{MINT, Mainz INTegral package}},}\ } (\bibinfo {year} {2015-2023}),\
  \bibinfo {note} {{Johannes Gutenberg-Universität Mainz,
  unpublished}}\BibitemShut {NoStop}%
\bibitem [{\citenamefont {Matthews}\ \emph {et~al.}(2020)\citenamefont
  {Matthews}, \citenamefont {Cheng}, \citenamefont {Harding}, \citenamefont
  {Lipparini}, \citenamefont {Stopkowicz}, \citenamefont {Jagau}, \citenamefont
  {Szalay}, \citenamefont {Gauss},\ and\ \citenamefont
  {Stanton}}]{matthews2020coupled}%
  \BibitemOpen
  \bibfield  {author} {\bibinfo {author} {\bibfnamefont {D.~A.}\ \bibnamefont
  {Matthews}}, \bibinfo {author} {\bibfnamefont {L.}~\bibnamefont {Cheng}},
  \bibinfo {author} {\bibfnamefont {M.~E.}\ \bibnamefont {Harding}}, \bibinfo
  {author} {\bibfnamefont {F.}~\bibnamefont {Lipparini}}, \bibinfo {author}
  {\bibfnamefont {S.}~\bibnamefont {Stopkowicz}}, \bibinfo {author}
  {\bibfnamefont {T.-C.}\ \bibnamefont {Jagau}}, \bibinfo {author}
  {\bibfnamefont {P.~G.}\ \bibnamefont {Szalay}}, \bibinfo {author}
  {\bibfnamefont {J.}~\bibnamefont {Gauss}}, \ and\ \bibinfo {author}
  {\bibfnamefont {J.~F.}\ \bibnamefont {Stanton}},\ }\bibfield  {title}
  {\enquote {\bibinfo {title} {Coupled-cluster techniques for computational
  chemistry: The {CFOUR} program package},}\ }\href {\doibase
  https://doi.org/10.1063/5.0004837} {\bibfield  {journal} {\bibinfo  {journal}
  {J. Chem. Phys.}\ }\textbf {\bibinfo {volume} {152}},\ \bibinfo {pages}
  {214108} (\bibinfo {year} {2020})}\BibitemShut {NoStop}%
\bibitem [{\citenamefont {Lawson}\ \emph {et~al.}(1979)\citenamefont {Lawson},
  \citenamefont {Hanson}, \citenamefont {Kincaid},\ and\ \citenamefont
  {Krogh}}]{Lawson1979}%
  \BibitemOpen
  \bibfield  {author} {\bibinfo {author} {\bibfnamefont {C.~L.}\ \bibnamefont
  {Lawson}}, \bibinfo {author} {\bibfnamefont {R.~J.}\ \bibnamefont {Hanson}},
  \bibinfo {author} {\bibfnamefont {D.~R.}\ \bibnamefont {Kincaid}}, \ and\
  \bibinfo {author} {\bibfnamefont {F.~T.}\ \bibnamefont {Krogh}},\ }\bibfield
  {title} {\enquote {\bibinfo {title} {Basic linear algebra subprograms for
  fortran usage},}\ }\href {\doibase 10.1145/355841.355847} {\bibfield
  {journal} {\bibinfo  {journal} {ACM Trans. Math. Softw.}\ }\textbf {\bibinfo
  {volume} {5}},\ \bibinfo {pages} {308–323} (\bibinfo {year}
  {1979})}\BibitemShut {NoStop}%
\bibitem [{\citenamefont {Stanton}\ \emph {et~al.}()\citenamefont {Stanton},
  \citenamefont {Gauss}, \citenamefont {Cheng}, \citenamefont {Harding},
  \citenamefont {Matthews},\ and\ \citenamefont {Szalay}}]{cfour}%
  \BibitemOpen
  \bibfield  {author} {\bibinfo {author} {\bibfnamefont {J.~F.}\ \bibnamefont
  {Stanton}}, \bibinfo {author} {\bibfnamefont {J.}~\bibnamefont {Gauss}},
  \bibinfo {author} {\bibfnamefont {L.}~\bibnamefont {Cheng}}, \bibinfo
  {author} {\bibfnamefont {M.~E.}\ \bibnamefont {Harding}}, \bibinfo {author}
  {\bibfnamefont {D.~A.}\ \bibnamefont {Matthews}}, \ and\ \bibinfo {author}
  {\bibfnamefont {P.~G.}\ \bibnamefont {Szalay}},\ }\href@noop {} {\enquote
  {\bibinfo {title} {{CFOUR, Coupled-Cluster techniques for Computational
  Chemistry, a quantum-chemical program package}},}\ }\bibinfo {note} {{W}ith
  contributions from {A}. {A}sthana, {A}.{A}. {A}uer, {R}.{J}. {B}artlett, {U}.
  {B}enedikt, {C}. {B}erger, {D}.{E}. {B}ernholdt, {S}. {B}laschke, {Y}. {J}.
  {B}omble, {S}. {B}urger, {O}. {C}hristiansen, {D}. {D}atta, {F}. {E}ngel,
  {R}. {F}aber, {J}. {G}reiner, {M}. {H}eckert, {O}. {H}eun, {M}. Hilgenberg,
  {C}. {H}uber, {T}.-{C}. {J}agau, {D}. {J}onsson, {J}. {J}us{\'e}lius, {T}.
  Kirsch, {M}.-{P}. {K}itsaras, {K}. {K}lein, {G}.{M}. {K}opper, {W}.{J}.
  {L}auderdale, {F}. {L}ipparini, {J}. {L}iu, {T}. {M}etzroth, {L}.{A}.
  {M}{\"u}ck, {D}.{P}. {O}'{N}eill, {T}. {N}ottoli, {J}. {O}swald, {D}.{R}.
  {P}rice, {E}. {P}rochnow, {C}. {P}uzzarini, {K}. {R}uud, {F}. {S}chiffmann,
  {W}. {S}chwalbach, {C}. {S}immons, {S}. {S}topkowicz, {A}. {T}ajti, {T.}
  Uhl\'{i}{\v{r}ov\'{a}, {J}. {V}{\'a}zquez, {F}. {W}ang, {J}.{D}. {W}atts,
  {P.} Yerg{\"u}n, {C}. {Z}hang, {X}. {Z}heng, and the integral packages
  {MOLECULE} ({J}. {A}lml{\"o}f and {P}.{R}. {T}aylor), {PROPS} ({P}.{R}.
  {T}aylor), {ABACUS} ({T}. {H}elgaker, {H}.{J}.{A}a. {J}ensen, {P}.
  {J}{\o}rgensen, and {J}. {O}lsen), and {ECP} routines by {A}. {V}. {M}itin
  and {C}. van {W}{\"u}llen. {F}or the current version, see
  http://www.cfour.de.}}\BibitemShut {Stop}%
\bibitem [{\citenamefont {Pritchard}\ \emph {et~al.}(2019)\citenamefont
  {Pritchard}, \citenamefont {Altarawy}, \citenamefont {Didier}, \citenamefont
  {Gibsom},\ and\ \citenamefont {Windus}}]{pritchard2019a}%
  \BibitemOpen
  \bibfield  {author} {\bibinfo {author} {\bibfnamefont {B.~P.}\ \bibnamefont
  {Pritchard}}, \bibinfo {author} {\bibfnamefont {D.}~\bibnamefont {Altarawy}},
  \bibinfo {author} {\bibfnamefont {B.}~\bibnamefont {Didier}}, \bibinfo
  {author} {\bibfnamefont {T.~D.}\ \bibnamefont {Gibsom}}, \ and\ \bibinfo
  {author} {\bibfnamefont {T.~L.}\ \bibnamefont {Windus}},\ }\bibfield  {title}
  {\enquote {\bibinfo {title} {A new basis set exchange: An open, up-to-date
  resource for the molecular sciences community},}\ }\href {\doibase
  10.1021/acs.jcim.9b00725} {\bibfield  {journal} {\bibinfo  {journal} {J.
  Chem. Inf. Model.}\ }\textbf {\bibinfo {volume} {59}},\ \bibinfo {pages}
  {4814--4820} (\bibinfo {year} {2019})}\BibitemShut {NoStop}%
\bibitem [{\citenamefont {Feller}(1996)}]{feller1996a}%
  \BibitemOpen
  \bibfield  {author} {\bibinfo {author} {\bibfnamefont {D.}~\bibnamefont
  {Feller}},\ }\bibfield  {title} {\enquote {\bibinfo {title} {The role of
  databases in support of computational chemistry calculations},}\ }\href
  {\doibase 10.1002/(SICI)1096-987X(199610)17:13<1571::AID-JCC9>3.0.CO;2-P}
  {\bibfield  {journal} {\bibinfo  {journal} {J. Comput. Chem.}\ }\textbf
  {\bibinfo {volume} {17}},\ \bibinfo {pages} {1571--1586} (\bibinfo {year}
  {1996})}\BibitemShut {NoStop}%
\bibitem [{\citenamefont {Schuchardt}\ \emph {et~al.}(2007)\citenamefont
  {Schuchardt}, \citenamefont {Didier}, \citenamefont {Elsethagen},
  \citenamefont {Sun}, \citenamefont {Gurumoorthi}, \citenamefont {Chase},
  \citenamefont {Li},\ and\ \citenamefont {Windus}}]{schuchardt2007a}%
  \BibitemOpen
  \bibfield  {author} {\bibinfo {author} {\bibfnamefont {K.~L.}\ \bibnamefont
  {Schuchardt}}, \bibinfo {author} {\bibfnamefont {B.~T.}\ \bibnamefont
  {Didier}}, \bibinfo {author} {\bibfnamefont {T.}~\bibnamefont {Elsethagen}},
  \bibinfo {author} {\bibfnamefont {L.}~\bibnamefont {Sun}}, \bibinfo {author}
  {\bibfnamefont {V.}~\bibnamefont {Gurumoorthi}}, \bibinfo {author}
  {\bibfnamefont {J.}~\bibnamefont {Chase}}, \bibinfo {author} {\bibfnamefont
  {J.}~\bibnamefont {Li}}, \ and\ \bibinfo {author} {\bibfnamefont {T.~L.}\
  \bibnamefont {Windus}},\ }\bibfield  {title} {\enquote {\bibinfo {title}
  {Basis set exchange: A community database for computational sciences},}\
  }\href {\doibase 10.1021/ci600510j} {\bibfield  {journal} {\bibinfo
  {journal} {J. Chem. Inf. Model.}\ }\textbf {\bibinfo {volume} {47}},\
  \bibinfo {pages} {1045--1052} (\bibinfo {year} {2007})}\BibitemShut {NoStop}%
\bibitem [{\citenamefont {Dunning}(1989)}]{dunning1989a}%
  \BibitemOpen
  \bibfield  {author} {\bibinfo {author} {\bibfnamefont {T.~H.}\ \bibnamefont
  {Dunning}},\ }\bibfield  {title} {\enquote {\bibinfo {title} {{Gaussian basis
  sets for use in correlated molecular calculations. I. The atoms boron through
  neon and hydrogen}},}\ }\href {\doibase 10.1063/1.456153} {\bibfield
  {journal} {\bibinfo  {journal} {J. Chem. Phys.}\ }\textbf {\bibinfo {volume}
  {90}},\ \bibinfo {pages} {1007--1023} (\bibinfo {year} {1989})}\BibitemShut
  {NoStop}%
\bibitem [{\citenamefont {Kendall}, \citenamefont {Dunning},\ and\
  \citenamefont {Harrison}(1992)}]{kendall1992a}%
  \BibitemOpen
  \bibfield  {author} {\bibinfo {author} {\bibfnamefont {R.~A.}\ \bibnamefont
  {Kendall}}, \bibinfo {author} {\bibfnamefont {T.~H.}\ \bibnamefont
  {Dunning}}, \ and\ \bibinfo {author} {\bibfnamefont {R.~J.}\ \bibnamefont
  {Harrison}},\ }\bibfield  {title} {\enquote {\bibinfo {title} {{Electron
  affinities of the first-row atoms revisited. Systematic basis sets and wave
  functions}},}\ }\href {\doibase 10.1063/1.462569} {\bibfield  {journal}
  {\bibinfo  {journal} {J. Chem. Phys.}\ }\textbf {\bibinfo {volume} {96}},\
  \bibinfo {pages} {6796--6806} (\bibinfo {year} {1992})}\BibitemShut {NoStop}%
\bibitem [{\citenamefont {Prascher}\ \emph {et~al.}(2011)\citenamefont
  {Prascher}, \citenamefont {Woon}, \citenamefont {Peterson}, \citenamefont
  {Dunning},\ and\ \citenamefont {Wilson}}]{prascher2011a}%
  \BibitemOpen
  \bibfield  {author} {\bibinfo {author} {\bibfnamefont {B.~P.}\ \bibnamefont
  {Prascher}}, \bibinfo {author} {\bibfnamefont {D.~E.}\ \bibnamefont {Woon}},
  \bibinfo {author} {\bibfnamefont {K.~A.}\ \bibnamefont {Peterson}}, \bibinfo
  {author} {\bibfnamefont {T.~H.}\ \bibnamefont {Dunning}}, \ and\ \bibinfo
  {author} {\bibfnamefont {A.~K.}\ \bibnamefont {Wilson}},\ }\bibfield  {title}
  {\enquote {\bibinfo {title} {{Gaussian basis sets for use in correlated
  molecular calculations. VII. Valence, core-valence, and scalar relativistic
  basis sets for Li, Be, Na, and Mg}},}\ }\href {\doibase
  10.1007/s00214-010-0764-0} {\bibfield  {journal} {\bibinfo  {journal} {Theor.
  Chem. Acc.}\ }\textbf {\bibinfo {volume} {128}},\ \bibinfo {pages} {69--82}
  (\bibinfo {year} {2011})}\BibitemShut {NoStop}%
\bibitem [{\citenamefont {Kj{\o}nstad}\ and\ \citenamefont
  {Koch}(2019)}]{kjonstad2019orbital}%
  \BibitemOpen
  \bibfield  {author} {\bibinfo {author} {\bibfnamefont {E.~F.}\ \bibnamefont
  {Kj{\o}nstad}}\ and\ \bibinfo {author} {\bibfnamefont {H.}~\bibnamefont
  {Koch}},\ }\bibfield  {title} {\enquote {\bibinfo {title} {An orbital
  invariant similarity constrained coupled cluster model},}\ }\href {\doibase
  https://doi.org/10.1021/acs.jctc.9b00702} {\bibfield  {journal} {\bibinfo
  {journal} {J. Chem. Theory Comput.}\ }\textbf {\bibinfo {volume} {15}},\
  \bibinfo {pages} {5386--5397} (\bibinfo {year} {2019})}\BibitemShut {NoStop}%
\bibitem [{\citenamefont {Hoffmann}\ and\ \citenamefont
  {Simons}(1988)}]{Hoffmann1988}%
  \BibitemOpen
  \bibfield  {author} {\bibinfo {author} {\bibfnamefont {M.~R.}\ \bibnamefont
  {Hoffmann}}\ and\ \bibinfo {author} {\bibfnamefont {J.}~\bibnamefont
  {Simons}},\ }\bibfield  {title} {\enquote {\bibinfo {title} {A unitary
  multiconfigurational coupled-cluster method: Theory and applications},}\
  }\href {\doibase 10.1063/1.454125} {\bibfield  {journal} {\bibinfo  {journal}
  {J. Chem. Phys.}\ }\textbf {\bibinfo {volume} {88}},\ \bibinfo {pages}
  {993–1002} (\bibinfo {year} {1988})}\BibitemShut {NoStop}%
\bibitem [{\citenamefont {Kjønstad}, \citenamefont {Angelico},\ and\
  \citenamefont {Koch}(2024)}]{Kjnstad2024}%
  \BibitemOpen
  \bibfield  {author} {\bibinfo {author} {\bibfnamefont {E.~F.}\ \bibnamefont
  {Kjønstad}}, \bibinfo {author} {\bibfnamefont {S.}~\bibnamefont {Angelico}},
  \ and\ \bibinfo {author} {\bibfnamefont {H.}~\bibnamefont {Koch}},\
  }\bibfield  {title} {\enquote {\bibinfo {title} {Coupled cluster theory for
  nonadiabatic dynamics: Nuclear gradients and nonadiabatic couplings in
  similarity constrained coupled cluster theory},}\ }\href {\doibase
  10.1021/acs.jctc.4c00276} {\bibfield  {journal} {\bibinfo  {journal} {J.
  Chem. Theory Comput.}\ ,\ \bibinfo {pages} {7080–7092}} (\bibinfo {year}
  {2024})}\BibitemShut {NoStop}%
\bibitem [{\citenamefont {Stoll}\ \emph {et~al.}(2025)\citenamefont {Stoll},
  \citenamefont {Angelico}, \citenamefont {Kjønstad},\ and\ \citenamefont
  {Koch}}]{Stoll2025}%
  \BibitemOpen
  \bibfield  {author} {\bibinfo {author} {\bibfnamefont {L.}~\bibnamefont
  {Stoll}}, \bibinfo {author} {\bibfnamefont {S.}~\bibnamefont {Angelico}},
  \bibinfo {author} {\bibfnamefont {E.~F.}\ \bibnamefont {Kjønstad}}, \ and\
  \bibinfo {author} {\bibfnamefont {H.}~\bibnamefont {Koch}},\ }\bibfield
  {title} {\enquote {\bibinfo {title} {Similarity constrained {CC2} for
  efficient coupled cluster nonadiabatic dynamics},}\ }\href {\doibase
  https://arxiv.org/abs/2504.11157} {\  (\bibinfo {year} {2025}),\
  https://arxiv.org/abs/2504.11157}\BibitemShut {NoStop}%
\end{thebibliography}%

\end{document}

\newpage

\includepdf[pages={1,2,3}]{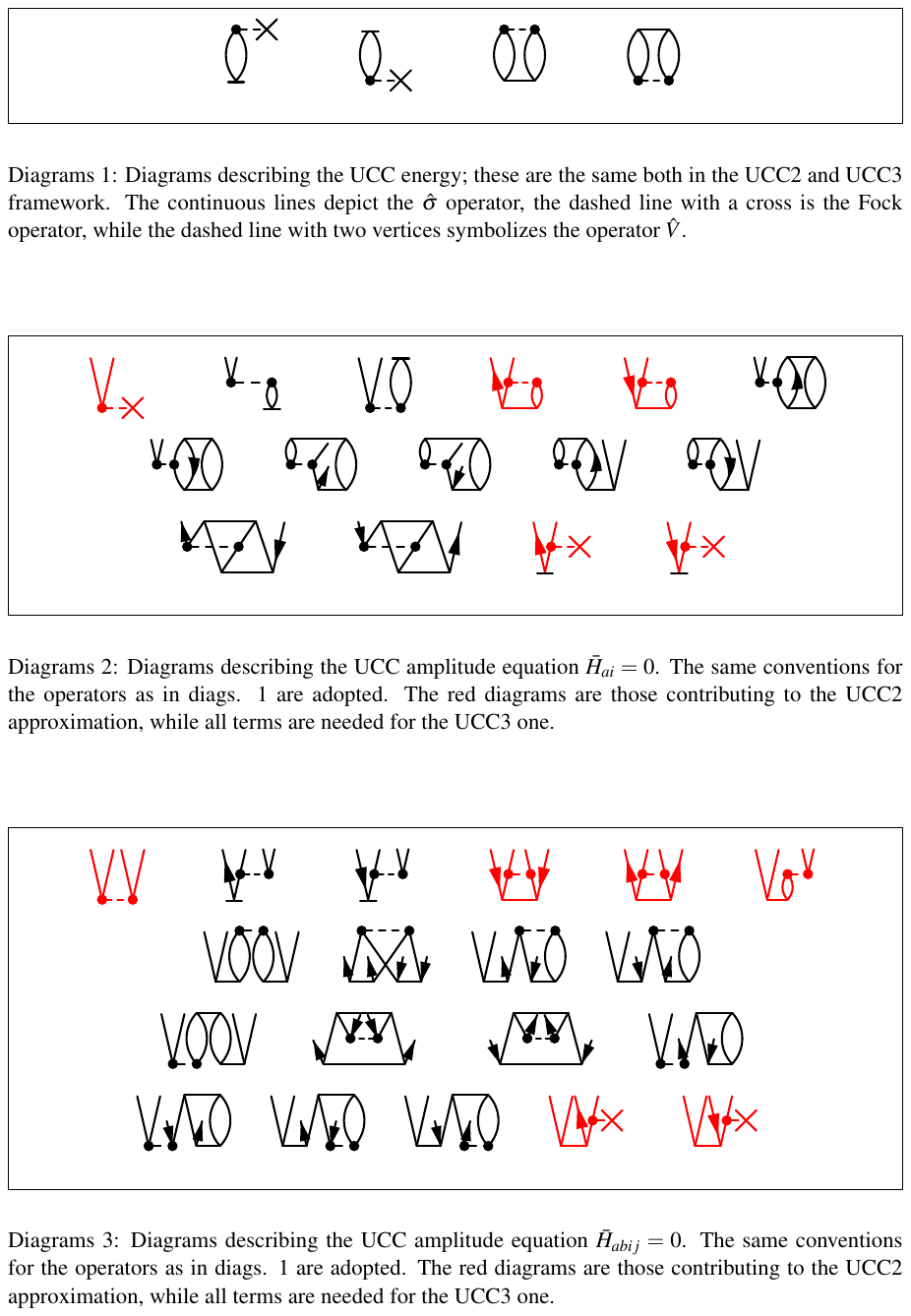}

\begin{figure*}
    \centering
    \begin{subfigure}{.45\columnwidth}
    \includegraphics[width=1.\linewidth]{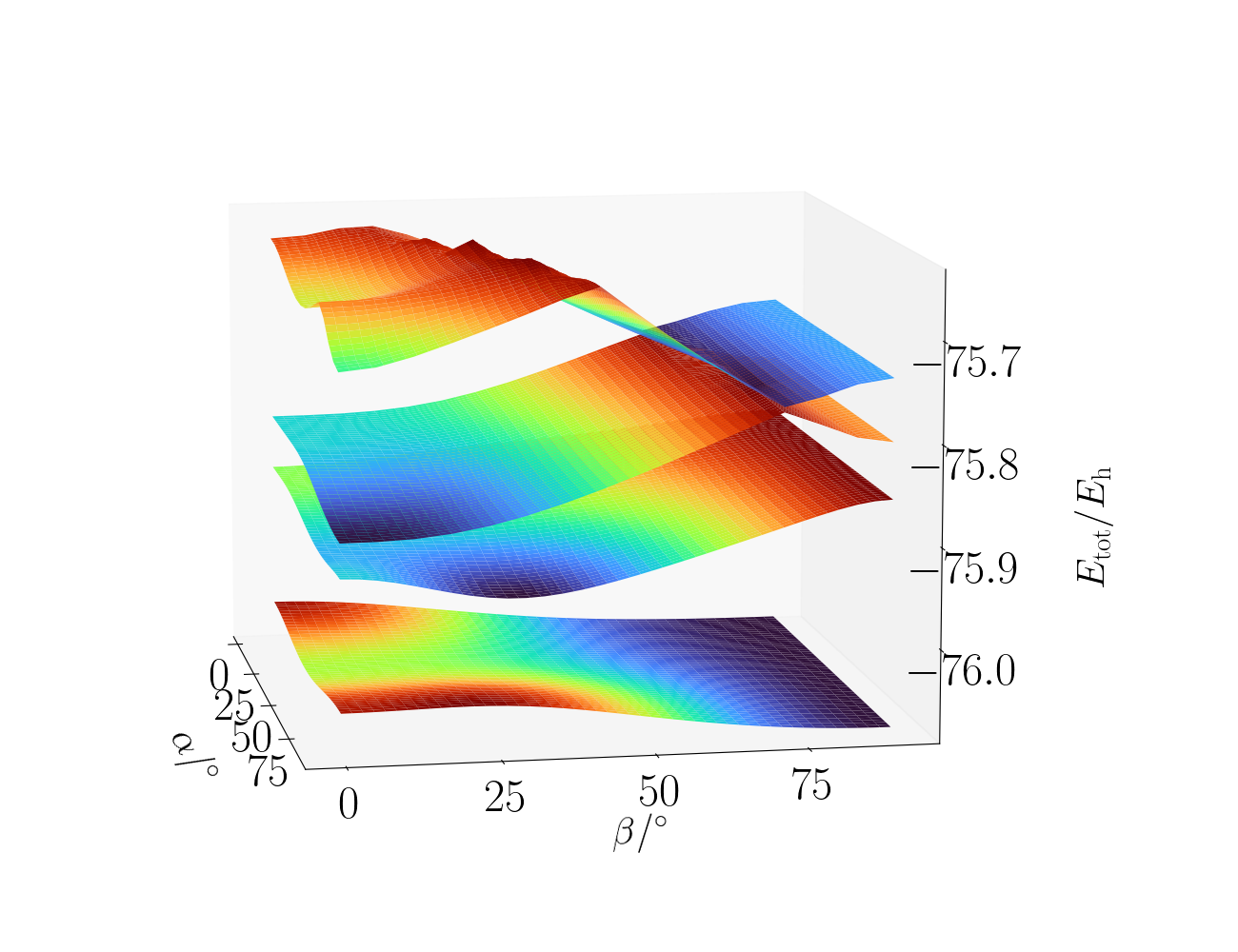}
    \caption{Four lowest-lying singlet states with CCSD.
        }
    \label{fig:ccsd_tot}
    \end{subfigure}
    \begin{subfigure}{.45\columnwidth}
    \includegraphics[width=1.\linewidth]{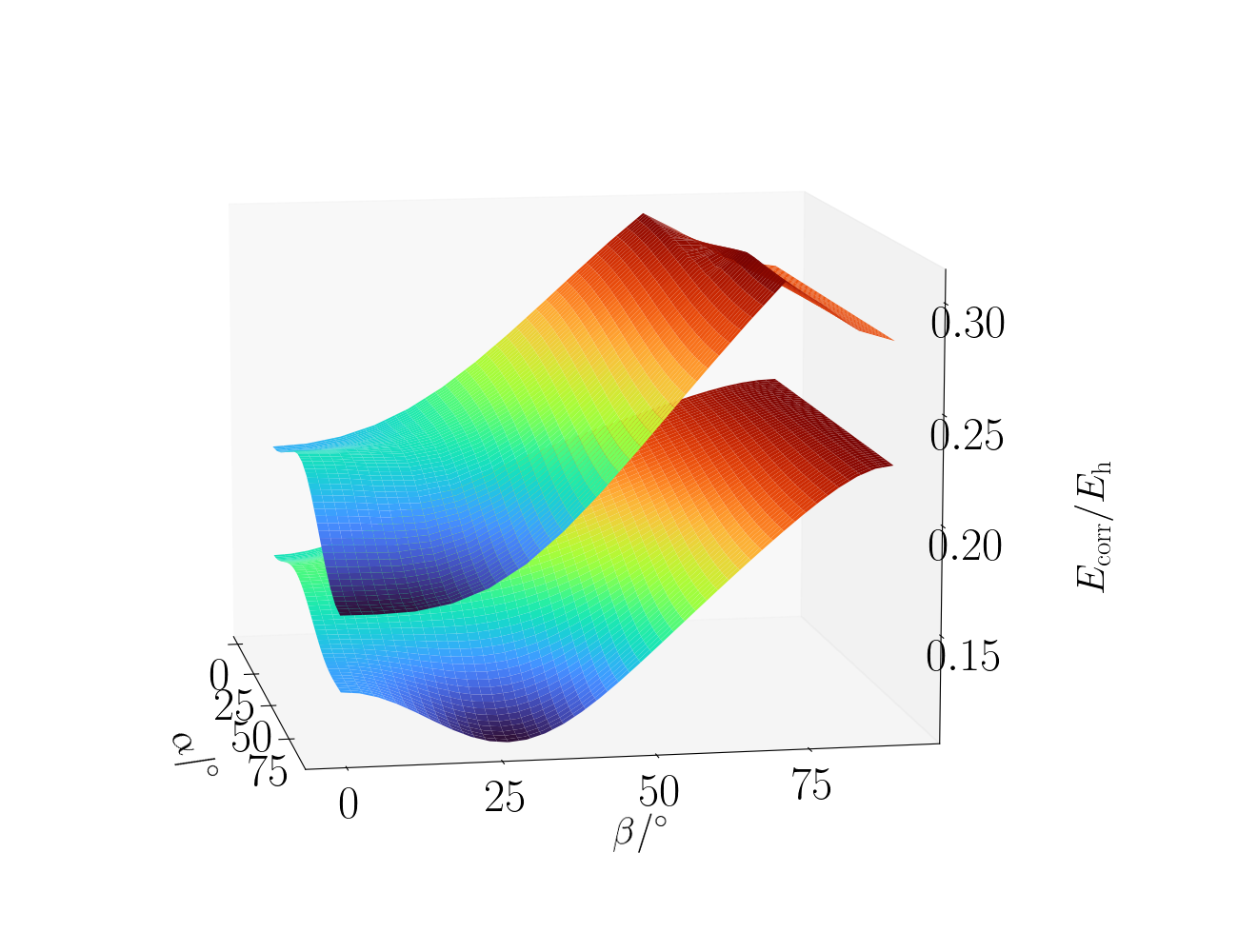}
    \caption{$\Psi_1$ and $\Psi_2$ with CCSD.
    }  
    \label{fig:ccsd_12}
    \end{subfigure}    
    \begin{subfigure}{.45\columnwidth}
    \includegraphics[width=1.\linewidth]{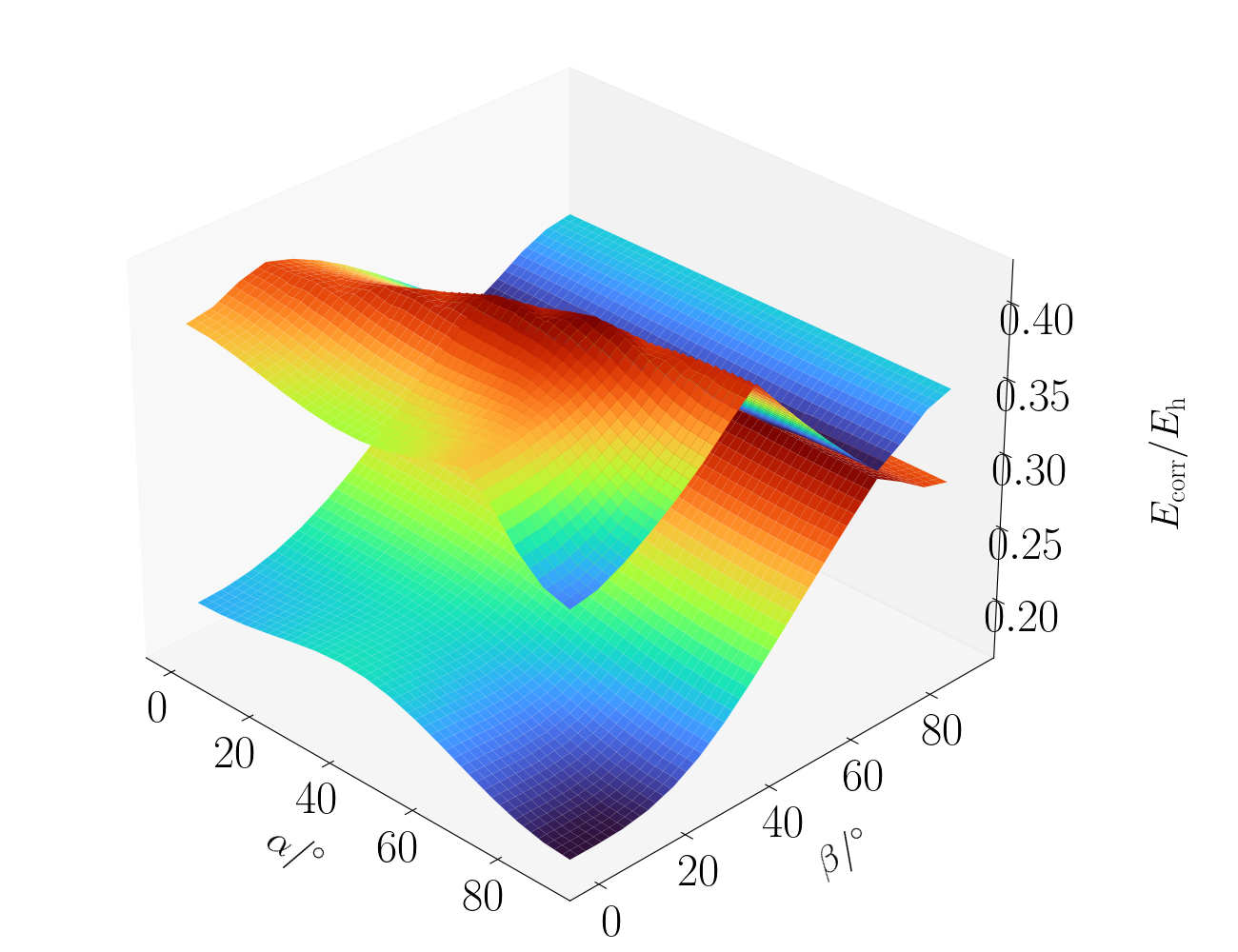}
    \caption{$\Psi_2$ and $\Psi_3$ with CCSD.
    }  
    \label{fig:ccsd_23}
    \end{subfigure}
    
    \caption{Real part of the energy surfaces of the ground and first three excited states of the water molecule in a magnetic field of B=0.5 B$_0$ as a function of its orientation, calculated at the ff-CCSD level of theory.}
    \label{fig:tot_h2o} 
\end{figure*}

\begin{table}[]
    \centering
\begin{tabular}{|l|r|r|r|r|}
    \hline
    i & CCSD          &            UCC3        &       CCSDT            &      UCC2\\
    \hline
         0    &        -38.0038356753   &       -38.0022837468   &     -38.0057417899     &    -38.0086394109\\
         1    &        -38.0020202151   &       -38.0004699229   &     -38.0039241590     &    -38.0068149783\\
         2    &        -37.9965831825   &       -37.9950377216   &     -37.9984806837     &    -38.0013515071\\
         3    &        -37.9875523495   &       -37.9860146922   &     -37.9894393363     &    -37.9922781333\\
         4    &        -37.9749731131   &       -37.9734458929   &     -37.9768458239     &    -37.9796423087\\
         5    &        -37.9589072204   &       -37.9573926560   &     -37.9607622825     &    -37.9635082534\\
         6    &        -37.9394310931   &       -37.9379309582   &     -37.9412655672     &    -37.9439549924\\
\hline 
    \end{tabular}
    \caption{Plotted points corresponding to the $1^1\Sigma^+/1^1\Sigma$ state in fig.~1a. The corresponding magnetic field is obtained by $B=0.05 \times i$.}
    \label{tab:alfa0_sigma}
\end{table}

         \begin{table}[]
    \centering
    \begin{tabular}{|l|r|r|r|r|}
    \hline
i   & CCSD            &          UCC3          &     CCSDT           &        UCC2\\
\hline
       0      &        -37.8846233169   &       -37.8871607984   &     -37.887682         &    -37.8711024891\\
       1      &        -37.8574438530   &       -37.8599808074   &     -37.860500         &    -37.8439119263\\
       2      &        -37.8259218131   &       -37.8284572110   &     -37.828974         &    -37.8123567087\\
       3      &        -37.7901056872   &       -37.7926385693   &     -37.793149         &    -37.7764856973\\
       4      &        -37.7500744313   &       -37.7526039425   &     -37.753108         &    -37.7363785092\\
       5      &        -37.7059347606   &       -37.7084601658   &     -37.708956         &    -37.6921428614\\
       6      &        -37.6578175305   &       -37.6603382082   &     -37.660824         &    -37.6439110021\\
\hline
    \end{tabular}
    \caption{Plotted points corresponding to the $1^1\Pi/1^1\Pi_{+}$ state in fig.~1a. The corresponding magnetic field is obtained by $B=0.05 \times i$.}
    \label{tab:alfa0_pi+}
\end{table}

       \begin{table}[]
    \centering
    \begin{tabular}{|l|r|r|r|r|}
    \hline
       i & CCSD            &          UCC3          &     CCSDT         &          UCC2\\
       \hline
       0      &        -37.8846233169   &       -37.8871607984   &     -37.8876813735     &    -37.8711024891\\
       1      &        -37.9074438530   &       -37.9099808074   &     -37.9105002556     &    -37.8939119263\\
       2      &        -37.9259218131   &       -37.9284572110   &     -37.9289740000     &    -37.9123567087\\
       3      &        -37.9401056872   &       -37.9426385693   &     -37.9431493519     &    -37.9264856973\\
       4      &        -37.9500744313   &       -37.9526039425   &     -37.9531076400     &    -37.9363785092\\
       5      &        -37.9559347606   &       -37.9584601658   &     -37.9589554083     &    -37.9421428614\\
       6      &        -37.9578175306   &       -37.9603382082   &     -37.9608240737     &    -37.9439110021\\
\hline
    \end{tabular}
    \caption{Plotted points corresponding to the $1^1\Pi/11\Pi_{-}$ state in fig.~1a. The corresponding magnetic field is obtained by $B=0.05 \times i$.}
    \label{tab:alfa0_pi-}
\end{table}

\begin{table}[]
    \centering
    \begin{tabular}{|l|r|r|r|r|}
    \hline
       i & CCSD            &          UCC3          &     CCSDT             &      UCC2 \\
       \hline
        0     &        -37.7151563326   &       -37.8051510322   &     -37.7501246972     &    \\
        1     &        -37.7625611120   &       -37.8525505314   &     -37.7975382891     &    \\
        2     &        -37.8047979528   &       -37.8947714683   &     -37.8398014280     &    \\
        3     &        -37.8419336833   &       -37.9318804539   &     -37.8769805049     &    \\
        4     &        -37.8740774319   &       -37.9639861703   &     -37.9091838611     &    \\
        5     &        -37.9013772879   &       -37.9912359258   &     -37.9365583717     &    \\
        6     &        -37.9240157228   &       -38.0138109612   &     -37.9592847783     &    \\
\hline
    \end{tabular}
   \caption{Plotted points corresponding to the $1^1\Delta/1^1\Delta_{-2}$ state in fig.~1a. The corresponding magnetic field is obtained by $B=0.05 \times i$.}
    \label{tab:alfa0_delta}
\end{table}

\begin{table}[]
    \centering
    \begin{tabular}{|l|r|r|r|r|}
    \hline
i & CCSD&UCC3&CCSDT&UCC2\\ \hline
         0     &      -38.0038356753   &         -38.0022837486 &   -38.005742    &     -38.0086394109 \\
         1     &      -38.0045018476   &         -38.0038013492 &   -38.006431    &     -38.0086599805 \\
         2     &      -38.0067938883   &         -38.0066442067 &   -38.008773    &     -38.0101095359 \\
         3     &      -38.0112521485   &         -38.0108956679 &   -38.013251    &     -38.0142038636 \\
         4     &      -38.0178571136   &         -38.0172247069 &   -38.019814    &     -38.0206579113 \\
         5     &      -38.0257389289   &         -38.0249694083 &   -38.027621    &     -38.0283942941 \\
         6     &      -38.0336686385   &         -38.0328522575 &   -38.035485    &     -38.0361721918 \\
\hline
    \end{tabular}
    \caption{Plotted points corresponding to the $1^1\Sigma^+/1^1A$ state in fig.~1b. The corresponding magnetic field is obtained by $B=0.05 \times i$.}
    \label{tab:Alfa30}
\end{table}

\begin{table}[]
    \centering
    \begin{tabular}{|l|r|r|r|r|}
    \hline
i & CCSD&UCC3&CCSDT&UCC2\\ \hline
        0      &      -37.8846233187   &         -37.8871608000 &   -37.887682    &     -37.8711024891 \\
        1      &      -37.9039161615   &         -37.9107559276 &   -37.906770    &     -37.8880628562 \\
        2      &      -37.9181158361   &         -37.9265242594 &   -37.920624    &     -37.8988140547 \\
        3      &      -37.9273653986   &         -37.9325485728 &   -37.929552    &     -37.9067264734 \\
        4      &      -37.9329939957   &         -37.9355591854 &   -37.934845    &     -37.9129183606 \\
        5      &      -37.9363174442   &         -37.9380534303 &   -37.937971    &     -37.9172299018 \\
        6      &      -37.9373739052   &         -37.9390533253 &   -37.939052    &     -37.9189718909 \\
\hline
    \end{tabular}
    \caption{Plotted points corresponding to the $1^1\Pi/2^1A$ state in fig.~1b. The corresponding magnetic field is obtained by $B=0.05 \times i$.}
    \label{tab:Alfa30}
\end{table}

\begin{table}[]
    \centering
    \begin{tabular}{|l|r|r|r|r|}
    \hline
i  & CCSD&UCC3&CCSDT&UCC2\\ \hline
          0    &      -37.8846233187   &         -37.8871608000 &   -37.887682    &     -37.8711024891 \\
          1    &      -37.8608133391   &         -37.8679647809 &   -37.863614    &     -37.8444910812 \\
          2    &      -37.8328643179   &         -37.8741224319 &   -37.835154    &     -37.8107376858 \\
          3    &      -37.8091320000   &         -37.8919530235 &   -37.845509    &     -37.7736617939 \\
          4    &      -37.8201548799   &         -37.8985644290 &   -37.858160    &     -37.7353215862 \\
          5    &      -37.8201613534   &         -37.8948746608 &   -37.859593    &     -37.6964945128 \\
          6    &      -37.8104888730   &         -37.8827368547 &   -37.851124    &     -37.6571472925 \\
\hline
    \end{tabular}
   \caption{Plotted points corresponding to the $1^1\Pi/3^1A$ state in fig.~1b. The corresponding magnetic field is obtained by $B=0.05 \times i$.}
    \label{tab:Alfa30}
\end{table}

\begin{table}[]
    \centering
    \begin{tabular}{|l|r|r|r|r|}
    \hline
i  & CCSD&UCC3&CCSDT&UCC2\\ \hline
        0      &      -37.7151563347   &         -37.8051510340 &   -37.750125  &  \\
        1      &      -37.7547435421   &         -37.8446625993 &   -37.789986  &  \\
        2      &      -37.7863631152   &         -37.8436559483 &   -37.822195  &  \\
        3      &      -37.8003090626   &         -37.8106840839 &   -37.802704  &  \\
        4      &      -37.7645135619   &         -37.7731107163 &   -37.766868  &  \\
        5      &      -37.7248950112   &         -37.7329093458 &   -37.727896  &  \\
        6      &      -37.6824868177   &         -37.6901932810 &   -37.686209  &  \\
        7      &      -37.6381577970   &         -37.6458560740 &   -37.642563  &  \\
        8      &      -37.6297480976   &         -37.6466231702 &   -37.638204  &  \\
        9      &      -37.6373438945   &         -37.6542551874 &   -37.645814  &  \\
        10     &      -37.6407400022   &         -37.6584346977 &   -37.649703  &  \\
        11     &      -37.6383888127   &         -37.6589727401 &   -37.649508  &  \\
        12     &      -37.6237750589   &         -37.6554077145 &   -37.643911  &  \\
        13     &      -37.5919123843   &         -37.6450077730 &   -37.627289  &  \\
        14     &      -37.5519787993   &         -37.6170877766 &   -37.594183  &  \\
        15     &      -37.5084251508   &         -37.5753940196 &   -37.552583  &  \\
        16     &      -37.4624612435   &         -37.5293553393 &   -37.507329  &  \\
        17     &      -37.4147230359   &         -37.4810420210 &   -37.459851  &  \\
        18     &      -37.3665160594   &         -37.4313202082 &   -37.411103  &  \\
        19     &      -37.3353882121   &         -37.3815142842 &   -37.363662  &  \\
        20     &      -37.3131717822   &         -37.3372248148 &   -37.328846  &  \\
    \hline
    \end{tabular}
    \caption{Plotted points corresponding to the $1^1\Delta/4^1A$ state in fig.~1b. The corresponding magnetic field is obtained by $B=0.05 \times i$.}
    \label{tab:Alfa30}
\end{table}

\begin{table}[]
    \centering
    \begin{tabular}{|l|r|r|r|r|}
    \hline
    i &   CCSD &UCC3 &CCSDT &UCC2\\ \hline
       0           &    -38.0038356753  &      -38.0022837468  &     -38.005742&        -38.0086394498\\
       1           &    -38.0090876946  &      -38.0084379458  &     -38.011051&        -38.0127453444\\
       2           &    -38.0222167876  &      -38.0213423942  &     -38.024221&        -38.0253563020\\
       3           &    -38.0386027837  &      -38.0376066831  &     -38.040571&        -38.0416188217\\
       4           &    -38.0549147043  &      -38.0538948265  &     -38.056826&        -38.0578563555\\
       5           &    -38.0692900974  &      -38.0682745367  &     -38.071150&        -38.0721657473\\
       6           &    -38.0807453786  &      -38.0797401891  &     -38.082565&        -38.0835621008\\
\hline
    \end{tabular}
    \caption{Plotted points corresponding to the $1^1\Sigma^+/1^1A$ state in fig.~1c. The corresponding magnetic field is obtained by $B=0.05 \times i$.}
    \label{tab:Alfa60}
\end{table}

\begin{table}[]
    \centering
    \begin{tabular}{|l|r|r|r|r|}
    \hline
    i &CCSD &UCC3 &CCSDT &UCC2\\ \hline
       0           &    -37.8846233169  &      -37.8871607982  &     -37.887682&       -37.8711024965\\
       1           &    -37.8949030678  &      -37.9018407787  &     -37.897550&       -37.8774880559\\
       2           &    -37.9018449357  &      -37.9059606423  &     -37.904094&       -37.8827532773\\
       3           &    -37.9067816940  &      -37.9092772836  &     -37.908748&       -37.8880234305\\
       4           &    -37.9096096743  &      -37.9114778092  &     -37.911390&       -37.8914707527\\
       5           &    -37.9097113093  &      -37.9113501593  &     -37.911388&       -37.8920556519\\
       6           &    -37.9066230473  &      -37.9081882843  &     -37.908250&       -37.8892934820\\
\hline
    \end{tabular}
    \caption{Plotted points corresponding to the $1^1\Pi/2^1A$ state in fig.~1c. The corresponding magnetic field is obtained by $B=0.05 \times i$.}
    \label{tab:Alfa60}
\end{table}

\begin{table}[]
    \centering
    \begin{tabular}{|l|r|r|r|r|}
    \hline
    i& CCSD &UCC3 &CCSDT &UCC2\\ \hline
       0           &    -37.8846233169  &      -37.8871607982  &     -37.887682&      -37.8711024965\\
       1           &    -37.8695596836  &      -37.8776248549  &     -37.871981&      -37.8494719963\\
       2           &    -37.8491172800  &      -37.8561681000  &     -37.850792&      -37.8224092316\\
       3           &    -37.8246039140  &      -37.8325654437  &     -37.826302&      -37.7952524410\\
       4           &    -37.7978496631  &      -37.8154538550  &     -37.801317&      -37.7670219107\\
       5           &    -37.7700356854  &      -37.7985298927  &     -37.778284&      -37.7367442442\\
       6           &    -37.7418239734  &      -37.7785126423  &     -37.756241&      -37.7040744678\\
\hline
    \end{tabular}
    \caption{Plotted points corresponding to the $1^1\Delta/4^1A$ state in fig.~1c. The corresponding magnetic field is obtained by $B=0.05 \times i$.}
    \label{tab:Alfa60}
\end{table}

\begin{table}[]
    \centering
    \begin{tabular}{|l|r|r|r|r|}
    \hline
    i& CCSD &UCC3 &CCSDT &UCC2\\ \hline
       0           &    -37.7151563329  &      -37.8051510322  &     -37.750125& \\
       1           &    -37.7344707001  &      -37.8226286122  &     -37.770132& \\
       2           &    -37.7425469435  &      -37.8250359079  &     -37.779421& \\
       3           &    -37.7400254660  &      -37.8154538550  &     -37.777966& \\
       4           &    -37.7283754587  &      -37.7904752575  &     -37.765829& \\
       5           &    -37.7091929613  &      -37.7575647355  &     -37.742966& \\
       6           &    -37.6838704576  &      -37.7215537069  &     -37.712158& \\
       7           &    -37.6538464151  &      -37.6838513950  &     -37.677100& \\
       8           &    -37.624980,     &      -37.6473176177  &     -37.640987& \\
       9           &    -37.5972988790  &      -37.6031077769  &     -37.613686& \\
       10          &    -37.5789222111  &      -37.5818659871  &     -37.598240& \\
       11          &    -37.5628697046  &      -37.5680592702  &     -37.575404& \\
       12          &    -37.5508375666  &      -37.5558572523  &     -37.558485& \\
       13          &    -37.5380840921  &      -37.5461275974  &     -37.545679& \\
       14          &    -37.5234189071  &      -37.5319743442  &     -37.531374& \\
       15          &    -37.5065462366  &      -37.5165100517  &     -37.515078& \\
       16          &    -37.4873154488  &      -37.4992560105  &     -37.496585& \\
       17          &    -37.4655838477  &      -37.4800114290  &     -37.475702& \\
       18          &    -37.4412300274  &      -37.4585625386  &     -37.452239& \\
       19          &    -37.4141995272  &      -37.4346779686  &     -37.426067& \\
       20          &    -37.3845365243  &      -37.4081349226  &     -37.397162& \\
         \hline
    \end{tabular}
    \caption{Plotted points corresponding to the $1^1\Pi/1^1\Pi_{+}$ state in fig.~1c. The corresponding magnetic field is obtained by $B=0.05 \times i$.}
    \label{tab:Alfa60}
\end{table}

\begin{table}[]
    \centering
    \begin{tabular}{|l|r|r|r|r|}
    \hline
    i &CCSD&UCC3&CCSDT&UCC2\\ \hline
 0     &      -38.0038356753    &    -38.0022837468  &     -38.005742 &    -38.0086394109 \\
 1     &      -38.0112243885    &    -38.0105148921  &     -38.013200 &    -38.0147641157 \\
 2     &      -38.0285131804    &    -38.0275470735  &     -38.030514 &    -38.0316423963 \\
 3     &      -38.0486202124    &    -38.0475786139  &     -38.050576 &    -38.0516607098 \\
 4     &      -38.0676108749    &    -38.0665633732  &     -38.069512 &    -38.0705987780 \\
 5     &      -38.0836962895    &    -38.0826570109  &     -38.085552 &    -38.0866412842 \\
 6     &      -38.0960746065    &    -38.0950448623  &     -38.097895 &    -38.0989834503 \\
\hline
    \end{tabular}
   \caption{Plotted points corresponding to the $1^1\Sigma^+/1^1A^\prime$ state in fig.~1d. The corresponding magnetic field is obtained by $B=0.05 \times i$.}
    \label{tab:Alfa90}
\end{table}

\begin{table}[]
    \centering
    \begin{tabular}{|l|r|r|r|r|}
    \hline
    i &CCSD&UCC3&CCSDT&UCC2\\ \hline
 0     &      -37.8846233169    &    -37.8871607982  &     -37.887682 &    -37.8711024891 \\
 1     &      -37.8861002182    &    -37.8902746641  &     -37.888980 &    -37.8705673920 \\
 2     &      -37.8894390248    &    -37.8923109636  &     -37.891953 &    -37.8730516482 \\
 3     &      -37.8924595066    &    -37.8945877828  &     -37.894653 &    -37.8760862763 \\
 4     &      -37.8934939860    &    -37.8952598175  &     -37.895464 &    -37.8773105660 \\
 5     &      -37.8916047070    &    -37.8931778116  &     -37.893425 &    -37.8756235982 \\
 6     &      -37.8863343689    &    -37.8877949221  &     -37.888051 &    -37.8705389307 \\
\hline
    \end{tabular}
    \caption{Plotted points corresponding to the $1^1\Pi/1^1A^{\prime\prime}$ state in fig.~1d. The corresponding magnetic field is obtained by $B=0.05 \times i$.}
    \label{tab:Alfa90}
\end{table}

\begin{table}[]
    \centering
    \begin{tabular}{|l|r|r|r|r|}
    \hline
    i &CCSD&UCC3&CCSDT&UCC2\\ \hline
 0     &      -37.8846233169    &    -37.8871607982  &     -37.887682 &    -37.8711024891 \\
 1     &      -37.8782692913    &    -37.8884838588  &     -37.880252 &    -37.8547059879 \\
 2     &      -37.8619096784    &    -37.8686589300  &     -37.863078 &    -37.8319465408 \\
 3     &      -37.8405687177    &    -37.8455108491  &     -37.841618 &    -37.8093057200 \\
 4     &      -37.8163047964    &    -37.8210039705  &     -37.817594 &    -37.7849308235 \\
 5     &      -37.7894501423    &    -37.7947787803  &     -37.791157 &    -37.7578249887 \\
 6     &      -37.7602801960    &    -37.7667507053  &     -37.762500 &    -37.7279763440 \\
\hline
    \end{tabular}
   \caption{Plotted points corresponding to the $1^1\Pi/2^1A^{\prime}$ state in fig.~1d. The corresponding magnetic field is obtained by $B=0.05 \times i$.}
    \label{tab:Alfa90}
\end{table}

\begin{table}[]
    \centering
    \begin{tabular}{|l|r|r|r|r|}
    \hline
    i &CCSD&UCC3&CCSDT&UCC2\\\hline
 0     &      -37.7151563329    &    -37.8051510322  &     -37.750125 & \\
 1     &      -37.7087625689    &    -37.7959564961  &     -37.744673 & \\
 2     &      -37.6916348748    &    -37.7725517117  &     -37.729849 & \\
 3     &      -37.6699589709    &    -37.7467538382  &     -37.710922 & \\
 4     &      -37.6490850179    &    -37.7229161829  &     -37.691239 & \\
 5     &      -37.6279605875    &    -37.6993952209  &     -37.670166 & \\
 6     &      -37.6049519624    &    -37.6743087311  &     -37.646764 & \\
 7     &      -37.5824979485    &    -37.6476552852  &     -37.621581 & \\
 8     &      -37.5800011860    &    -37.6226168533  &     -37.600374 & \\
 9     &      -37.5709294618    &    -37.6030775070  &     -37.585637 & \\
 10    &      -37.5478228414    &    -37.5813606651  &     -37.562949 & \\
 11    &      -37.5142776881    &    -37.5524494839  &     -37.531052 & \\
 12    &      -37.4765307489    &    -37.5187592807  &     -37.494404 & \\
 13    &      -37.4415637865    &    -37.4830350029  &     -37.456545 & \\
 14    &      -37.4170677924    &    -37.4471543161  &     -37.422952 & \\
 15    &      -37.3961410320    &    -37.4128594655  &     -37.399304 & \\
 16    &      -37.3742417092    &    -37.3831275758  &     -37.377190 & \\
 17    &      -37.3507706781    &    -37.3577312314  &     -37.353831 & \\
 18    &      -37.3256669873    &    -37.3323991139  &     -37.328924 & \\
 19    &      -37.2989056714    &    -37.3058419613  &     -37.302401 & \\
 20    &      -37.2704481746    &    -37.2777987158  &     -37.274212 & \\
    \hline
    \end{tabular}
   \caption{Plotted points corresponding to the $1^1\Delta/3^1A^\prime$ state in fig.~1d. The corresponding magnetic field is obtained by $B=0.05 \times i$.}
    \label{tab:Alfa90}
\end{table}

\begin{table}[]
    \centering
    \begin{tabular}{|l|r|r|r|r|r|r|r|}
    \hline
     B & $^1A^\prime$ & $^1E_{1}^\prime$ & $^1E_{2}^\prime$ & $^1A^{\prime\prime}$ & $^1E_{1}^{\prime\prime}$ & $^1E_{2}^{\prime\prime}$ \\ \hline
  0.00 &   -252.1615732  &    0.3541826   &   0.3622465     &  0.3622465   &    0.3311491   &     0.2989578   &   0.2989578\\
  0.05 &   -252.1512566  &    0.3580695   &   0.3667007     &  0.3564556   &    0.3262937   &     0.3045107   &   0.3031327\\
  0.10 &   -252.1202673  &    0.3607525   &   0.3560856     &  0.3532824   &    0.3253830   &     0.3114830   &   0.3148048\\
  0.15 &   -252.0686536  &    0.3518518   &   0.3377184     &  0.3463282   &    0.3275221   &     0.3138453   &   0.3271029\\
  0.20 &   -251.9969309  &    0.3317732   &   0.3148731     &  0.3342740   &    0.3247675   &     0.3145165   &   0.3367404\\
  0.25 &   -251.9059170  &    0.2959661   &   0.2930608     &  0.3186232   &    0.3142848   &     0.3183387   &   0.3152947\\
  0.30 &   -251.7965939  &    0.2524310   &   0.2711955     &  0.2958935   &    0.2982821   &     0.3247321   &   0.2826686\\
\hline
    \end{tabular}
    \caption{Real part of the energies of the ground and excited states of B(OH)$_3$, at the CC3 level of theory, plotted in figs.~8-9.}
    \label{tab:Boh3_cc3}
\end{table}

\begin{table}[]
    \centering
    \begin{tabular}{|l|r|r|r|r|r|r|r|}
    \hline
     B& $^1A^\prime$ & $^1E_{1}^\prime$ & $^1E_{2}^\prime$ & $^1A^{\prime\prime}$ & $^1E_{1}^{\prime\prime}$ & $^1E_{2}^{\prime\prime}$ \\ \hline
  0.00  &   -252.1338335  &    0.3563775   &   0.3643064     &  0.3643064   &    0.3325524   &     0.3018311   &   0.3018311 \\
  0.05  &   -252.1235252  &    0.3600837   &   0.3680537     &  0.3585482   &    0.3278731   &     0.3068540   &   0.3062639 \\
  0.10  &   -252.0925679  &    0.3620730   &   0.3581307     &  0.3552980   &    0.3272641   &     0.3138190   &   0.3178197 \\
  0.15  &   -252.0410165  &    0.3533838   &   0.3408729     &  0.3482647   &    0.3299106   &     0.3168528   &   0.3296974 \\
  0.20  &   -251.969379   &    0.3342771   &   0.3184873     &  0.3362435   &    0.3274795   &     0.3180631   &   0.3390395 \\
  0.25  &   -251.8784594  &    0.2997759   &   0.2965897     &  0.3206986   &    0.3175134   &     0.3220579   &   0.3193303 \\
  0.30  &   -251.7692329  &    0.2572341   &   0.2744005     &  0.2983150   &    0.3024007   &     0.3281348   &   0.2867486 \\
\hline
    \end{tabular}
    \caption{Real part of the energies of the ground and excited states of B(OH)$_3$, at the CCSD level of theory, plotted in figs.~8-9.}
    \label{tab:Boh3_ccsd}
\end{table}

\begin{table}[]
    \centering
    \begin{tabular}{|l|r|r|r|r|r|r|r|}
    \hline
    B & $^1A^\prime$ & $^1E_{1}^\prime$ & $^1E_{2}^\prime$ & $^1A^{\prime\prime}$ & $^1E_{1}^{\prime\prime}$ & $^1E_{2}^{\prime\prime}$ \\ \hline
  0.00 &   -252.12968    &    0.3665601   &   0.3737871     & 0.3737871   &    0.3388748   &    0.3081290   & 0.3081290 \\
  0.05 &   -252.11938    &    0.3696815   &   0.3771565     & 0.3677284   &    0.3341629   &    0.3127302   & 0.3124876 \\
  0.10 &   -252.08842    &    0.3703942   &   0.3662751     & 0.3635343   &    0.3327917   &    0.3190020   & 0.3236196 \\
  0.15 &   -252.03687    &    0.3604577   &   0.3479972     & 0.3554441   &    0.3342886   &    0.3213822   & 0.3347876 \\
  0.20 &   -251.96524    &    0.3402480   &   0.3247925     & 0.3422714   &    0.3310593   &    0.3219120   & 0.3428780 \\
  0.25 &   -251.87432    &    0.3052962   &   0.3017865     & 0.3253862   &    0.3206898   &    0.3249537   & 0.3220260 \\
  0.30 &   -251.76509    &    0.2620728   &   0.2783526     & 0.3018212   &    0.3052415   &    0.3298871   & 0.2893802 \\
\hline
    \end{tabular}
    \caption{Energies of the ground and excited states of B(OH)$_3$, at the UCC3 level of theory, plotted in figs.~8-9.}
    \label{tab:Boh3_ucc3}
\end{table}

\begin{table}[]
    \centering
    \begin{tabular}{|l|r|r|r|r|r|r|r|}
    \hline
     B& $^1A^\prime$ & $^1E_{1}^\prime$ & $^1E_{2}^\prime$ & $^1A^{\prime\prime}$ & $^1E_{1}^{\prime\prime}$ & $^1E_{2}^{\prime\prime}$ \\ \hline
  0.00  &   -252.1338335  &    0.3563775   &  0.3643064     & 0.36430643    &   0.3325524    &    0.3018311   &  0.3018311 \\
  0.05  &   -252.1235974  &    0.3601461   &  0.3685908     & 0.35780056    &   0.3283769    &    0.3070667   &  0.3052452 \\
  0.10  &   -252.0928975  &    0.3629264   &  0.3573669     & 0.35368339    &   0.3272861    &    0.3130295   &  0.3158459 \\
  0.15  &   -252.0418074  &    0.3539828   &  0.3385164     & 0.34645595    &   0.3289071    &    0.3148656   &  0.3276228 \\
  0.20  &   -251.9707959  &    0.3332930   &  0.3153941     & 0.33487801    &   0.3256802    &    0.3153212   &  0.3286663 \\
  0.25  &   -251.8806127  &    0.2973552   &  0.2930715     & 0.31991674    &   0.3148537    &    0.3186736   &  0.2978657 \\
  0.30  &   -251.7721939  &    0.2537368   &  0.2713456     & 0.29815431    &   0.2989116    &    0.3246462   &  0.2659881 \\
\hline
    \end{tabular}
    \caption{Energies of the ground and excited states of B(OH)$_3$, at the CISD level of theory, plotted in figs.~8-9.}
    \label{tab:Boh3_cisd}
\end{table}

\begin{sidewaystable}
    \centering
    \begin{tabular}{|l|r|r|r|r|r|r|r|}
    \hline
 B & $^1A^\prime$ & $^1E_{1}^\prime$ & $^1E_{2}^\prime$ & $^1A^{\prime\prime}$ & $^1E_{1}^{\prime\prime}$ & $^1E_{2}^{\prime\prime}$ \\
 \hline
   0.00 &        0.000000000    &      0.000000000      &  3.45180E-06      &   -3.45330E-06  &    -5E-10          &   -1.190740E-05     &  1.190740E-05  \\
   0.05 &        0.000001095    &      -1.03408E-05     &  3.55369E-06      &   1.28712E-05   &    6.78483E-06     &   9.858962E-06      &  -1.095680E-05 \\
   0.10 &        0.000002108    &      -1.09544E-05     &  6.08940E-06      &   1.43772E-05   &    2.1114E-06      &   3.589400E-06      &  -5.079300E-06 \\
   0.15 &        0.000003080    &      -1.44628E-05     &  7.18550E-06      &   1.01040E-05   &    -5.4302E-06     &   -6.276000E-06     &  2.294100E-06  \\
   0.20 &        0.000004154    &      -3.36426E-05     &  8.78800E-06      &   4.93150E-06   &    -1.80657E-05    &   -1.652190E-05     &  1.140900E-06  \\
   0.25 &        0.000005458    &      -3.69304E-05     &  7.21630E-06      &   -2.38800E-06  &    -2.94385E-05    &   -2.482420E-05     &  5.316430E-05  \\
   0.30 &        0.000006975    &      -3.47217E-05     &  4.33000E-08      &   -1.58484E-05  &    -3.54342E-05    &   -3.387290E-05     &  5.679730E-05  \\
\hline
    \end{tabular}
   \caption{Imaginary part of the energies of the ground and excited states of B(OH)$_3$, at the CC3 level of theory, plotted in figs.~8-9.}
    \label{tab:Boh3_im_cc3}
\end{sidewaystable}

\begin{sidewaystable}
    \centering
    \begin{tabular}{|l|r|r|r|r|r|r|r|}
    \hline
        B& $^1A^\prime$ & $^1E_{1}^\prime$ & $^1E_{2}^\prime$ & $^1A^{\prime\prime}$ & $^1E_{1}^{\prime\prime}$ & $^1E_{2}^{\prime\prime}$ \\  
        \hline
   0.00 &        0.00000000     &      0.000000000      &   4.66578E-05     &   -4.66578E-05  &    0.000000000     &    5.13331E-05      &  -0.0000513331 \\  
   0.05 &        -1.5163E-06    &      -2.91564E-05     &   5.62757E-05     &    9.4869E-06   &     5.62188E-05    &    4.88148E-05      &  -0.0000384837 \\
   0.10 &        -3.1322E-06    &       0.000030696     &   0.000094154     &    7.97992E-05  &     6.24372E-05    &    4.46173E-05      &   0.0000015281 \\  
   0.15 &        -4.8396E-06    &       9.87508E-05     &   0.000136967     &    0.000157409  &     5.42475E-05    &    3.85138E-05      &   0.0000659032 \\
   0.20 &        -6.6378E-06    &       0.00010224      &   0.000167311     &    0.000213847  &     3.02004E-05    &    3.85667E-05      &   0.0001122227 \\  
   0.25 &        -8.7336E-06    &       8.08258E-05     &   0.00020155      &    0.000241509  &     2.09522E-05    &    5.57919E-05      &  -0.0001357661 \\
   0.30 &        -1.11985E-05   &       8.15121E-05     &   0.000221318     &    0.000230545  &     4.11463E-05    &    8.21586E-05      &  -0.0001288078 \\
\hline
    \end{tabular}
    \caption{Imaginary part of the energies of the ground and excited states of B(OH)$_3$, at the CCSD level of theory, plotted in figs.~8-9.}
    \label{tab:Boh3_im_ccsd}
\end{sidewaystable}